%% file: vision.tex
\ifx\deliverable\undefined
\documentclass[a4paper,twoside,twocolumn,DIV=17,BCOR=5mm]{scrreprt}
\else
\documentclass[a4paper,headheight=20mm]{scrreprt}
\usepackage[left=30mm,right=30mm,top=34mm,bottom=30mm]{geometry}
\fi

\usepackage[T1]{fontenc}
\usepackage[utf8]{inputenc}
\usepackage{typearea}
\usepackage{lmodern}
\usepackage{amsmath}
\usepackage{microtype}
\usepackage{tabularx}
\usepackage{booktabs}
\usepackage[table]{xcolor}
\usepackage{comment}
\PassOptionsToPackage{cmyk}{xcolor}
\usepackage{tikz}
\usetikzlibrary{fit,shapes}
\usepackage{pgfplots}
\usepackage{pgfplotstable}
\usepackage{listings}
\usepackage[datesep=.,showzone=false,showseconds=false,style=ddmmyy]{datetime2}
\usepackage{textcomp}
\usepackage[style=numeric,backend=biber,bibencoding=utf-8,sorting=none,maxnames=8,giveninits]{biblatex} 
\usepackage{scrpage2}
\usepackage{gensymb}
\usepackage{ifthen}
\usepackage{euler}
\usepackage{siunitx}

\ifx\draft\undefined\else
\fi

\ifx\print\undefined\else
\usepackage[cam,noinfo,width=216truemm,height=303truemm,center]{crop}
\fi

\ifx\deliverable\undefined
\usepackage{gentium}
\usepackage{nimbusmononarrow}
\usepackage{Rosario}
\else
\usepackage[default,scale=0.95]{opensans}
\fi
\setkomafont{pageheadfoot}{\sffamily\small}
\setkomafont{pagenumber}{\sffamily\small}

\ifx\deliverable\undefined
\setkomafont{section}{\Large\color{eublue}}
\setkomafont{subsection}{\large\color{eublue}}
\else
\setkomafont{section}{\Large}
\setkomafont{subsection}{\large}
\fi

\definecolor{eublue}{cmyk}{1.0,0.8,0.0,0.0}
\definecolor{gray}{cmyk}{0.0,0.0,0.0,0.2}
\ifx\deliverable\undefined
\definecolor{background}{cmyk}{0.3,0.1,0.0,0.0}
\definecolor{table}{cmyk}{0.1,0.0,0.0,0.0}
\else
\definecolor{background}{cmyk}{0.0,0.0,0.0,0.0}
\definecolor{table}{cmyk}{0.0,0.0,0.0,0.0}
\fi
\definecolor{vdarkgray}{cmyk}{0.0,0.0,0.0,0.5}

\usepackage{hyperref}
\hypersetup{
colorlinks = true,
linkcolor = black,
citecolor = black,
urlcolor = black,
filecolor = black,
pdfcreator = {pdflatex},
pdfproducer = {LaTeX with hyperref}
}

\newcommand{\Vrule}{~\rule[-.2\baselineskip]{0.5pt}{\baselineskip}~}

\newcommand{\tabularz}[3][nohead]{%
    \def\coltypes{}
    \foreach \t in {#2} {%
        \ifx\deliverable\undefined
            \global\edef\coltypes{\coltypes\t}
        \else
            \global\edef\coltypes{\coltypes|\t}
        \fi
    }
    \ifx\deliverable\undefined\else
        \global\edef\coltypes{\coltypes|}
    \fi
    \expandafter\newcolumntype\expandafter{\expandafter L\expandafter}\expandafter{\coltypes}
    \pgfplotstableread[
        format=inline,
        col sep = &,
        row sep=\\,
        header=false,
    ]{#3}\tabledata
    \pgfplotstableset{
        begin table={\rmfamily\small\begin{tabularx}{\textwidth}{L}},
        end table={\end{tabularx}},
        skip coltypes=true,
        format=inline,
        string type,
        every head row/.style={
            output empty row,
        },
        every even row/.style={
            before row={\rowcolor{background}},
        },
        every odd row/.style={
            before row={\rowcolor{table}},
        },
    }
    \ifx\deliverable\undefined
        \ifthenelse{\equal{#1}{header}}{%
            \pgfplotstableset{
                every column/.style={
                    postproc cell content/.append code={
                        \ifnum\pgfplotstablerow=0
                            \pgfkeysalso{@cell content/.add={\textbf\bgroup\sffamily}{\egroup}}
                        \else\fi
                    },
                },
            }
        }{%
            \pgfplotstableset{
                every even row/.style={
                    before row={\rowcolor{background}},
                },
                every odd row/.style={
                    before row={\rowcolor{table}},
                },
                every column/.style={
                    postproc cell content/.append code={}
                },
            }
        }
    \else 
        \ifthenelse{\equal{#1}{header}}{%
            \pgfplotstableset{
                every row no 0/.style={
                    before row={\hline\rowcolor{eublue}},
                },
                every column/.style={
                    postproc cell content/.append code={
                        \ifnum\pgfplotstablerow=0
                            \pgfkeysalso{@cell content/.add={\textbf\bgroup\color{white}}{\egroup}}
                        \else\fi
                    },
                },
            }
        }{%
            \pgfplotstableset{
                every row no 0/.style={
                    before row={\hline},
                },
                every column/.style={
                    postproc cell content/.append code={}
                },
            }
        }
        \pgfplotstableset{
            begin table={\sffamily\footnotesize\begin{tabularx}{\textwidth}{L}},
            every even row/.style={
                before row={\hline},
            },
            every odd row/.style={
                before row={\hline},
            },
            every last row/.style={
                after row={\hline},
            },
        }
    \fi
    \pgfplotstabletypeset\tabledata
}

\ifx\deliverable\undefined

\else
\fi

\automark[chapter]{chapter}
\ifx\draft\undefined
    \ifoot[]{}
\else
    \ifoot[\normalfont\small\texttt{DRAFT \DTMnow}]{\normalfont\small\texttt{DRAFT \DTMnow}}
\fi
\ifx\deliverable\undefined
    \ohead{}
    \lefoot[\pagemark \Vrule {\color{eublue}\bfseries\leftmark}]{\pagemark \Vrule {\color{eublue}\bfseries\leftmark}}
    \rofoot[{\color{eublue}\bfseries\rightmark}\Vrule \pagemark]{{\color{eublue}\bfseries\rightmark}\Vrule \pagemark}
\else
    \ihead[]{}
    \ofoot[]{}
    \ohead[\scalebox{0.15}{\includegraphics{images/logo}}]{\scalebox{0.15}{\includegraphics{images/logo}}}
\fi

\defbibheading{bibliography}[\refname]{\subsection*{#1}}

\bibliography{../references}
\AtEveryBibitem{\clearlist{language}} 
\AtEveryBibitem{\clearfield{note}}    
\AtEveryBibitem{\clearlist{issn}}     
\AtEveryBibitem{\clearfield{isbn}}
\AtEveryBibitem{\clearlist{editor}}

\AtEveryBibitem{%
    \clearfield{urlyear}%
    \clearfield{urlmonth}%
    \clearfield{urlday}%
    \clearfield{isbn}%
    \clearfield{issn}%
    \clearlist{publisher}%
    \clearname{editor}%
\ifentrytype{online}{
}{}
}

\KOMAoptions{parskip=half}

\begin{document}

\begin{titlepage}
\ifx\deliverable\undefined
    \begin{tikzpicture}[overlay,remember picture]
        \setlength{\fboxsep}{0pt}

        \fill[table] ([xshift=-2mm]current page.north west) rectangle node {} ++([xshift=4mm]\paperwidth,-\paperheight);

        \fill[eublue] ([xshift=-2mm,yshift=2mm]current page.north west) rectangle node[white,align=center,font=\Large\scshape] {
            Foundation for a Centre of Excellence in\\ High-Performance Computing Systems
        } ++([xshift=4mm,yshift=-2mm]\paperwidth,-2.55);

    \node at (-8.0,-17.25) {\scalebox{3.3}{\includegraphics{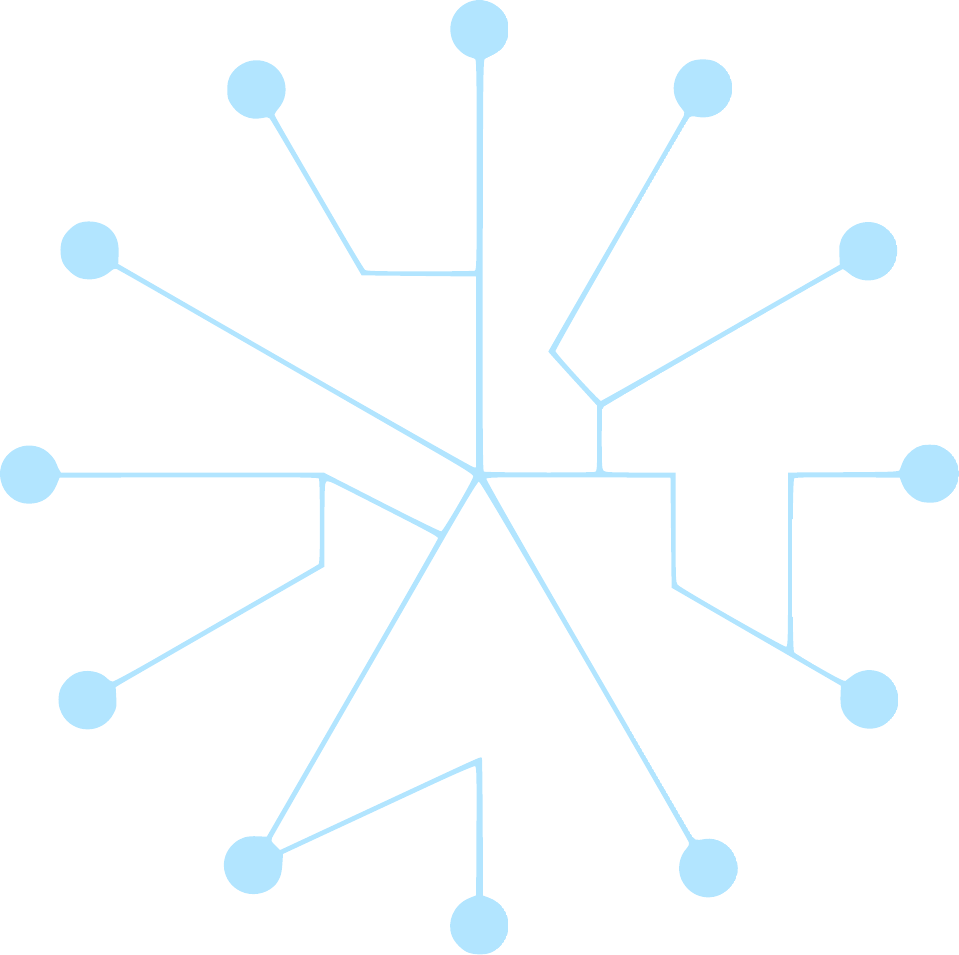}}};

        \fill[background] ([xshift=-2mm,yshift=8.7cm]current page.south west) rectangle node[eublue,align=center,font=\fontsize{1.3cm}{1.6cm}\selectfont\sffamily\bfseries] {
            Eurolab-\hskip2pt\raisebox{0.4ex}{4}\hskip2pt-HPC Long-Term Vision\\ on High-Performance Computing\\
            {\normalfont\Large\sffamily\color{black} Editors: Theo Ungerer, Paul Carpenter}\\ ~
        } ++([xshift=4mm]\paperwidth,-8.7);

    \fill[eublue] ([xshift=-2mm,yshift=1.8cm]current page.south west) rectangle node[gray,align=center,font=\scriptsize\sffamily] {
        Funded by the European Union Horizon 2020 Framework Programme (H2020-EU.1.2.2. - FET Proactive)
    } ++([xshift=4mm,yshift=-2mm]\paperwidth,-1.8);

        \node[anchor=south east] at ([yshift=0.3cm,xshift=-0.25cm]current page.south east) {\fcolorbox{background}{eublue}{\includegraphics[scale=0.15,trim=10 10 10 10,clip]{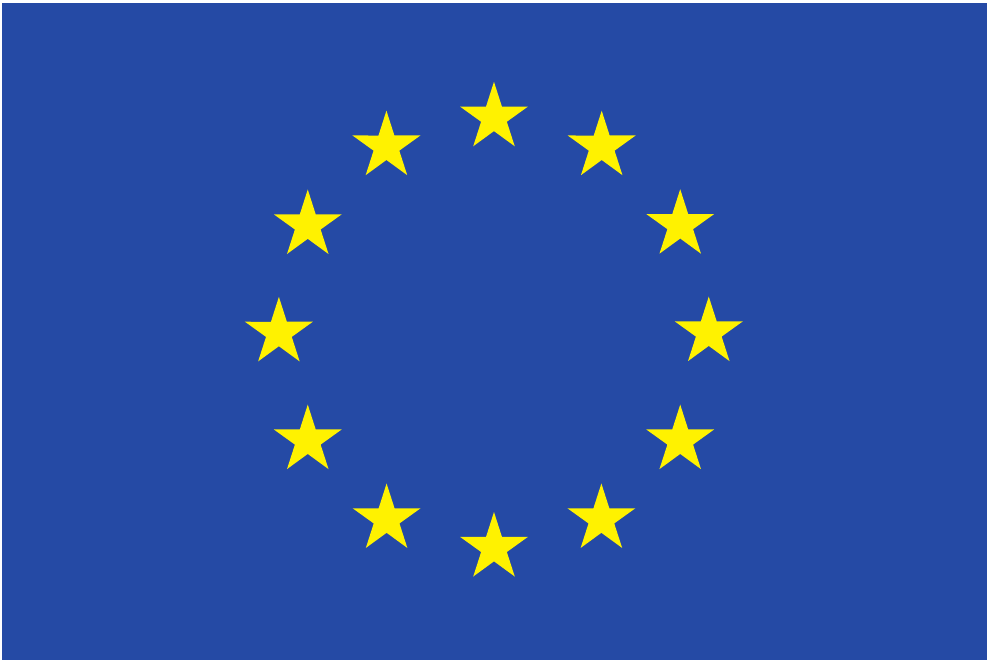}}};

    \ifx\draft\undefined\else
        \node[background,align=center,anchor=center,rotate=45] at (0,-0.30\paperheight) {\fontsize{3cm}{3cm}\selectfont\ttfamily DRAFT\\[0.5cm]
            {\large\ttfamily V3.1 ~~ 2017-08-03}
        };
    \fi

    \end{tikzpicture}
    {\centering
        \vfill
        \vskip5\baselineskip
    }
    \clearpage
\else
    {\centering
            {\Large\scshape\strut Foundation for a Centre of Excellence in\\\strut High-Performance Computing Systems}
            \vskip3\baselineskip
            {\Huge\bfseries\strut D2.2 ~ The Eurolab-4-HPC\\ Centre of Excellence Portfolio}
        \vfill
        \scalebox{1.0}{\includegraphics{images/logo}}\par
        \vskip5\baselineskip
        }
    \clearpage
\fi
\end{titlepage}

\input{0_prefix}

\cleardoublepage
\input{0_executive_summary}

\cleardoublepage
\tableofcontents
\cleardoublepage

\begin{refsection}
\input{1_introduction}
\end{refsection}
\printbibliography[section=1]

\begin{refsection}
\input{2_evolutionary_applications}

\end{refsection}
\printbibliography[section=2]

\cleardoublepage
\begin{refsection}
\input{3_data_center}
\end{refsection}
\printbibliography[section=3]

\cleardoublepage
\begin{refsection}
\input{4_disruptive_technologies}

\end{refsection}
\printbibliography[section=4]

\input{4_2_sustaining_technology}
\begin{refsection}
\input{4_2_1_cmos_scaling}
\end{refsection}
\printbibliography[section=5]

\clearpage
\begin{refsection}
\input{4_2_2_die_stacking}

\end{refsection}
\printbibliography[section=6]

\clearpage
\input{4_3_disruptive_technologies}
\begin{refsection}
\input{4_3_1_nvm}

\end{refsection}
\printbibliography[section=7]

\clearpage
\begin{refsection}
\input{4_3_2_photonics}

\end{refsection}
\printbibliography[section=8]

\clearpage
\input{4_4_alternative_ways}
\begin{refsection}
\input{4_4_1_resistive}
\end{refsection}
\printbibliography[section=9]

\clearpage
\begin{refsection}
\input{4_4_2_neuromorphic}

\end{refsection}
\printbibliography[section=10]

\clearpage
\begin{refsection}
\input{4_4_3_quantum}

\end{refsection}
\printbibliography[section=11]

\clearpage
\input{4_5_beyond_cmos}

\begin{refsection}
\input{4_5_1_nanotubes}
\end{refsection}
\printbibliography[section=12]

\clearpage
\begin{refsection}
\input{4_5_2_graphene}

\end{refsection}
\printbibliography[section=13]

\clearpage
\begin{refsection}
\input{4_5_3_diamond}

\end{refsection}
\printbibliography[section=14]

\cleardoublepage
\begin{refsection}
\input{5_impact}
\end{refsection}
\printbibliography[section=15]

\cleardoublepage
\begin{refsection}
\input{6_vertical_callenges}

\end{refsection}
\printbibliography[section=16]

\cleardoublepage
\begin{refsection}
\input{7_system_software}

\end{refsection}
\printbibliography[section=17]

\ifx\deliverable\undefined
\else
\input{8_eurolab4hpc}
\fi



\end{document}

%% file: 0_prefix.tex

\ifx\deliverable\undefined\else
\tabularz[header]{l,X}{%
        Document identifier     & EUROLAB4HPC-DEL-D2.2 \\
        Deliverable lead        &   UAU and BSC \\
        Related work package    &   WP2         \\
        Author(s)               &   Theo Ungerer (TU), Paul Carpenter (PC)  \\
        Main Contributor(s)     &   Theo Ungerer, Paul Carpenter, Avi Mendelson, Babak Falsafi, Dietmar Fey \\
        Due data of deliverable &   2017-08-31  \\
        Actual submission date  &   \\
        Reviewed by             &   Koen de Bosschere, Luca Benini\\
        Approved by             &   \\
        Dissemination level     &   PU  \\
        Website                 &   \url{http://eurolab4hpc.eu} \\
        Call                    &   H2020-FETHPC-2014   \\
        Grant agreement no.     &   67610   \\
        Funding scheme          &   CSA -- Coordination and Support Action  \\
        Project start date      &   2015-09-01  \\
        Duration                &   24 months   \\
}

\vskip4\baselineskip

\tabularz[header]{l,p{3cm},p{2cm},X,l}{
        Rev N   &   Description &   Author      &   Reviewers   &   Date        \\
        2.1     &   Preliminary &   TU et al.   &   --          &   2017-07-19  \\
        3.0     &   Internal Reviewer's Version   & TU et al.  &   Koen de Bosschere, Luca Benini    &   2017-07-20  \\
        3.1     &   Pre-final   &   TU et al.             &  all             &  2017-07-28             \\
        4.0     &   Final       &   TU et al.   &   --          &  2017-08-09             \\
}

\clearpage
\fi

\section*{Overall Editors and Authors}

Prof. Dr. Theo Ungerer, University of Augsburg\\
Dr. Paul Carpenter, BSC, Barcelona

\vskip4\baselineskip

\section*{Authors}
    \tabularz{p{4cm},p{4cm},X}{
        Nader Bagherzadeh           &   University of California, Irvine            &   Die Stacking and 3D-Chips \\
        Sandro Bartolini            &   Univeristy of Siena                         &   Photonics \\
        Luca Benini                 &   ETH Zürich                                  &   Die Stacking and 3D-Chips \\
        Koen Bertels                &   Delft University of Technology              &   Quantum Computing \\
        Fran\c{c}ois Bodin          &   University of Rennes                        &   Overall Comments \\
        Jose Manuel García Carrasco &   University of Murcia                        &   Photonics \\
        Koen De Bosschere           &   Ghent University                            &   Overall Comments \\
        Marc Duranton               &   CEA LIST DACLE                              &   Overall Comments \\
        Babak Falsafi               &   Ecole Polytechnique Federale de Lausanne    &   Data Centre and Cloud Computing, Green ICT and Resiliency \\
        Dietmar Fey                 &   University of Erlangen-Nuremberg            &   Memristors \\
        Said Hamdioui               &   Delft University of Technology              &   Quantum and Resistive Computing \\
        Christian Hochberger        &   Technical University of Darmstadt           &   Nanotubes and Nanowires, Graphene, Diamond Transistors \\
        Avi Mendelson               &   Technion                                    &   Diamond Computing, Hardware Impact \\
        Benjamin Pfundt             &   University of Erlangen-Nuremberg            &   3D Stacking, Memristors, Resistive Computing \\
        Ulrich Rückert              &   University of Bielefeld                     &   Neuromorphic Computing \\
        Igor Zacharov               &   Eurotech                                    &   Green ICT and Resiliency \\
}

\vfill
\section*{Compiled by}
\tabularz{X,p{4cm}}{
    Rico Amslinger, Martin Frieb, Florian Haas, Christian Mellwig, Jörg Mische,\newline Alexander Stegmeier, Sebastian Weis & University of Augsburg \\
}
\vfill

\parbox[t][2\baselineskip][t]{\textwidth}{
We also acknowledge the members of the Working Groups during the first year, as
well as the numerous people that provided valuable feedback at the
roadmapping workshops at HiPEAC~CSW and HPC~Summit, to HiPEAC and EXDCI for hosting
the workshops and Xavier Salazar for the organizational support.
}

%% file: 0_executive_summary.tex
\chapter*{Executive Summary}
\addtocentrydefault{chapter}{}{Executive Summary}
\markboth{Executive Summary}{Executive Summary}

\ifx\deliverable\undefined\else
This deliverable reports on the activities in EuroLab-4-HPC WP2 Research during the final year (M13 to M24) of the project. It contains the EuroLab-4-HPC Long-Term Vision for High Performance Computing, as well as a final chapter proposing topics for the future EuroLab-4-HPC Centre of Excellence portfolio.  
The Long-Term Vision also exists as a separate public document that will be printed and distributed within the HPC community.
\fi

Radical changes in computing are foreseen for the next decade. The US IEEE society wants to ``reboot computing'' and the HiPEAC Vision 2017 sees the time to ``re-invent computing'', both by challenging its basic assumptions. This document presents the ``EuroLab-4-HPC Long-Term Vision on High-Performance Computing'' of August 2017, a road mapping effort within the EC CSA\footnote{European Commission Community and Support Action} Eurolab-4-HPC that targets potential changes in hardware, software, and applications in High-Performance Computing (HPC). 

The objective of the Eurolab-4-HPC vision is to provide a long-term roadmap from 
2023 to 2030 for High-Performance Computing (HPC). Because of the long-term perspective and its speculative nature, the authors started with an assessment of future computing technologies that could influence HPC hardware and software. The proposal on research topics is derived from the report and discussions within the road mapping expert group. We prefer the term ``vision'' over ``roadmap'', firstly because timings are hard to predict given the long-term perspective, and secondly because EuroLab-4-HPC will have no direct control over the realization of its vision.

\subsection*{The Big Picture}
High-performance computing (HPC) typically targets scientific and engineering simulations with numerical programs mostly based on floating-point computations. We expect the continued scaling of such scientific and engineering applications to continue well beyond Exascale computers. As just one example, the NASA CFD roadmap from 2014 envisions scaling of Computational Fluid Dynamics to Zetascale by 2030\footnote{CFD Vision 2030 Study: A Path to Revolutionary Computational Aerosciences. NASA/CR-2014-218178}.

However, three trends are changing the landscape for high-performance computing and supercomputers. The first trend is the emergence of data analytics complementing simulation in scientific discovery. While simulation still remains a major pillar for science, there are massive volumes of scientific data that are now gathered by sensors augmenting data from simulation available for analysis.
High-Performance Data Analysis (HPDA) will complement simulation in future HPC applications.

The second trend is the emergence of cloud computing and warehouse-scale computers (also known as data centres). Data centres consist of low-cost volume processing, networking and storage servers, aiming at cost-effective data manipulation at unprecedented scales. The scale at which they host and manipulate (e.g., personal, business) data has led to fundamental breakthroughs in data analytics. 

There are a myriad of challenges facing massive data analytics including management of highly distributed data sources, and tracking of data provenance, data validation, mitigating sampling bias and heterogeneity, data format diversity and integrity, integration, security, privacy, sharing, visualization, and massively parallel and distributed algorithms for incremental and/or real-time analysis.

Large datacentres are fundamentally different from traditional supercomputers in their design, operation and software structures. Particularly, big data applications in data centres and cloud computing centres require different algorithms and differ significantly from traditional HPC applications such that they may not require the same computer structures. 

With modern HPC platforms being increasingly built using volume servers (90\% of the systems in the June 2017 TOP500 list are based on Intel Xeon), there are a number of features that are shared among warehouse-scale computers and modern HPC platforms, including dynamic resource allocation and management, high utilization, parallelization and acceleration, robustness and infrastructure costs. These shared concerns will serve as incentives for the convergence of the platforms.

There are, meanwhile, a number of ways that traditional HPC systems differ from modern warehouse-scale computers: efficient virtualization, adverse network topologies and fabrics in cloud platforms, low memory and storage bandwidth in volume servers. HPC customers must adapt to co-exist with cloud services; warehouse-scale computer operators must innovate technologies to support the workload and platform at the intersection of commercial and scientific computing. 

It is unclear whether a convergence of HPC with big data applications will arise. Investigating hardware and software structures targeting such a convergence is of high research and commercial interest. 
However, some HPC applications will be executed more economically on data centres. Exascale and post-Exascale supercomputers could become a niche for HPC applications. 

The third trend arises from Deep Neural Networks (DNN) for back propagation learning of complex patterns, which emerged as new technique penetrating different application areas. DNN learning requires high performance and is often run on high-performance supercomputers. Recent GPU accelerators are seen as very effective for DNN computing by their enhancements, e.g. support for 16-bit floating-point and tensor processing units.
It is widely assumed that it will be applied in future autonomous cars thus opening a very large market segment for embedded HPC. DNNs will also be applied in engineering simulations traditionally running on HPC supercomputers. 

Embedded high-performance computing demands are upcoming needs. It may concern smartphones but also applications like autonomous driving, requiring on-board high-performance computers. In particular the trend from current advanced ADAS (automatic driving assistant systems) to piloted driving (2018–2020) and to fully autonomous cars in the next decade will increase on-board performance requirements and may even be coupled with high-performance supercomputers in the Cloud. The target is to develop systems that adapt more quickly to changing environments, opening the door to highly automated and autonomous transport, capable of eliminating human error in control, guidance and navigation and so leading to more safety. High-performance computing devices in cyber-physical systems will have to fulfil further non-functional requirements such as timeliness, (very) low energy consumption, security and safety.
However, further applications will emerge that may be unknown today or that receive a much higher importance than expected today. 

Power and thermal management is considered as highly important and will continue its preference in future. Post-Exascale computers will target more than 1 Exaflops with less than 30 MW power consumption requiring processors with a much better performance per Watt than available today.  On the other side, embedded computing needs high performance with low energy consumption. The power target at the hardware level is widely the same, a high performance per Watt.

In addition to mastering the technical challenges, reducing the environmental impact of upcoming computing infrastructures is also an important matter. Reducing CO\textsubscript{2} emissions and overall power consumption should be pursued. A combination of hardware techniques, such as new processor cores, accelerators, memory and interconnect technologies, and software techniques for energy and power management will need to be cooperatively deployed in order to deliver energy-efficient solutions.

Because of the foreseeable end of CMOS scaling, new technologies are under development, such as, for example, Die Stacking and 3D Chip Technologies, Non-volatile Memory (NVM) Technologies, Photonics, Resistive Computing, Neuromorphic Computing, Quantum Computing, Nanotubes, Graphene, and diamond-based transistors. Since it is uncertain if/when some of the technologies will mature, it is hard to predict which ones will prevail.

The particular mix of technologies that achieve commercial success will strongly impact the hardware and software architectures of future HPC systems, in particular the processor logic itself, the (deeper) memory hierarchy, and new heterogeneous accelerators.

There is a clear trend towards more complex systems, which is expected to continue over the next decade. These developments will significantly increase software complexity, demanding more and more intelligence across the programming environment, including compiler, run-time and tool intelligence driven by appropriate programming models. Manual optimization of the data layout, placement, and caching will become uneconomic and time consuming, and will, in any case, soon exceed the abilities of the best human programmers.

If accurate results are not necessarily needed, another speedup could emerge from more efficient special execution units, based on analog, or even a mix between analog and digital technologies. Such developments would benefit from more advanced ways to reason about the permissible degree of inaccuracy in calculations at run time.  Furthermore, new memory technologies like memristors may allow on-chip integration, enabling tightly-coupled communication between the memory and the processing unit. With the help of memory computing algorithms, data could be pre-processed ``in-'' or ``near-'' memory. 

But it is also possible that new hardware developments reduce software complexity. New materials like graphene, nanotubes and diamonds could be used to run processors at much higher frequencies than are currently possible, and with that, may even enable a significant increase in the performance of single-threaded programs. 

Optical networks on die and Terahertz-based connections may eliminate the need for preserving locality since the access time to local storage may not be as significant in future as it is today. Such advancements will lead to storage-class memory, which features similar speed, addressability and cost as DRAM combined with the non-volatility of storage.  In the context of HPC, such memory may reduce the cost of checkpointing or eliminate it entirely.  

The adoption of neuromorphic, resistive and/or quantum computing as new accelerators may have a dramatic effect on the system software and programming models.  It is currently unclear whether it will be sufficient to offload tasks, as on GPUs, or whether more dramatic changes will be needed. By 2030, disruptive technologies may have forced the introduction of new and currently unknown abstractions that are very different from today. Such new programming abstractions may include domain-specific languages that provide greater opportunities for automatic optimization. Automatic optimization requires advanced techniques in the compiler and runtime system. We also need ways to express non-functional properties of software in order to trade various metrics: performance vs. energy, or accuracy vs. cost, both of which may become more relevant with near threshold, approximate computing or accelerators.

Nevertheless, today's abstractions will continue to evolve incrementally and will continue to be used well beyond 2030, since scientific codebases have very long lifetimes, on the order of decades. 

Execution environments will increase in complexity requiring more intelligence, e.g., to manage, analyse and debug millions of parallel threads running on heterogeneous hardware with a diversity of accelerators, while dynamically adapting to failures and performance variability. Spotting anomalous behavior may be viewed as a big data problem, requiring techniques from data mining, clustering and structure detection. This requires an evolution of the incumbent standards such as OpenMP to provide higher-level abstractions. An important question is whether and to what degree these fundamental abstractions may be impacted by disruptive technologies.

\subsection*{The Work Needed}
As new technologies require major changes across the stack, a vertical funding approach is needed, from applications and software systems through to new hardware architectures and potentially down to the enabling technologies. We see HP Lab's memory-driven computing architecture ``The Machine'' as an exemplary project that proposes a low-latency NVM (Non-Volatile Memory) based memory connected by photonics to processor cores. Projects could be based on multiple new technologies and similarly explore hardware and software structures and potential applications. Required research will be interdisciplinary. Stakeholders will come from academic and industrial research.

\subsection*{The Opportunity}
The opportunity may be development of competitive new hardware/software technologies based on upcoming new technologies to advantageous position European industry for the future. Target areas could be High-Performance Computing and Embedded High-Performance devices. The drawback could be that the chosen base technology may not be prevailing but be replaced by a different technology. For this reason, efforts should be made to ensure that aspects of the developed hardware architectures, system architectures and software systems could also be applied to alternative prevailing technologies. For instance, several NVM technologies will bring up new memory devices that are several magnitudes faster than current Flash technology and the developed system structures may easily be adapted to the specific prevailing technologies, even if the project has chosen a different NVM technology as basis. 

\subsection*{EC Funding Proposals}
The Eurolab4HPC vision recommends the following funding opportunities for topics beyond Horizon 2020 (ICT):
\begin{itemize}
    \item Convergence of HPC and HPDA:
        \begin{itemize}
            \item Data Science, Cloud computing and HPC: Big Data meets HPC
            \item Inter-operability and integration
            \item Limitations of clouds for HPC
            \item Edge Computing: local computation for processing near sensors
        \end{itemize}
    \item Impact of new NVMs: 
        \begin{itemize}
            \item Memory hierarchies based on new NVMs
            \item Near- and in-memory processing: pre- and post-processing in (non-volatile) memory
            \item HPC system software based on new memory hierarchies 
            \item Impact on checkpointing and reciliency
        \end{itemize}
    \item Programmability:
        \begin{itemize}
            \item Hide new memory layers and HW accelerators from users by abstractions and intelligent programming environments
            \item Monitoring of a trillion threads
            \item Algorithm-based fault tolerance techniques within the application as well as moving fault detection burden to the library, e.g. fault-tolerant message-passing library
        \end{itemize}
    \item Green ICT and Energy
        \begin{itemize}
            \item Integration of cooling and electrical subsystem
            \item Supercomputer as a whole system for Green ICT
        \end{itemize}
\end{itemize}

As remarked above, projects should be interdisciplinary, from applications and software systems through hardware architectures and, where relevant, enabling hardware technologies.

%% file: 1_introduction.tex
\chapter{Introduction}

Upcoming application trends and disruptive VLSI technologies will change the way computers will be programmed and used as well as the way computers will be designed. New application trends such as High-Performance Data Analysis (HPDA) and deep-learning will induce changes in High-Performance Computing; disruptive technologies will change the memory hierarchy, hardware accelerators and even potentially lead to new ways of computing. The HiPEAC Vision 2017\footnote{\url{www.hipeac.net/publications/vision}} sees the time to revisit the basic concepts: The US wants to ``reboot computing'', the HiPEAC Vision proposes to ``re-invent computing'' by challenging basic assumptions such as binary coding, interrupts, layers of memory, storage and computation.

Exascale does not merely refer to a LINPACK $R_{\mathrm{max}}$ of 1 Exaflops. The PathForward definition of a capable Exascale system is a good one, as it focuses on scientific problems rather than benchmarks, as well as raising the core challenges of power consumption and resiliency:  ``a supercomputer that can solve science problems 50X faster (or more complex) than on the 20 Petaflop systems (Titan and Sequoia) of today in a power envelope of 20-30 megawatts, and is sufficiently resilient that user intervention due to hardware or system faults is on the order of a week on average''~\cite{Hemsoth2016missing}.

This document has been funded by the EC CSA Eurolab-4-HPC (Sept. 2015 – August 2017) project. It outlines a long-term vision for excellence in European High-Performance Computing research, with a timescale beyond Exascale computers, i.e. a timespan of approximately 2023-2030.  

\section{Current Proposals for Exascale Machines}

\paragraph{USA:} Today's leading organizations are using machine learning-based tools to automate decision processes, and they are starting to experiment with more advanced uses of artificial intelligence (AI) for digital transformation. Corporate investment in artificial intelligence is predicted to triple in 2017, becoming a \$100 billion market by 2025.~\cite{Wellers2017}

The U.S. Department of Energy --- and the hardware vendors it partners with --- are set to enliven the Exascale effort with nearly a half billion dollars in research, development, and deployment investments.  The push is led by the DoE's Exascale Computing Project and its extended PathForward program landing us in the 2021 -- 2022 timeframe with ``at least one'' Exascale system. This roadmap was confirmed in June 2017 with a DoE announcement that backs six HPC companies as they create the elements for next-generation systems. The vendors on this list include Intel, Nvidia, Cray, IBM, AMD, and Hewlett Packard Enterprise (HPE).~\cite{hemsoth2017american}

\paragraph{China} currently possesses the fastest supercomputer in the world, called the Sunway TaihuLight. The supercomputer is theoretically capable of 124.5 Petaflops of performance, making it the first computer system to surpass 100 Petaflops. Interestingly, this supercomputer contains entirely Chinese-made processing chips. In January, China said it would soon have the world's first Exascale supercomputer prototype up and running. China said it will have a completed Exascale supercomputer by 2020.~\cite{tilley2017}.

This year, China is aiming for breakthroughs in high-performance processors and other key technologies to build the world's first prototype Exascale supercomputer, the Tianhe-3, said Meng Xiangfei, the director of application at the National Super Computer Tianjin Center. ``The prototype is expected to be completed in early 2018. Tianhe-3 will be made entirely in China, from processors to operating system. It will be stationed in Tianjin and fully operational by 2020, earlier than the US plan for its Exascale supercomputer''.~\cite{zhihao2017}.

The Exascale supercomputer will be able to analyse smog distribution on a national level, while current models can only handle a district. Tianhe-3 also could simulate earthquakes and epidemic outbreaks in more detail, allowing swifter and more effective government responses. The new machine also will be able to analyse gene sequence and protein structures in unprecedented scale and speed. That may lead to new discoveries and more potent medicine, he said.~\cite{zhihao2017}.

\paragraph{Japan:} The successor to the K supercomputer, which is being developed under the Flagship2020 program, will use ARM-based processors and these chips will be at the heart of a new system built by Fujitsu for RIKEN (Japan's Institute of Physical and Chemical Research) that would break the Exaflops barrier by 2020.~\cite{morgan2016japan}.

\paragraph{European Community:} EC President Juncker has declared that the European Union has to be competitive in the international arena with regard to the USA, China, Japan and other stakeholders, in order to enhance and promote the European industry in the public as well as the private sector related to HPC.~\cite{emmen2017}

The first step will be ``Extreme-Scale Demonstrators'' (EsDs) that should provide pre-Exascale platforms deployed by HPC centres and used by Centres of Excellence for their production of new and relevant applications. Such demonstrators are planned by ETP4HPC Initiative and included in the EC LEIT-ICT 2018 calls. At project end, the EsDs will have a high TRL (Technical Readiness Level) that will enable stable application  production at reasonable scale.~\cite{etp4hpcsra}

The EuroHPC Initiative is based on a Memorandum of Understanding that was signed on March 23, 2017 in Rome. Its plans for the creation of two pre-Exascale machines, followed by the delivery of two machines that are actually Exascale. There are a lot of things to consider such as the creation of the micro-processor with European technology and the integration of the micro-processor in the European Exascale machines~\cite{emmen2017}.  IPCEI (Important Project of Common European Interest) is another parallel initiative, related to EuroHPC. The IPCEI for HPC at the moment involves France, Italy, Spain, and Luxembourg but it is also open to other countries in the European Union. If all goes according to plan, the first pre-Exascale machine will be released by 2022 -- 2023. By 2024 -- 2025, the Exascale machines will be delivered.~\cite{emmen2017}

\begin{figure*}[ht]
    \centering
\begin{tikzpicture}
    {\sffamily\small
    \ifx\deliverable\undefined
    \fill[background] (-5,3.5) rectangle ++(17,-12);
    \else
    \fi
    \setlength{\fboxsep}{0pt}
    \node[align=left] (us) at (-0.5,0.5) {\parbox[t][7cm][c]{7cm}{
        \begin{center}\fbox{\includegraphics[scale=0.046]{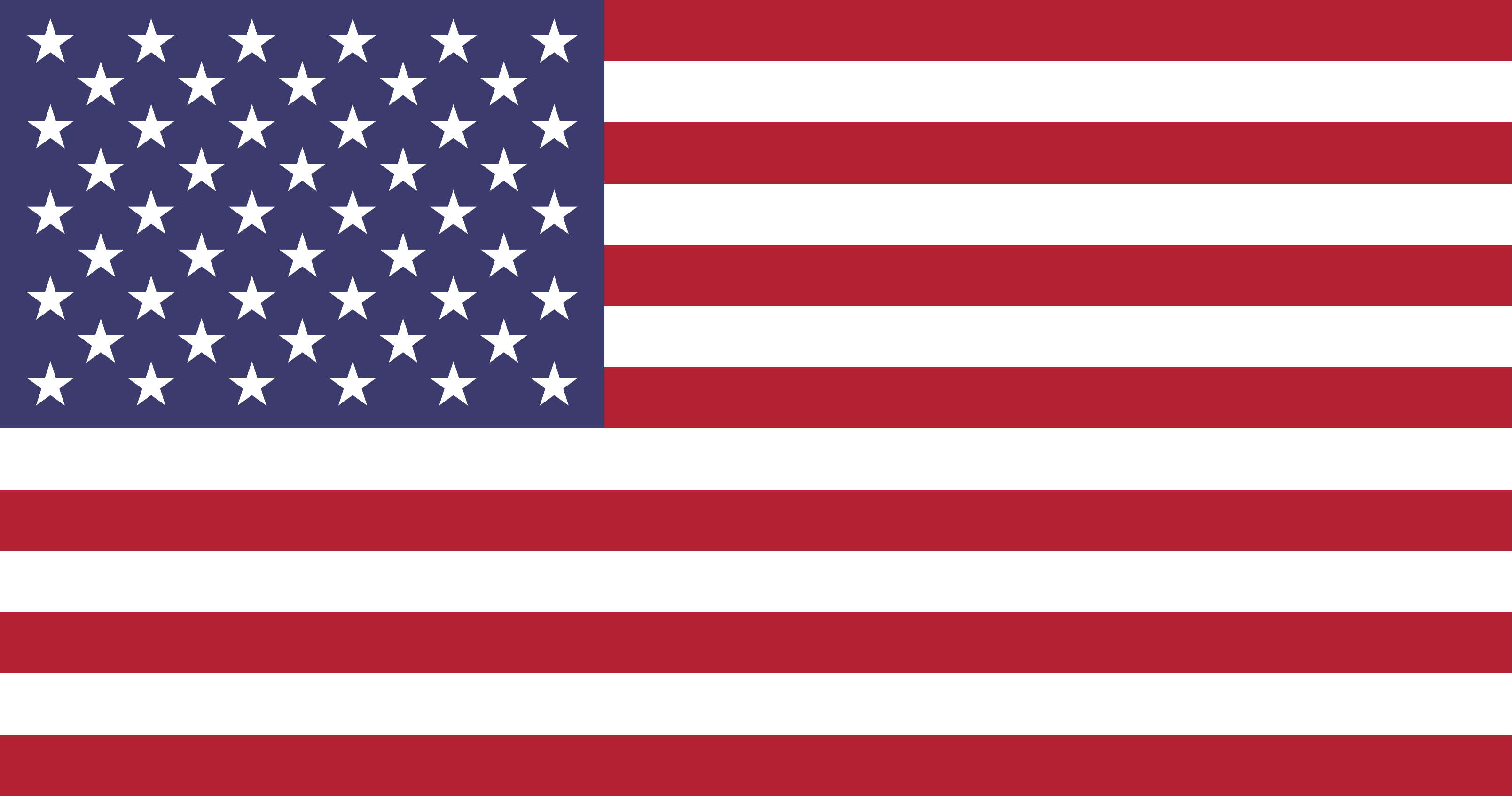}}
        \parbox[b][0.8cm][c]{1cm}{~~\textbf{U.S.}}\end{center}
        \begin{itemize}
    \setlength\itemsep{0pt}
    \setlength\parskip{0pt}
            \item Sustained ES: 2023
            \item Peak ES: 2021
            \item Vendors: U.S.
            \item Processors: U.S.
            \item Initiatives: NSCI/ECP
            \item Cost: \$\,300 -- \$\,500\,M per system, plus heavy R\&D investments
        \end{itemize}
    }};
    \node[align=left] (eu) at (7,0.5) {\parbox[t][7cm][c]{7cm}{
        \begin{center}\fbox{\includegraphics[scale=0.119]{images/eu}}
        \parbox[b][0.8cm][c]{1cm}{~~\textbf{EU}}\end{center}
        \begin{itemize}
    \setlength\itemsep{0pt}
    \setlength\parskip{0pt}
            \item Sustained ES: 2023 -- 2024
            \item Peak ES: 2021
            \item Vendors: U.S., Europe
            \item Processors: U.S., ARM
            \item Initiatives: PRACE, ETP4HPC
            \item Cost: \$\,300 -- \$\,350\,M per system, plus heavy R\&D investments
        \end{itemize}
    }};
    \node[align=left] (cn) at (-0.5,-5.5) {\parbox[t][5.5cm][t]{7cm}{
        \begin{center}\fbox{\includegraphics[scale=0.05]{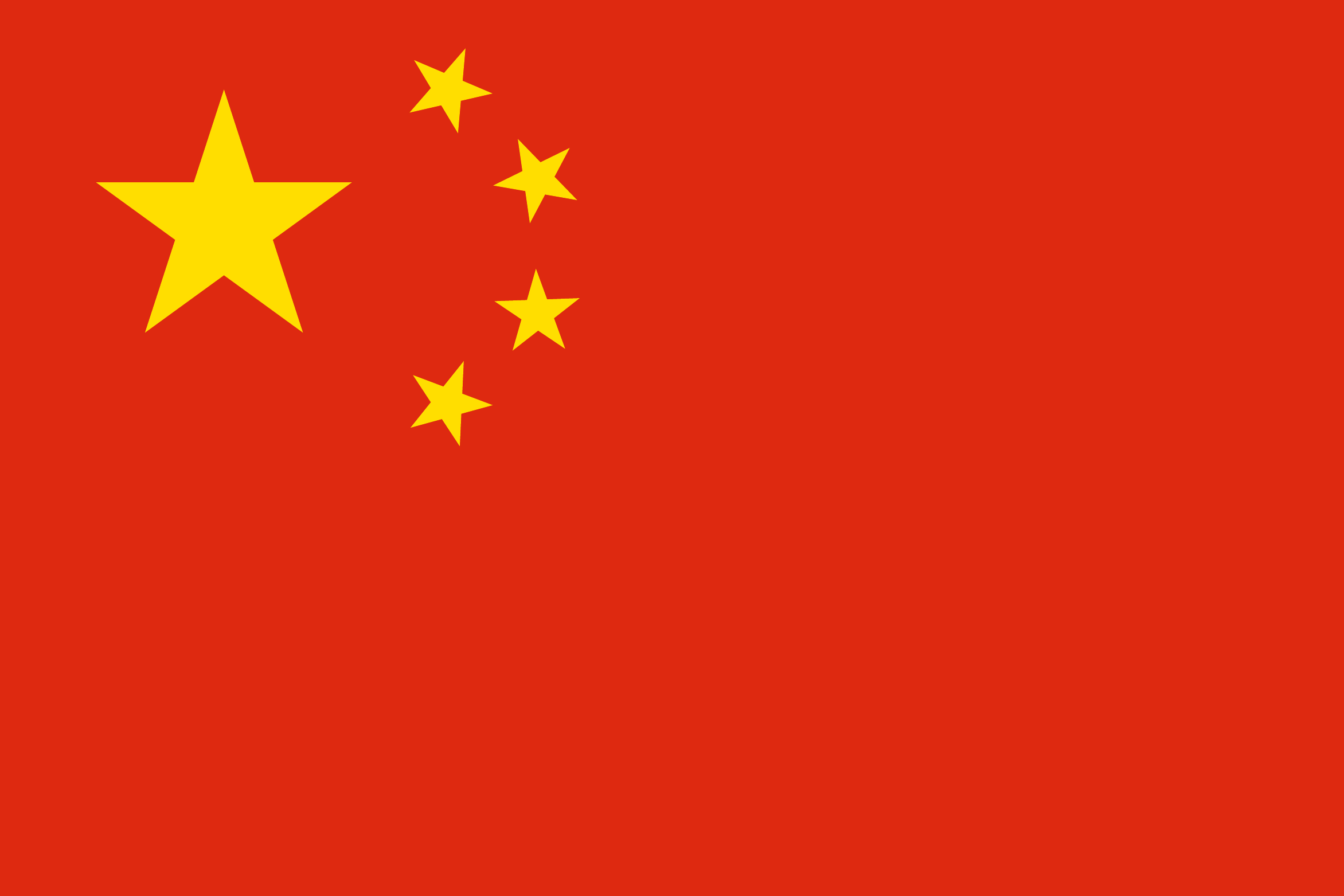}}
        \parbox[b][0.8cm][c]{2cm}{~~\textbf{China}}\end{center}
        \begin{itemize}
    \setlength\itemsep{0pt}
    \setlength\parskip{0pt}
            \item Sustained ES: 2023
            \item Peak ES: 2020
            \item Vendors: Chinese
            \item Processors: Chinese (plus U.S.?)
            \item 13\textsuperscript{th} 5-Year Plan
            \item Cost: \$\,350 -- \$\,500\,M per system, plus heavy R\&D investments
        \end{itemize}
    }};
    \node[align=left] (jp) at (7,-5.5) {\parbox[t][5.5cm][t]{7cm}{
        \begin{center}\fbox{\includegraphics[scale=0.05]{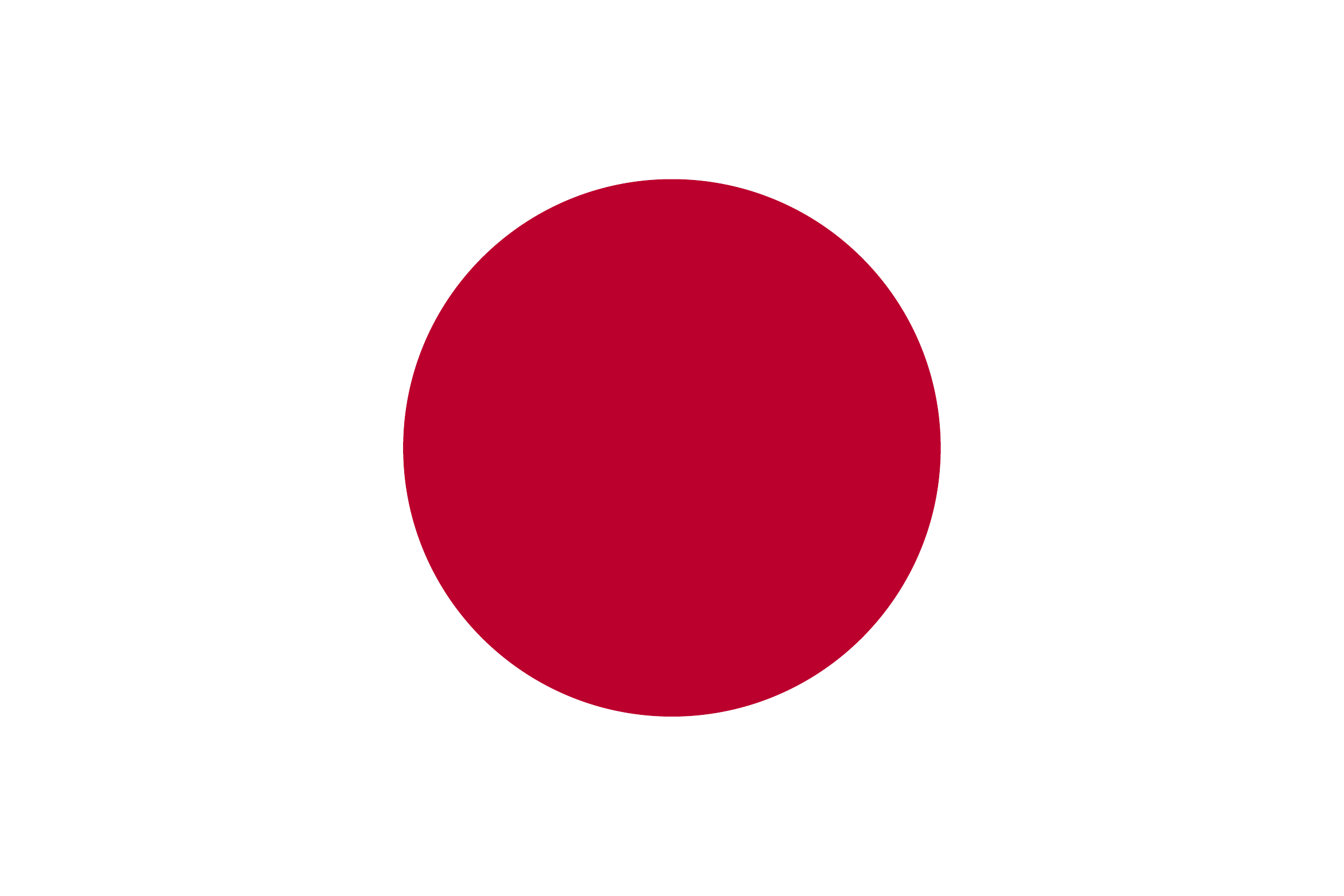}}
        \parbox[b][0.8cm][c]{2cm}{~~\textbf{Japan}}\end{center}
        \begin{itemize}
    \setlength\itemsep{0pt}
    \setlength\parskip{0pt}
            \item Sustained ES: 2023 -- 2024
            \item Peak ES: Not planned
            \item Vendors: Japanese
            \item Processors: Japanese
            \item Cost: \$\,600 -- \$\,850\,M, this includes both 1 system and the R\&D costs.  Will also do many smaller size systems
        \end{itemize}
    }};
    }
    \draw[thick] (-3.5,-2.25) -- ++(14,0);
    \draw[thick] (3.5,3) -- ++(0,-11);
\end{tikzpicture}
\caption{Summary of Exascale race as seen by Hyperion at April 20, 2017~\cite{russell2017}}
\label{fig-intro-hyperion}
\end{figure*}

Partly other time lines are shown in the summary on Exascale race as seen by Hyperion at April 20, 2017~\cite{russell2017} (see Figure~\ref{fig-intro-hyperion}).

\section{Related Roadmapping Initiatives}
The Eurolab-4-HPC vision complements existing efforts such as the ETP4HPC Strategic Research Agenda (SRA). ETP4HPC is an industry-led initiative to build a globally competitive HPC system value chain. Development of the EuroLab-4-HPC vision is aligned with ETP4HPC SRA in its latest version to be scheduled for September 2017.  SRA 2017 is targeting a roadmap towards Exascale computers that spans until approximately 2022, whereas the Eurolab-4-HPC vision targets the speculative period beyond Exascale, so approximately 2023 – 2030.
The EuroLab-4-HPC vision is developed in close collaboration with the ``HiPEAC Vision'' of HiPEAC CSA that features the broader area of ``High Performance and Embedded Architecture and Compilation''. The EuroLab-4-HPC vision complements the HiPEAC Vision 2017 document with a stronger focus on disruptive technologies and HPC.
The current state of available roadmaps that are adjacent to the Eurolab-4-HPC vision is shown in the table below:

\begin{table*}[ht]
    \caption{Current state of available roadmaps adjacent to the Eurolab-4-HPC vision.}
    \label{tbl-intro-roadmaps}
        \tabularz[header]{p{2.5cm},X,p{2.1cm},p{1.1cm},p{3cm}}{
            &   Goal    &   Timespan    &   SWOT/\newline Political  & Scope \\
        HiPEAC Vision   &   Steer European academic research (driven by industry)   &   Short: 3 years,\newline Mid: 6 years,\newline Long: > 2020  &   Yes &   HPC + embedded \\
        ETP4HPC SRA/EXDCI       & Strengthening European (industrial) HPC ecosystem         &   6 years\newline (2014 to 2020)  &   Yes &   HPC except applications \\
PRACE Scientific Case   &
(Academic) need for European HPC infrastructure &
8 years (2012 to 2020)  &
Yes &
HPC applications    \\
EESI (European Exascale Software Initiative)    &
Development of efficient Exascale applications  &
5 to 10 years   &
No  &
Exascale applications   \\
BDVA (Big Data Value Association)   &
Big Data technologies roadmap   &
2020    &
--   &
Big data    \\
Rethink Big &
Roadmap for European Technologies in Hardware and Networking for Big Data   &
--    &
--   &
Big data    \\
ECSEL MASRIA    &
European leadership in enabling and industrial technologies. Competitive EU ECS industry.   &
2015 roadmap to about 2025  &
Yes &
Electronic components and systems (ECS) \\
Next Generation Computing Roadmap   &
Strengthening European industry &
2014: 10 to 15 years    &
--    &
HPC 
extensively covered \\
\textbf{Eurolab-4-HPC}   &
\textbf{Academic excellence in HPC}  &
\textbf{2023 -- 2030}   &
\textbf{No}  &
\textbf{Whole HPC stack }\\
}
\end{table*}

\section{Working Towards the Eurolab-4-HPC Roadmap/Vision}
The Eurolab-4-HPC vision has been developed as a research roadmap with a substantially longer-term time-window than most of the roadmaps shown above. Since the beginning, it has been our target to stick to technical matters and provide an academic research perspective. Because targeting the post-Exascale era with a horizon of approximately 2022 -- 2030 will be highly speculative, we proceeded as follows:
\begin{enumerate}
    \item Select disruptive technologies that may be technologically feasible in the next decade.
    \item Assess the potential hardware architectures and their characteristics.
    \item Assess what that could mean for the different working groups (WG) topics (concerns all WGs).
\end{enumerate}

The vision roughly follows the structure:
``\emph{IF} technology ready \emph{THEN} foreseeable impact on WG topic could be ...''

The first task performed was to select potentially disruptive technologies and summarize its potential for the next decade with the help of experts in a ``Report on Disruptive Technologies''.  The report has reached a stage of maturity and its impact on hardware and software is provided in working group zero, which is the basis for all other working groups. 

\begin{enumerate}
    \setcounter{enumi}{-1}
    \item Impact of disruptive technologies\\
(Theo Ungerer, University of Augsburg, Germany)
\end{enumerate}

Aside from a working group zero on disruptive technologies, we defined five more working groups and assigned working group leaders:

\begin{enumerate}
    \item New technologies and hardware architectures\\
        (Avi Mendelson, Technion, Haifa)
    \item System software and programming environment\\
        (Paul Carpenter, BSC, Barcelona)
    \item HPC application requirements\\
        (Paul Carpenter, BSC, Barcelona)
    \item Vertical challenges: Green ICT, energy and resiliency\\
        (Bastian Koller and Axel Tenschert, HLRS, Stuttgart) 
    \item Convergence of HPC, with IoT and the Cloud\\
        (Babak Falsafi, EPFL, Lausanne)
\end{enumerate}

Altogether about 46 contributors signed up to work on the vision.

The timescale of the first year concerned:

2016, April 30: We delivered an input to the EC consultation process regarding ``game changing technology''\footnote{\url{ec.europa.eu/futurium/en/content/fet-proactive}}.

2016, August 31: preliminary roadmap available\footnote{\url{www.eurolab4hpc.eu/roadmap/}}

The preliminary roadmap deliverable was well received by EC reviewers as well as in HPC public shown by an article in The Next Platform of October 12, 2016~\cite{hemsoth2016postexa}.

The EC Reviewer Comments were
\begin{itemize}
    \item enhance with proposals of what EC should be funding
    \item integrate/combine with EXDCI/ETP4HPC SRA.
\end{itemize}

Other observations were that the working groups proved not very effective and the WG sections in preliminary roadmap are not well aligned with each other.

The Mission for the Second Year concerned:
\begin{itemize}
    \item form a single \emph{expert working group}:\\
        Avi Mendelson, Luca Benini, Babak Falsafi, Sandro Bartolini, Dietmar Fey, Marc Duranton, Fran\c{c}ois Bodin, Simon McIntosh–Smith, Igor Zacharov, Paul Carpenter, Theo Ungerer
    \item revisit Disruptive Technologies and implications of current roadmap
    \item harmonize, restructure and revise the different roadmap sections
    \item recommend potential EC funding opportunities
\end{itemize}

The working schedule for the second year was:
\begin{itemize}
    \item 2017-01-23 DTHPC: Workshop on Disruptive Technologies in high-Performance Computing in the Next Decade, talk on roadmap
    \item 2017-03-17  Kickoff Telco of expert working group
    \item 2017-03-20  Discuss at ETP4HPC SRA Kickoff
    \item 2017-04-28  HiPEAC CSW Zagreb: Workshop: ``Towards the Eurolab-4-HPC Long-term Roadmap on High-performance Computing in Years 2022 -- 2030''
    \item 2017-05-03 Talk ``Potential Impact of Future Disruptive Technologies on Embedded Multicore Computing'', at AK Multicore, Munich 
    \item 2017-05-04 same talk at PARS Workshop, Hagen
    \item 2017-05-17 at HPC Summit:  Roadmapping talk at Workshop of EuroLab-4-HPC: the Future of High-Performance
    \item 2017-05-29+30: 1\textonehalf~ day Roadmap meeting of expert working group at EPFL Lausanne 
    \item 2017-06: Experts prepare inputs
    \item 2017-07-31: Final Roadmap done
    \item 2017-08-31: Final Roadmap deliverable due
\end{itemize}

\section{Document Structure}
The rest of this document is structured as follows: The next section provides some insights in evolutionary, i.e. scaling of engineering HPC applications, and potentially new upcoming applications. Section~\ref{sec-datacenter} covers data centres and cloud computing, eventually leading from HPC to HPDA. Section~\ref{sec-disruptive} focuses on Disruptive Technologies followed by section~\ref{sec-impact} that summarizes the Potential Long-Term Impacts of Disruptive Technologies for HPC Hardware and Software in separate subsections. Section~\ref{sec-vertical} covers Green ICT and Resiliency as Vertical Challenges, and finally Section~\ref{sec-system} focuses on System Software and Programming Environment.

%% file: 2_evolutionary_applications.tex
\chapter{Evolutionary and New Upcoming Applications}

Industrial and scientific applications are the \emph{raison d'\^{e}tre} of high-performance computing. Production HPC systems should meet the needs of the users, and they must anticipate future evolutionary and disruptive changes in these requirements. This section collects the main requirements of HPC users, including applications, numerical libraries, and algorithms.  The focus is on the impact of HPC requirements on HPC computing systems, rather than the applications themselves. 

We expect a continuous scaling of existing HPC engineering applications, but also combining HPC engineering applications with data analysis (from HPC to HPDA) and AI techniques. It is important to note that scientific applications have very long lifetimes, measured in decades, which is much longer than other software domains, and dramatically longer than hardware. We also see new applications for HPC and HPDA that may influence future HPC systems. Such new applications can only partly be predicted.

The top three challenges for support of future applications in HPC and HPDA are:
\begin{itemize}
    \item Programming environments: the need for suitable abstractions between the application and the underlying complex hardware and storage,
    \item System software and management: scalable and smart runtime systems,
    \item Big data and usage models: smart processing, visualization, quantification of uncertainties; post-processing takes more time than computation.
\end{itemize}

\section{Strong Scaling Evolutionary HPC Applications}
The workflows of HPC applications are becoming more complex, moving from code coupling, resilience, and reproducibility to add application integrating multiphysics, multiphase models, data assimilations, data analytics, and edge computing.

\subsection*{Importance of Legacy Codes}

GPGPU-programming, PGAS and other programming paradigms have not become as widespread as envisioned. Fortran, C++, OpenMP, MPI are still dominating.

The cost of code is so huge that rewriting or re-architecturing an application is almost unfeasible, particularly due to the existence of community code and code developed by ISVs (independent software vendors), and due to specific training of users.

In practice, changes happen on extreme events (e.g. stop working) not in anticipation and cannot be anticipated without access to the new technologies (e.g. NVM).

The larger the code is, the more difficult it becomes to change since the number of possible failures due to source modifications increases dramatically with size as well as the time to run validation tests.

The older the code is, the higher is the probability that the original authors are not around anymore and that nobody really masters the innards of the source code.

The cost of evolution is also a slowing factor. A code developer produces around 10,000 lines of validated code (LOC) per year and a LOC costs between 10\,€ and 100\,€~\cite{prace2012}.

\section{Potentially New HPC Applications}
We expect a \emph{pull} by New Applications and SW Technologies:
\begin{itemize}
    \item High Performance Data Analysis (HPDA): data mining and analysis of big data
    \item Pre- and post-processing, and data assimilation
    \item Integrate simulation, big data and machine learning
    \item Machine Learning: deep learning/neuromorphic in engineering simulation
    \item Internet of Things: real-time and interactive analysis and visualisation (Industry 4.0, smart cities, connected autonomic cars)
    \item New expert programming languages (DSLs)
    \item Approximate computing (concerns both software and hardware)
\end{itemize}

\paragraph{HPDA} will be covered in detail in Section~\ref{sec-datacenter} ``Data Centre and Cloud Computing''. 

\paragraph{Machine Learning} is a very popular approach for Artificial Intelligence that trains systems to learn how to make decisions and predict results on their own. Deep learning is a machine learning technique inspired by the neural learning process of the human brain. Deep learning uses deep neural networks (DNNs), so called because of their deep layering of many connected artificial neurons (sometimes called \emph{perceptrons}), which can be trained with enormous amounts of input data to quickly solve complex problems with high accuracy. Once a neural network is trained, it can be deployed and used to identify and classify objects or patterns in a process known as \emph{inference}.

Most neural networks consist of multiple layers of interconnected neurons. Each neuron and layer contributes towards the task that the network has been trained to execute. For example; AlexNet, the Convolutional Neural Network (CNN) that won the 2012 ImageNet competition, consists of eight layers, 650,000 interconnected neurons, and almost 60 million parameters. Today, the complexity of neural networks has increased significantly, with recent networks such as deep residual networks (for example ResNet-152) having more than 150 layers, and millions more connected neurons and parameters.~\cite{nvidiav100}

Today's leading organizations are using machine learning-based tools to automate decision processes, and they're starting to experiment with more-advanced uses of artificial intelligence (AI) for digital transformation. Corporate investment in artificial intelligence is predicted to triple in 2017, becoming a \$\,100 billion market by 2025.~\cite{Wellers2017}

Deep learning brings a shift in how we approach massive-scale simulations. The early applications of deep learning in using an approximation approach to HPC is taking experimental or supercomputer simulation data and using it to train a neural network, then turning that network around in inference mode to replace or augment a traditional simulation.  Such an approach is incredibly promising, but raises the question of understandability and confidence in the results, unless the neural network is able to explain the reasons for its output. Ultimately, by allowing the simulation to become the training set, the Exascale-capable resources can be used to scale a more informed simulation, or simply be used as the hardware base for a massively scalable neural network.

On the software side, it means that pre- and post-processing data can be trained and certain parts of the application can be scrapped in favour of AI (or numerical approaches can click on at a certain point using trained data). Either way, applications will have to change.~\cite{hemsoth2017shift}

\paragraph{Internet of Things (IoT)} is also having an impact on traditional high-performance computing because of a number of industrial applications that have historically adopted embedded technologies but can benefit from higher performance. Sensors and cyber-physical systems are prominent examples of embedded technologies that require managing and analysing massive amounts of data. In these applications, the embedded systems must collaborate hand-in-hand to filter and analyse data locally due to the massive scale of the data generated prior to consulting with a cloud service for high quality decisions.

\paragraph{New expert programming languages (DSLs)}  Domain-specific languages (DSLs) are computer languages specialized to particular application domains, in contrast to general-purpose languages.~\cite{wikidsl}

\paragraph{Approximate computing} is a computation which returns a possibly inaccurate result rather than a guaranteed accurate result, for a situation where an approximate result is sufficient for a purpose. One example of such situation is for a search engine where no exact answer may exist for a certain search query and hence, many answers may be acceptable. Similarly, occasional dropping of some frames in a video application can go undetected due to perceptual limitations of humans~\cite{approximate}. Scientific domains such as weather and climate prediction have had success using lower precision calculations: 32 bits and even potentially 16 or 8 bits~\cite{dueben2014}. 

Approximate computing trades off computation quality with effort expended, and as rising performance demands confront plateauing resource budgets, approximate computing has become not merely attractive, but even imperative. A survey of techniques for approximate computing is provided by~\cite{Mittal2016}.

\section{HPC Application Requirements}
\subsection{Need for More Performance}
There is no doubt that all user communities see a continued demand for ever-more computational performance well beyond Exascale.  In addition, many users highlight increasing challenges related to data storage and processing. More quantitative details on future computational requirements are in the U.S Advanced Scientific Computing Advisory Committee (ASCAC) report~\cite{ascacreport2010} and the 2012 PRACE Scientific Case~\cite{prace2012}.

\subsection{Adapting Applications for Scalability and Heterogeneity}
HPC applications need to be adapted for Exascale systems and beyond.  It will be some time after the introduction of the first Exaflops machine before more than a handful of applications are able to take full advantage of such a machine.  The biggest issues relate to scalability (identifying and managing sufficient levels of parallelism), heterogeneity (including accelerators), and parallel I/O. Scientific codebases have very long lifetimes, on the order of decades, over which they have earned their users' trust~\cite{ascac2015}. For this reason, HPC application developers are reluctant to rewrite their software, and are keen to follow an incremental path~\cite{eesi2final}. 

There is strong interest in higher-level programming abstractions to provide independence and portability from the details of particular hardware implementations and execution environments, including varying degrees of parallelism, application-specific designs, heterogeneity, accelerators, and complex (deeper) memory hierarchies~\cite{eesi2enabling}. Compilers and runtime systems should perform complex transformations such as overlapping computations and communications~\cite{eesi2final}, auto-tuning~\cite{eesi2enabling}, scheduling and load balancing (especially difficult with multi-scale multiphysics codes). New abstractions are needed to improve parallel I/O. Domain-Specific Languages (DSLs) help by separating domain science from the programming challenges~\cite{eesi2enabling}. Much more research is needed in these areas, but from the application point-of-view, the main barriers to their adoption are lack of standardization or long-term support in compilers and libraries~\cite{eesi2report}, as well as difficulties in the interoperability of multiple programming models in large codebases. Regarding accelerators, there are currently too many incompatible programming interfaces, e.g. CUDA, OpenCL, OpenACC, and OpenMP 4.0, and consolidation on an open, vendor-neutral and widely used standard is needed~\cite{eesi2enabling}. 

There are serious difficulties with performance analysis and debugging, and existing techniques based on printf, logging and trace visualization will soon be intractable. Existing debuggers are good for small problems, but more work is needed to (graphically) track variables to find out where the output first became incorrect, especially for bugs that are difficult to reproduce. Performance analysis tools require lightweight data collection using sampling, folding and other techniques, so as not to increase execution time or disturb application performance (leading to non-representative analysis). There is a need for both superficial on-the-fly analysis and in-depth AI and deep learning analytics. As compilers and runtime systems become more complex, there will be a growing gap between runtime behaviour and the changes in the application's source code required to improve performance—although this does not yet seem to be a significant problem.

There is a concern that future systems will have worse performance stability and predictability, due to complex code transformations, dynamic adapting for energy and faults, dynamically changing clock speeds, and migrating work~\cite{ascac2015}. This is problematic when predictability is required, e.g., for real-time applications such as weather forecasting and for making proposals for access to HPC resources (since proposals need an accurate prediction of application performance scalability). 

\subsection{Need for Co-Design}
Application communities are keen to be involved in co-design activities, in order to ensure appropriate future system designs, with appropriate memory capacities, memory hierarchies, networks and topologies and storage systems well suited to a class of applications~\cite{eesi2enabling}. Users need early access to prototypes, in order to test algorithm performance and provide feedback to system designers~\cite{ascac2015}.  LINPACK performance is seen as non-representative of real world performance.  Long-term partnerships are needed between vendors, HPC centres, research institutes and universities, as is being done in the U.S. ExMatEx (extreme materials), CESAR (advanced reactors) and ExaCT (combustion in turbulence) co-design centres.

\subsection{Extreme Data}
A new paradigm for scientific discovery is emerging due to the exponentially increasing volumes of data generated by HPC simulations and collected from telescopes, colliders, and other scientific instruments or sensors~\cite{ascac2013}.  From the application point of view, the major problem is how to extract new knowledge or insights from the data~\cite{connolly,ascac2013}. Specific problems related to computing systems are \emph{managing data} (streaming data processing, archiving, curation, metadata, provenance, distribution and access), \emph{data analytics} (statistical streaming data analysis, machine learning on high-dimensional data), \emph{data-intensive simulation} (large-scale multi-physics and multi-scale simulations), \emph{data-driven inversion and assimilation} (high-dimensional Bayesian inference, e.g., Full Waveform Inversion for oil and gas), and \emph{statistics and stochastic methods} (direct-inverse uncertainties and extreme event statistics)~\cite{vilotte2016}.  Users may wish to continue using a trusted (but inefficient) algorithm that has worked well on smaller data volumes~\cite{idcbigdata}.

Data movement is a major problem, including distributing data among scientists worldwide at acceptable cost and movement across infrastructure from the point of generation or collection. There is a need for \emph{in situ} analytics and data reduction, with pre-processing, simulation, post-processing and visualization executed on the same HPC cluster. This requires batch and interactive workloads to coexist and it needs  interoperable file formats~\cite{eesi2final} and means of communication between HPC and analytics, such as databases or object stores.

More details on the convergence of HPC and big data are given in Section 3.

\subsection{Interactivity and Usage Models}
There are two broad categories of HPC usage. Capability computing refers to very large jobs that use (almost) the entire machine, e.g., brain simulation, or high-resolution turbulence model, and such a job must complete in the minimum time. Capacity (or throughput) computing refers to a large number of concurrent jobs, with a trade-off between minimising individual job execution time and maximising overall throughput.  Capacity computing currently uses perhaps a few thousand cores per job, and it is commonly used for large ensembles of moderate-scale computations, e.g., for weather or climate simulations (in order to understand the distribution of possible outcomes) and for design space exploration.

There is increasing interest in ``real time'' and interactive supercomputing. High priority simulations are needed for extreme weather and mission-critical jobs (e.g. at NASA). Interactive jobs are also needed, as described above, for in situ visualization, as well as for computational steering: changing parameters in a simulation model as it runs, and changing resolutions in certain places of importance.  Interactive and batch jobs should adapt to dynamic resource availability~\cite{eesi2enabling}, which is an opportunity for new algorithms and programming models.

Finally, there is an opportunity to execute HPC workloads in the cloud, especially for SMEs and to support real time or high priority jobs.  There have been some pilots, that show problems with the cost model, data security~\cite{prace2012} and privacy (e.g. for medical data), licencing problems and data transfer costs.

\subsection{Other Application Issues}
\paragraph{Resiliency} is a vertical problem, and Application-Based Fault Tolerance (ABFT) techniques handle detectable, correctable and silent errors inside the application.  Some algorithms have better fault tolerance than others, for example iterative solvers, which are widely used in Computational Fluid Dynamics (CFD) and other areas tolerate errors (or approximations like analog computing) by executing more iterations.  

\paragraph{Energy Minimization} Since energy consumption is a major concern, users require better tools to measure the energy consumption.  More importantly, they also need to be incentivized to minimise their energy use.

\paragraph{Other Application Issues} outside the scope of this roadmap (because they can be dealt with inside the application communities themselves) include: development of ultra-scalable solvers based on hierarchical algorithms~\cite{eesi2report}, mesh generation~\cite{eesi2report}, verification and validation and uncertainty quantification (VVUQ)~\cite{eesi2report}, difficulty of coupling models at different scales, etc.~\cite{ascac2013}, parallelization in time~\cite{eesi2report}, methods to extract information/understanding from large quantities of scientific data~\cite{connolly}, parallelization in time~\cite{eesi2report}.

%% file: 3_data_center.tex
\chapter{Data Centre and Cloud Computing}
\label{sec-datacenter}

\section{Convergence of HPC and Cloud Computing}
High-performance computing refers to technologies that enable
achieving a high-level computational capacity as compared to a
general-purpose computer \cite{supercomputer}. High-performance
computing in recent decades has been widely adopted for both
commercial and research applications including but not limited to
high-frequency trading, genomics, weather prediction, oil
exploration. Since inception of high-performance computing, these
applications primarily relied on simulation as a third paradigm for
scientific discovery together with empirical and theoretical science.

The technological backbone for simulation has been high-performance
computing platforms (also known as supercomputers) which are
specialized computing instruments to run simulation at maximum speed
with lesser regards to cost. Historically these platforms were
designed with specialized circuitry and architecture ground up with
maximum performance being the only goal. While in the extreme such
platforms can be domain-specific \cite{anton-computer}, supercomputers
have been historically programmable to enable their use for a broad
spectrum of numerically-intensive computation. To benefit from the
economies of scale, supercomputers have been increasingly relying on
commodity components starting from microprocessors in the eighties and
nineties, to entire volume servers with only specialized interconnects
\cite{crayxc} taking the place of fully custom-designed platforms
\cite{np-sc-strategy-shifts}.

In the past decade, there have been two trends that are changing the
landscape for high-performance computing and supercomputers. The first
trend is the emergence of data analytics as the fourth paradigm
\cite{ms-the-fourth-paradigm} complementing simulation in scientific
discovery. The latter is often related to as High-Performance Data
Analytics (HPDA). While simulation still remains as a major pillar for
science, there are massive volumes of scientific data that are now
gathered by instruments, sensors augmenting data from simulation
available for analysis. The Large Hadron Collider and the Square
Kilometre Array are just two examples of scientific experiments that
generate in the order of Petabytes of data a day. This recent trend
has led to the emergence of data science and data analytics as a
significant enabler not just for science but also for humanities.

The second trend is the emergence of cloud computing and
warehouse-scale computers (also known as data centres)
\cite{41606}. Today, the backbone of IT and the ``clouds'' are data
centres that are utility-scale infrastructure. Data centres consist of
low-cost volume processing, networking, and storage servers aiming at
cost-effective data manipulation at unprecedented scales. Data centre
owners prioritize capital and operating costs (often measured in
performance per watt) over ultimate performance. Typical high-end
data centres draw around 20 MW, occupy an area equivalent to 17 times a
football field and incur a 3 billion Euros in investment. While
data centres are primarily designed for commercial use, the scale at
which they host and manipulate (e.g., personal, business) data has led
to fundamental breakthroughs in both data analytics and data
management. By pushing computing costs to unprecedented low limits and
offering data and computing services at a massive scale, the clouds
will subsume much of embarrassingly parallel scientific workloads in
high-performance computing, thereby pushing custom infrastructure for
the latter to a niche.

\section{Massive Data Analytics}
We are witnessing a second revolution in IT, at the centre of which is
data. The emergence of e-commerce and massive data analytics for
commercial use in search engines, social networks and online shopping
and advertisement has led to wide-spread use of massive data analytics
(in the order of Exabytes) for consumers. Data now also lies at the
core of the supply-chain for both products and services in modern
economies. Collecting user input (e.g., text search) and documents
online not only has led to ground-breaking advances in language
translation but is also in use by investment banks mining blogs to
identify financial trends. The IBM Watson experiment is a major
milestone in both natural language processing and decision making to
showcase a question answering system based on advanced data analytics
that won a quiz show against human players.

The scientific community has long relied on generating (through
simulation) or recording massive amounts of data to be analysed
through high-performance computing tools on supercomputers. Examples
include meteorology, genomics, connectomics (connectomes:
comprehensive maps of connections within an organism's nervous
system), complex physics simulations, and biological and environmental
research. The proliferation of data analytics for commercial use on
the internet, however, is paving the way for technologies to collect,
manage and mine data in a distributed manner at an unprecedented scale
even beyond conventional supercomputing applications.

Sophisticated analytic tools beyond indexing and rudimentary
statistics (e.g., relational and semantic interpretation of underlying
phenomena) over this vast repository of data will not only serve as
future frontiers for knowledge discovery in the commercial world but
also form a pillar for scientific discovery \cite{NAP18374}. The
latter is an area where commercial and scientific applications
naturally overlap, and high-performance computing for scientific
discovery will highly benefit from the momentum in e-commerce.

There are a myriad of challenges facing massive data analytics
including management of highly distributed data sources, and tracking
of data provenance, data validation, mitigating sampling bias and
heterogeneity, data format diversity and integrity, integration,
security, sharing, visualization, and massively parallel and
distributed algorithms for incremental and/or real-time analysis.

With respect to algorithmic requirements and diversity, there are a
number of basic operations that serve as the foundation for
computational tasks in massive data analytics (often referred to as
\emph{dwarfs} \cite{Asanovic:EECS-2006-183} or \emph{giants}
\cite{NAP18374}). They include but are not limited to: basic
statistics, generalized n-body problems, graph analytics, linear
algebra, generalized optimization, computing integrals and data
alignment. Besides classical algorithmic complexity, these basic
operations all face a number of key challenges when applied to massive
data related to streaming data models, approximation and sampling,
high-dimensionality in data, skew in data partitioning, and sparseness
in data structures. These challenges not only must be handled at the
algorithmic level, but should also be put in perspective given
projections for the advancement in processing, communication and
storage technologies in platforms.

Many important emerging classes of massive data analytics also have
real-time requirements. In the banking/financial markets, systems
process large amounts of real-time stock information in order to
detect time-dependent patterns, automatically triggering operations in
a very specific and tight timeframe when some pre-defined patterns
occur. Automated algorithmic trading programs now buy and sell
millions of dollars of shares time-sliced into orders separated by 1\,ms.
Reducing the latency by 1\,ms can be worth up to \$\,100 million a
year to a leading trading house. The aim is to cut microseconds off
the latency in which these systems can reach to momentary variations
in share prices \cite{algo-trading}.

\section{Warehouse-Scale Computers}
Large-scale internet services and cloud computing are now fuelled by
large data centres which are a warehouse full of computers. These
facilities are fundamentally different from traditional supercomputers
and server farms in their design, operation and software structures
and primarily target delivering a negotiated level of internet service
performance at minimal cost. Their design is also holistic because
large portions of their software and hardware resources must work in
tandem to support these services \cite{41606}.

High-performance computing platforms are also converging with
warehouse scale computers primarily due to the growth rate in cloud
computing and server volume in the past decade. James Hamilton, Vice
President and Distinguished Engineer at Amazon and the architect of
their data centres commented on the growth of Amazon Web Services (AWS)
stating in 2014 that ``every day AWS adds enough new server capacity to
support Amazon's global infrastructure when it was a \$7B annual
revenue enterprise (in 2004)''.

Silicon technology trends such as the end of Dennard Scaling
\cite{5967003} and the slowdown and the projected end of density
scaling \cite{te-after-moores-law} are pushing computing towards a new
era of platform design tokened ISA: (1) technologies for tighter
integration of components (from algorithms to infrastructure), (2)
technologies for specialization (to accelerate critical services), and
(3) technologies to enable novel computation paradigms for
approximation. These trends apply to all market segments for digital
platforms and reinforce the emergence and convergence of volume
servers in warehouse-scale computers as the building block for
high-performance computing platforms.

With modern high-performance computing platforms being increasingly
built using volume servers, there are a number of salient features
that are shared among warehouse-scale computers and modern
high-performance computing platforms including dynamic resource
allocation and management, high utilization, parallelization and
acceleration, robustness and infrastructure costs. These shared
concerns will serve as incentive for the convergence of the platforms.

There are also a number of ways that traditional high-performance
computing ecosystems differ from modern warehouse-scale computers
\cite{hpc-cloud-bad}. With performance being a key criterion, there are a number of
challenges facing high-performance computing on warehouse-scale
computers. These include but are not limited to efficient
virtualization, adverse network topologies and fabrics in cloud
platforms, low memory and storage bandwidth in volume servers,
multi-tenancy in cloud environments, and open-source deep software
stacks as compared to traditional supercomputer custom stacks. As
such, high-performance computing customers must adapt to co-exist with
cloud services given these challenges, while warehouse-scale computer
operators must innovate technologies to support the workload and
platform at the intersection of commercial and scientific computing.

\section{Cloud-Embedded HPC and Edge Computing}
The emergence of data analytics for sciences and warehouse scale
computing will allow much of the HPC that can run on massively
parallel volume servers at low cost to be embedded in the clouds,
pushing infrastructure for HPC to the niche. While the cloud vendors
primarily target a commercial use of large-scale IT services and may
not offer readily available tools for HPC, there are a myriad of
opportunities to explore technologies that enable embedding HPC into
public clouds.

Large-scale scientific experiments also will heavily rely on edge
computing. The amount of data sensed and sampled is far beyond any
network fabric capabilities for processing in remote sites. For
example, in the Large Hadron Collider (LHC) in CERN, beam collisions
occur every 25 ns, which produce up to 40 million events per
second. All these events are pipelined with the objective of
distinguishing between interesting and non-interesting events to
reduce the number of events to be processed to a few hundreds events~\cite{supersymmetry}. These endeavours will need custom solutions with
proximity to sensors and data to enable information extraction and
hand in hand collaboration with either HPC sites or cloud-embedded HPC
services.


%% file: 4_disruptive_technologies.tex
\chapter{Disruptive Technologies in Hardware/VLSI}
\label{sec-disruptive}

\section{Introduction}
Roadmapping beyond the upcoming Exascale machines (2023 -- 2030) is extremely speculative. The basic idea of the Eurolab-4-HPC vision is therefore to assess potentially disruptive technologies and summarize their impacts on HPC hardware as \emph{IF~...~THEN~...} statements, i.e. \emph{IF} disruptive technology will be available \emph{THEN} potential impact on hardware could be.

We survey the current state of research and development and its potential for the future of the following VLSI/hardware technologies: 
\begin{itemize}
    \setlength\itemsep{0.25\baselineskip}
    \setlength\parskip{0pt}
    \item CMOS scaling
    \item Die stacking and 3D chip technologies
    \item Non-volatile Memory (NVM) technologies
    \item Photonics
    \item Resistive Computing
    \item Neuromorphic Computing
    \item Quantum Computing
    \item Nanotubes and Nanowires
    \item Graphene 
    \item Diamond Transistors
\end{itemize}
\vskip\baselineskip

To sort the different technologies we define three different types of disruptive technology innovations besides Sustaining Technology. The above technologies are filed as follows:

\begin{itemize}

    \item \emph{Sustaining:}  An innovation that does not principally affect existing HPC. Innovations improving HPC hardware in ways that were generally expected.
        \vspace{-1.5ex}
        \begin{itemize}
            \item CMOS scaling and Die stacking, see section~\ref{sec-sustaining}
        \end{itemize}

    \item \emph{Disruptive technologies that create a new line of HPC hardware} in a way generally expected. 
        \vspace{-1.5ex}
        \begin{itemize}
            \item NVM and Photonics, see section~\ref{sec-disruptive-hw}
        \end{itemize}

    \item \emph{Disruptive technologies that potentially create alternative ways of computing}.
        \vspace{-1.5ex}
        \begin{itemize}
            \item Resistive, Neuromorphic, and Quantum Computing, see section~\ref{sec-disruptive-alt}
        \end{itemize}

    \item \emph{Disruptive technologies that potentially replace CMOS} for processor logic.
        \vspace{-1.5ex}
        \begin{itemize}
            \item Nanotube, Graphene, and Diamond technologies, see section~\ref{sec-disruptive-beyond}
        \end{itemize}
\end{itemize}

We summarize potential long-term impacts of Disruptive Technologies on HPC hardware in section~\ref{sec-impact} of the vision. Such impacts could concern the processor logic, the memory hierarchy, and potential hardware accelerators.

%% file: 4_2_sustaining_technology.tex
\section{Sustaining Technology (improving HPC HW in ways that are generally expected)}
\label{sec-sustaining}

%% file: 4_2_1_cmos_scaling.tex
\subsection{Continuous CMOS scaling}

Continuing Moore's Law and managing power and performance tradeoffs remain as the key drivers of the International Technology Roadmap For Semiconductors 2015 Edition (ITRS 2015) grand challenges. Silicon scales according to the Semiconductor Industry Association's ITRS 2.0, Executive Summary 2015 Edition~\cite{itrs-exec} to 11/10\,nm in 2017, 8/7\,nm in 2019, 6/5\,nm in 2021, 4/3\,nm in 2024, 3/2.5\,nm in 2027, and 2/1.5\,nm in 2030 for MPUs or ASICs.

DRAM half pitch (i.e., half the distance between identical features in an array) is projected to scale down to 10\,nm in 2025 and 7.7\,nm in 2028 allowing up to 32\,GBits per chip.  However, DRAM scaling below 20\,nm is very challenging. DRAM products are approaching fundamental limitations as scaling DRAM capacitors is becoming very difficult in 2D structures. It is expected that these limits will be reached by 2024 and after this year DRAM technology will saturate at the 32\,Gbit level unless some major breakthrough will occur~\cite{itrs-report-dram}. The same report foresees that by 2020 the 2D Flash topological method will reach a practical limit with respect to cost effective realization of pitch dimensions. 3D stacking is already extensively used to scale flash memories by 3D flash memories.

Process downscaling results in increasing costs below 10\,nm: the cost per wafer increases from one technology node to the next~\cite{hipeac-vision-2015}. The ITRS roadmap does not guarantee that silicon-based CMOS will extend that far because transistors with a gate length of 6nm or smaller are significantly affected by quantum tunneling.

CMOS scaling depends on fabs that are able to manufacture chips at the highest technology level. Only four such fabs are remaining worldwide: GlobalFoundries, TSMC, Intel and Samsung.

\subsubsection{Current State}
Current (July 2017) high-performance multiprocessors feature 14- to 16-nm technology. 14-nm FinFET technology is available by Intel (Intel Kaby Lake) and GlobalFoundries. 10-nm manufacturing process is expected for 2nd half of 2017 or beginning of 2018 by Intel (Cannonlake processor), Intel's difficulties and changed plans show the continuing challenges with keeping pace with Moore's law. Samsung and TSMC also apply 10-nm technology in 2017.

Samsung revealed in March 2017 that it had shipped over 70 thousand wafers processed using its first-generation 10\,nm FinFET fabrication process (10LPE). This manufacturing process allowed the company to make its chips 30\% smaller compared to ICs made using its 14LPE process as well as reducing power consumption by 40\% (at the same frequency and complexity) or increasing their frequency by 27\% (at the same power and complexity). Samsung applies that process to the company's own Exynos 9 Octa 8895 as well as Qualcomm's Snapdragon 835 seen in the Samsung Galaxy S8~\cite{anandtech-samsung-tsmc}.

\subsubsection{Company Roadmaps}
R\&D has begun for 5\,nm by all four remaining fabs TSMC, GlobalFoundries, Intel and Samsung and also beyond towards 3\,nm. Both 5\,nm and 3\,nm present a multitude of unknowns and challenges. Regardless, based on the roadmaps from various chipmakers, Moore's Law continues to slow as process complexities and costs escalate at each new chip generation. 

\textbf{Intel} plans 7\,nm FinFET for production in early to mid-2020, according to industry sources. Intel's 5\,nm production is targeted for early 2023, sources said, meaning its traditional 2-year process cadence is extending to roughly 2.5 to 3 years~\cite{semieng-uncertainty}.

\textbf{TSMC} plans to ship 5\,nm in 2020, which is also expected to be a FinFET. In reality, though, TSMC's 5\,nm will likely be equivalent in terms of specs to Intel's 7\,nm, analysts said~\cite{semieng-uncertainty}.

TSMC will be starting risk production of their 7\,nm process in early 2017 and is already actively in development of 5\,nm process technology as well. Furthermore, TSMC is also in development of 3nm process technology. Although 3\,nm process technology already seems so far away, TSMC is further looking to collaborate with academics to begin developing 2\,nm process technology~\cite{cpcr-already}.

\textbf{Samsung}'s newest foundry process technologies and solutions introduced at the annual Samsung Foundry Forum include 8\,nm, 7\,nm, 6\,nm, 5\,nm, 4\,nm in its newest process technology roadmap~\cite{samsung-process-roadmap}. However, no time scale is provided.

\textbf{GlobalFoundries} decided to skip 10\,nm in favor for its next generation 7\,nm manufacturing technology, which is planned to start mass production of commercial chips in the second half of 2018.  It is targeting high-performance components, such as CPUs, GPUs and SoCs for various applications (mobile, PC, servers, etc.)~\cite{anandtech-globalfoundries-roadmap}.

Compared to GlobalFoundries' current leading-edge 14LPP fabrication technology, the initial DUV-only (deep ultraviolet) 7\,nm process promises over 50\% area reduction as well as over 60\% power reduction (at the same frequency and complexity) or over 30\% performance improvement (at the same power and complexity). In practice, this means that in an ideal scenario GlobalFoundries will be able to double the amount of transistors per chip without increasing its die size while improving its performance per watt characteristics~\cite{anandtech-globalfoundries-roadmap}.

Initially GlobalFoundries will continue to use DUV argon fluoride (ArF) excimer lasers with 193\,nm wavelength with its 7\,nm production process, but over time it hopes to insert extreme ultraviolet lithography (EUV) tools with 13.5\,nm wavelength into production flow. GlobalFoundries does not reveal timeframes for its 7\,nm with EUV, but it is safe to say that EUV will be used in 2019 at the earliest. Also Samsung, Intel, and TSMC have confirmed their intentions to pursue a DUV-only 7\,nm process technology~\cite{anandtech-globalfoundries-roadmap}.

\subsubsection{Research Perspective}
``It is difficult to shed a tear for Moore's Law when there are so many interesting architectural distractions on the systems horizon''~\cite{nextplatform-neuromorphic}. However, silicon technology scaling will still continue and research in silicon-based hardware is still prevailing, in particular targeting specialized and heterogeneous processor structures and hardware accelerators. 

However, each successive process shrink becomes more expensive and therefore each wafer will be more expensive to deliver. One trend to improve the density on chips will be 3D integration also of logic. Hardware structures that mix silicon-based logic with new NVM technology are upcoming and intensely investigated.  A revolutionary DRAM/SRAM replacement will be needed~\cite{itrs-exec}.

As a result, non-silicon extensions of CMOS, using III–V materials or Carbon nanotube/nanowires, as well as non-CMOS platforms, including molecular electronics, spin-based computing, and single-electron devices, have been proposed~\cite{itrs-exec}.

For a higher integration density, new materials and processes will be necessary. Since there is a lack of knowledge of the fabrication process of such new materials, the reliability might be lower, which may result in the need of integrated fault-tolerance mechanisms~\cite{itrs-exec}.

Research in CMOS process downscaling and building fabs is driven by industry, not by academic research. Availability of such CMOS chips will be a matter of costs and not only of availability of technology.

%% file: 4_2_2_die_stacking.tex
\subsection{Die Stacking and 3D-Chip}

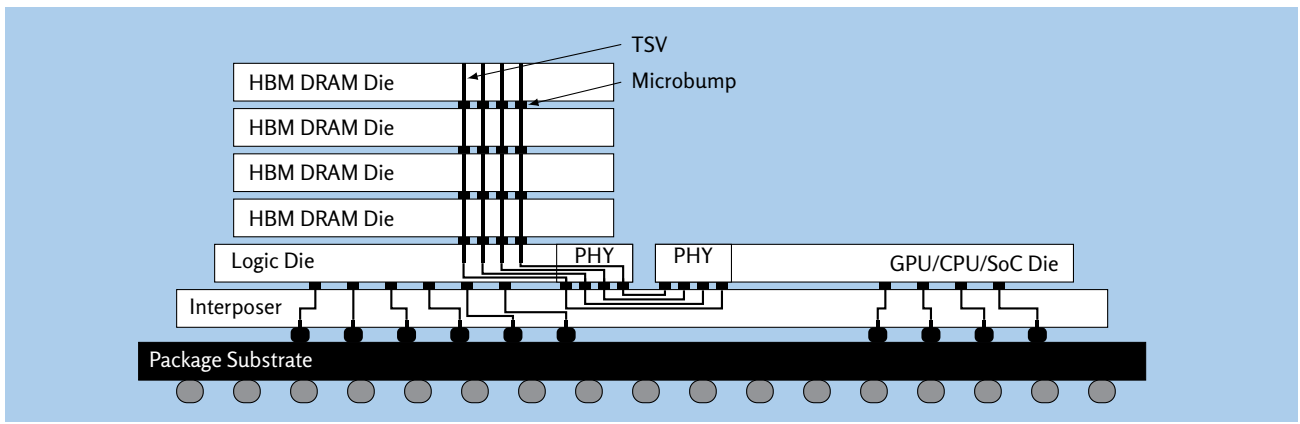
\begin{figure*}[ht]
    \centering
    \begin{tikzpicture}
        {\sffamily
        \ifx\deliverable\undefined
        \fill[background] (-1.0,1.75) rectangle ++(17,5.5);
        \else
        \fi
        \newcommand{\hbmdie}[1]{%
            \draw[black,fill=white]   (2,#1) rectangle node[black,anchor=east,align=left,xshift=-0.25cm,font=\scriptsize] {HBM DRAM Die} ++(5,0.5);
            \fill[black]    (5,#1) rectangle ++(0.05,0.5);
            \fill[black]    (5.25,#1) rectangle ++(0.05,0.5);
            \fill[black]    (5.5,#1) rectangle ++(0.05,0.5);
            \fill[black]    (5.75,#1) rectangle ++(0.05,0.5);
            \fill[black]    (4.95,#1) rectangle ++(0.15,-0.1);
            \fill[black]    (5.2,#1) rectangle ++(0.15,-0.1);
            \fill[black]    (5.45,#1) rectangle ++(0.15,-0.1);
            \fill[black]    (5.7,#1) rectangle ++(0.15,-0.1);
        }
        \newcommand{\mbump}[1]{%
            \fill[black]    (#1) rectangle ++(0.15,-0.1);
        }
        \newcommand{\bbump}[1]{%
            \fill[black,rounded corners=2pt]    (#1) rectangle ++(0.25,-0.2);
            \fill[black]    ([xshift=0.1cm]#1) rectangle ++(0.05,0.1);
        }
        \newcommand{\lbump}[1]{%
            \draw[black,fill=vdarkgray,rounded corners=4pt]    (#1) rectangle ++(0.35,-0.3);
        }
        \hbmdie{6}
        \hbmdie{5.4}
        \hbmdie{4.8}
        \hbmdie{4.2}
        \node[font=\scriptsize,anchor=west] (tsv) at (7.1,6.75) {TSV};
        \node[font=\scriptsize,anchor=west] (mib) at (7.1,6.25) {Microbump};
        \draw[-latex] (tsv.west) -- (5.07,6.3);
        \draw[-latex] (mib.west) -- (5.87,5.95);

        \draw[black,fill=white]   (1.75,3.6) rectangle node[black,anchor=east,align=left,xshift=-1.3cm,font=\scriptsize] {Logic Die} ++(5.5,0.5);
        \fill[black]    (5,3.85) rectangle ++(0.05,0.25);
        \fill[black]    (5.25,3.85) rectangle ++(0.05,0.25);
        \fill[black]    (5.5,3.85) rectangle ++(0.05,0.25);
        \fill[black]    (5.75,3.85) rectangle ++(0.05,0.25);
        \mbump{3.0,3.6}
        \mbump{3.5,3.6}
        \mbump{4.0,3.6}
        \mbump{4.5,3.6}
        \mbump{5.0,3.6}
        \mbump{5.5,3.6}

        \draw[black,fill=white]   (6.25,3.6) rectangle node[black,yshift=0.1cm,font=\scriptsize] {PHY} ++(1,0.5);
        \mbump{6.3,3.6}
        \mbump{6.55,3.6}
        \mbump{6.80,3.6}
        \mbump{7.05,3.6}

        \draw[black,fill=white]   (7.55,3.6) rectangle node[black,anchor=west,align=right,xshift=0.2cm,font=\scriptsize] {GPU/CPU/SoC Die} ++(5.5,0.5);
        \mbump{10.5,3.6}
        \mbump{11.0,3.6}
        \mbump{11.5,3.6}
        \mbump{12.0,3.6}

        \draw[black,fill=white]   (7.55,3.6) rectangle node[black,yshift=0.1cm,font=\scriptsize] {PHY} ++(1,0.5);
        \mbump{7.6,3.6}
        \mbump{7.85,3.6}
        \mbump{8.10,3.6}
        \mbump{8.35,3.6}

        \draw[black,fill=white]   (1.25,3.0) rectangle node[black,anchor=east,align=left,xshift=-4.6cm,font=\scriptsize] {Interposer} ++(12.25,0.5);
        \bbump{2.75,3.0};
        \bbump{3.45,3.0};
        \bbump{4.15,3.0};
        \bbump{4.85,3.0};
        \bbump{5.55,3.0};
        \bbump{6.25,3.0};
        \bbump{10.35,3.0};
        \bbump{11.05,3.0};
        \bbump{11.75,3.0};
        \bbump{12.45,3.0};

        \draw[black,fill=black]   (0.75,2.3) rectangle node[white,anchor=east,align=left,xshift=-4.2cm,font=\scriptsize] {Package Substrate} ++(13.25,0.5);
        \lbump{1.25,2.3};
        \lbump{2.0,2.3};
        \lbump{2.75,2.3};
        \lbump{3.5,2.3};
        \lbump{4.25,2.3};
        \lbump{5.0,2.3};
        \lbump{5.75,2.3};
        \lbump{6.5,2.3};
        \lbump{7.25,2.3};
        \lbump{8.0,2.3};
        \lbump{8.75,2.3};
        \lbump{9.5,2.3};
        \lbump{10.25,2.3};
        \lbump{11.0,2.3};
        \lbump{11.75,2.3};
        \lbump{12.5,2.3};
        \lbump{13.25,2.3};

        \draw[thick] (5.025,3.85) -- ++(0,-0.19) -- ++(1.35,0) -- ++(0,-0.1);
        \draw[thick] (5.275,3.85) -- ++(0,-0.14) -- ++(1.35,0) -- ++(0,-0.2);
        \draw[thick] (5.525,3.85) -- ++(0,-0.09) -- ++(1.35,0) -- ++(0,-0.2);
        \draw[thick] (5.775,3.85) -- ++(0,-0.04) -- ++(1.35,0) -- ++(0,-0.3);

        \draw[thick] (6.375,3.65) -- ++(0,-0.40) -- ++(2.05,0) -- ++(0,0.24);
        \draw[thick] (6.625,3.65) -- ++(0,-0.34) -- ++(1.55,0) -- ++(0,0.18);
        \draw[thick] (6.875,3.65) -- ++(0,-0.28) -- ++(1.05,0) -- ++(0,0.12);
        \draw[thick] (7.125,3.65) -- ++(0,-0.22) -- ++(0.55,0) -- ++(0,0.06);

        \draw[thick] (3.075,3.6) -- ++(0,-0.35) -- ++(-0.2,0) -- ++(0,-0.15);
        \draw[thick] (3.575,3.6) -- ++(0,-0.35) -- ++(0.0,0) -- ++(0,-0.15);
        \draw[thick] (4.075,3.6) -- ++(0,-0.35) -- ++(0.2,0) -- ++(0,-0.15);
        \draw[thick] (4.575,3.6) -- ++(0,-0.35) -- ++(0.4,0) -- ++(0,-0.15);
        \draw[thick] (5.075,3.6) -- ++(0,-0.45) -- ++(0.6,0) -- ++(0,-0.05);
        \draw[thick] (5.575,3.6) -- ++(0,-0.4) -- ++(0.8,0) -- ++(0,-0.1);

        \draw[thick] (10.575,3.6) -- ++(0,-0.35) -- ++(-0.1,0) -- ++(0,-0.15);
        \draw[thick] (11.075,3.6) -- ++(0,-0.35) -- ++(0.1,0) -- ++(0,-0.15);
        \draw[thick] (11.575,3.6) -- ++(0,-0.35) -- ++(0.3,0) -- ++(0,-0.15);
        \draw[thick] (12.075,3.6) -- ++(0,-0.35) -- ++(0.5,0) -- ++(0,-0.15);
        }
    \end{tikzpicture}
    \caption{High Bandwidth Memory utilizing an active silicon Interposer~\cite{amdhbm}}
    \label{fig-diestacking}
\end{figure*}

Die Stacking and 3D chip integration denote the concept of stacking integrated circuits (e.g. processors and memories) vertically in multiple layers. 3D packaging assembles vertically stacked dies in a package, e.g., system-in-package (SIP) and package-on-package (POP). 

Die stacking can be achieved by connecting separately manufactured wafers or dies vertically either via wafer-to-wafer, die-to-wafer, or even die-to-die. The mechanical and electrical contacts are realized either by wire bonding as in SIP and POP devices or microbumps. SIP is sometimes listed as a 3D stacking technology, although it should be better denoted as 2.5D technology.

An evolution of SiP approach consists of stacking multiple dies (called chiplets) on a large interposer that provides connectivity among chiplets and to the package. The interposer can be passive or active.   A passive interposer, often implemented with an organic  material to reduce cost, provides multiple levels of metal interconnects and vertical vias for inter-chiplet connectivity and for redistribution of connections to the package. It also provides micropads for the connection of  the chiplets on top.   Active silicon interposers offer the additional possibility to include logic and circuits in the interposer itself. This more advanced and high cost integration approach is much more flexible than passive interposers, but it is also much more challenging for design, manufacturing, test and thermal management. Hence it is not yet widespread in commercial products, with the exception of three-dimensional DRAM memories (e.g. High-Bandwidth Memory (HBM, cf. Fig.~\ref{fig-diestacking}) and Hybrid Memory Cube (HMC)) where the bottom layer is active and hosts the physical interface of the memory to the external system.  

The advantages of 3D technology based on Interposers are numerous: Firstly, short communication distance between dies, thus reducing communication load and then reducing communication power consumption. Secondly, the possibility of stacking dies from various heterogeneous technologies, like stacking memory on top of logic like Flash, non-volatile memories, or even photonic devices, in order to benefit of the best technology where it fits best. And thirdly, an improved system yield and cost by partitioning the system in a divide and conquer approach: multiple similar dies are fabricated, tested and sorted before the final 3D assembly, instead of fabricating ultra large dies with much reduced yield. The main challenges to the diffusion of these technologies are manufacturing cost (setup and yield optimization) and thermal management since cooling high-performance die stacks requires complex packages, thermal coupling materials and heat spreaders. 

Die stacking can also be achieved by stacking active layers vertically on a single wafer in a monolithic approach. Such kind of 3D chip integration does not use micro-pads or Through-Silicon Vias (TSVs) for communication, but it uses multi-level interconnects between layers, with a much finer pitch than that allowed by TSVs.  The main challenge in monolithic integration is to ensure that elementary devices (transistors) have similar quality level and performance in all the silicon layers. This is a very challenging goal since the manufacturing process is not identical for all the layers (low temperature processes are needed for the layers grown on top of the bulk layer). 

Some advanced solutions for vertical communication do not require ohmic contact in metal, i.e. capacitive and inductive coupling as well as short-range RF communication solutions that are proposed instead do not require a flow of electrons passing through a continuous metal connection. These approaches are usable both in die-stacked and monolithically integrated ICs, but the modulation and demodulation circuits do take space and vertical connectivity density is currently not better than that of TSVs. 

\subsubsection{Current State}
The monolithic approach of die stacking is already used in 3D Flash memories from Samsung and also for smart sensors. Commercial prototypes of 3D technology date back until 2004 when Tezzaron released a 3D IC microcontroller~\cite{tezzaron2016}. Intel evaluated chip stacking for a Pentium 4 already in 2006~\cite{Black2006}. Recent multicore designs using Tezzaron's technology include the 64 core 3D-MAPS (3D MAssively Parallel processor with Stacked memory) research prototype from 2012~\cite{kim20123dmaps} and the Centip3De with 64 ARM Cortex-M3 Cores also from 2012~\cite{fick2012centipede}. Fabs are able to handle 3D packages (e.g.~\cite{amkor3d}). In 2011 IBM announced 3D chip production process~\cite{ibm3dchip}. Intel announced "3D XPoint" memory in 2015 (assumed to be 10x the capacity of DRAM and 1000x faster than NAND Flash~\cite{inteloptane}). Intel/Micro 3D-Xpoint memory is now available as Optane-SSDs DC P4800X-SSD as 375-Gbyte since March 2017 and stated to be 2.5 to 77 times better than NAND-SSDs. 

Both NVIDIA and AMD already exploit the high-bandwidth and low latencies given by 3D stacked memories for a high-dense memory processor, called High-Bandwidth Memory (HBM). AMD's GPUs based on the Fiji architecture with HBM are available since 2015, and NVIDIA released Pascal-based GPUs in 2016~\cite{nvidiapascal}. A direction towards future 3D stacking of memory dies on processor dies is the Hybrid Memory Cube from Micron~\cite{micron-hmc}. It stacks multiple DRAM dies and a separate layer for a controller which is vertically linked with the DRAM dies. This interposer approach is used in high end FPGAs to reduce cost.

\subsubsection{Perspective}
3D NAND Flash may be prevailing. 3D Flash memories may enable SSDs with up to 10 TB of capacity in the short term~\cite{registernand}. In 2007, earliest potential was seen in memory stacks for mobile applications~\cite{lu20073d}. It is to expect that 3D chip technology will widely enter the market for mainstream architectures within the next 5 years. Representative for this current development are, e.g., Intel's Xeon Phi Knights Landing processors equipped with package-integrated DRAMs in 2016 as a result of their cooperation with Micron.  While memories are already exploiting the full spectrum of 3D integration options, logic processes lag behind in this respect, mostly because of cost and thermal reasons. Since memories are much cooler than logic (due to the lower activity and operating frequency) their 3D integration is thermally sustainable and cost-effective today. 

It is also to be expected that in a long-term perspective the technology will be expanded progressively from 3D packaging technologies towards real 3D chip stacking and possibly towards 3D ICs in 3D packages in order to profit from all the benefits such technology will offer in particular for HPC architectures. 

The main challenge in establishing this 3D chip stacking technology is gaining control of the thermal problems that have to be overcome to realize reliably very dense 3D interconnections. This requires the availability of appropriate design tools, which are explicitly supporting 3D layouts. Both topics represent an important issue for research in the next 10 to 15 years. 

\subsubsection{Impact on Hardware}
3D stacking has a series of beneficial impacts on the hardware in general and on the possibilities how to design future processor-memory-architectures in particular. Wafers can be partitioned into smaller dies because comparatively long horizontally running links are relocated to the third dimension and thus enable smaller form factors. 3D stacking also enables heterogeneity, by integrating layers, manufactured in different processes, e.g., different memory technologies, like SRAM, DRAM, Spin-transfer-torque RAM (STT–RAM) and also memristor technologies, which would be incompatible among each other in monolithic circuits. Due to short connection wires, reduction of power consumption is to be expected. Simultaneously, a high communication bandwidth between layers connected with TSVs can be expected leading to particularly high processor-to-memory bandwidth.

The last-level caches will probably be the first to be affected by 3D stacking technologies when they will enter logic processes. 3D caches will increase bandwidth and reduce latencies by a large cache memory stacked on top of logic circuitry. In a further step it is consequent to expand 3D chip integration also to main memory in order to make a strong contribution in reducing decisively the current memory wall which is one of the strongest obstructions in getting more performance in HPC systems. Furthermore, possibly between 2026 and 2030, 3D arithmetic units will undergo the same changes ending up in complete 3D many-core microprocessors, which are optimized in power consumption due to reduced wire lengths.

A collateral but very interesting trend is 3D stacking of sensors. A technology was presented by Olympus in which more than 4 million microbumps have been used for stacking a 16 megapixel array sensor directly on top of a circuit implementing a global shutter control logic. Sony used TSV technology to combine image sensors directly with column-parallel analogue-digital-converters and logic circuits~\cite{kondo2015cmos,sonysensor}. This trend will open the opportunity for fabricating fully integrated systems that will also include sensors and their analogue-to-digital interfaces. 

\subsubsection{Funding Perspectives}
The main issue is that 3D as a technology requires heavy  industrial investment because it ultimately is a problem of reaching cost effective volume production. So this is probably beyond what can be funded by research money.  However, more and more hardware devices will use 3D technology, so even system-level design will need to become 3D-aware. So definitely the EU needs to invest in research on how to develop components and systems based on 3D technology. Even if 3D technology will not be developed in EU at production level, EU should invest in research to design effective components and computing systems that use 3D technology.


%% file: 4_3_disruptive_technologies.tex
\section{Disruptive Technology in Hardware/VLSI (innovation that creates a new line of HPC hardware superseding existing HPC techniques)}
\label{sec-disruptive-hw}

%% file: 4_3_1_nvm.tex
\subsection{Non-volatile Memory (NVM) Technologies}
\label{sec-nvm}

The computer architecture development in the last 1.5 decades was primarily characterized by energy driven advancement (better performance/Watt ratio) and by the transition from single- to multi-/many-core and to heterogeneous architectures consisting of a multi-core processor and an accelerator. New Non-volatile Memory (NVM) technologies will strongly influence the memory hierarchy and potentially lead to Resistive Computing and new Neuromorphic Computing chips (see section~\ref{sec-neuromorphic}).

Currently NAND Flash is the most common NVM technology, which finds its usages on SSDs, memory cards and memory sticks. NAND Flash uses floating-gate transistors for storing single bits. This technology is facing a big challenge, because scaling down decreases the endurance and performance significantly \cite{shimpi2013samsung}. Hence the importance of other NVM technologies increases. Using emerging NVM technologies in computing systems is a further step towards energy-aware measures for future HPC architectures.

Resistive memories, i.e. memristors, are an emerging class of non-volatile memory technology. A memristor is defined by Leon Chua's system theory as a memory device with a hysteresis loop that is pinched i.e. their I–U (current–voltage) curve goes to the zero point of the coordinate system. To this memristor class belong PCM, ReRAM, CBRAM and STT-RAMs.  The memristors electrical resistance is not constant but depends on the previously applied voltage and the resulting current. The device remembers its history—the so-called non-volatility property: when the electric power supply is turned off, the memristor remembers its most recent resistance until it is turned on again \cite{wpMemristor}. 

Among the most prominent memristor candidates and close to commercialization are phase change memory (PCM) \cite{lee2010phase,lam2008cell,lee2009architecting,qureshi2009scalable,zhou2009durable}, metal oxide resistive random access memory (RRAM or ReRAM) \cite{xu2015overcoming,xu2011design}, and conductive bridge random access memory (CBRAM) \cite{wong2014conductive}. 

PCM can be integrated in the CMOS process and the read/write latency is only by tens of nanoseconds slower than DRAM whose latency is roughly around 100ns. The write endurance is hundreds of millions of writes per cell at current processes. This is why PCM is currently positioned only as a Flash replacement \cite{lee2009architecting}. RRAM offers a simple cell structure which enables reduced processing costs. The endurance can be more than 50 million cycles and the switching energy is very low \cite{govoreanu201110}. RRAM can deliver 100x lower read latency and 20x faster write performance compared to NAND Flash \cite{Crossbar}. CBRAM can also write with relatively low energy and with high speed. The read/write latencies are close to DRAM.

Spintronics is the technology of manipulating the spin state of electrons. Instead of using the electrons charge, spin states can be utilized as a substitute in logical circuits or in traditional memory technologies like SRAM. A Spin-transfer torque magnetic random-access memory (STT-RAM) \cite{apalkov2013spin} memory cell stores data in a magnetic tunnel junction (MTJ). Each MTJ is composed of two ferromagnetic layers (free and reference layers) and one tunnel barrier layer (MgO). If the magnetization direction of the magnetic fixed reference layer and the switchable free layer is anti-parallel, resp. parallel, a high, resp. a low, resistance is adjusted, representing a digital "0" or "1".  Recently it was reported that by adjusting intermediate magnetization angles in the free layer 16 different states can be stored in one physical cell, enabling to realize multi-cell storages in MTJ technology \cite{bernard2016spintronic}.
 
The read latency and read energy of STT-RAM is expected to be comparable to that of SRAM. The expected 3x higher density and 7x less leakage power consumption in the STT-RAM makes it suitable for replacing SRAMs to build large NVMs. However, a write operation in an STT-RAM memory consumes 8x more energy and exhibits a 6x longer latency than a SRAM. Therefore, minimizing the impact of inefficient writes is critical for successful applications of STT-RAM \cite{noguchi2013250}.

NRAM, short for Nano RAM is a proprietary technology of Nantero. The RAM uses a fabric of carbon nanotubes (CNT) for saving bits. The resistive state of the CNT fabric determines, whether a one or a zero is saved in a memory cell. The resistance depends on if the CNTs are in contact with each other. With the help of a small voltage, the CNTs can be brought into contact or be separated. Reading out a bit means to measure the resistance. Nantero claims that their technology features the same read- and write latencies as DRAM, has a high endurance and reliability even in high temperature environments and is low power with essentially zero power consumption in standby mode.  Furthermore NRAM is compatible with existing CMOS fabs without needing any new tools or processes, and it is scalable even to below 5\,nm \cite{Nantero}.

\subsubsection{Current state}
 IBM announced MLC-PCM technology replacing Flash and to use them e.g. as storage class memory (SCM) to fill the latency gap between DRAM main memory and the hard disk based background memory. 
Intel and Micron announced the new Breakthrough Memory 3D XPoint Technology \cite{evangelho2015intel} as revolutionary Flash replacement. Their Optane-SSDs DC P4800X-SSD with 375-Gbyte is available since March 2017 and said to be 2.5 to 77 times better that NAND-SSDs. It is widely assumed but not confirmed that Optane is based on PCM. Intel and Micron expect that the X-Point technology could become the dominating technology as an alternative to RAM devices offering in addition NVM property in the next ten years.
Adesto is currently offering CBRAM technology in their serial memory chips \cite{Adesto}.

Nantero together with Fujitsu announced a Multi-GB-NRAM memory in Carbone-Nanotube-Technique expected for 2018. Everspin announced Spin-Torque-Transfer-MRAMs (STT) in perpendicular Magnetic Tunnel Junction Process (pMTJ) as 256-MBit-MRAMs und 1 GB-SSDs expected in 2017. IBM also developed a neuromorphic core with a 64-K-PCM-cell as Synaptic-Array (256 Axons $\times$ 256 Dendrite) to implement SNNs (Spiking Neural Networks) \cite{hilson2015ibm}. The circuit-level performance, energy, and area model of the emerging non-volatile memory simulator NVSim \cite{dong2014nvsim} allows the investigation of architectural structures for future NVM based high-performance computers.

\subsubsection{Perspective}
It is foreseeable, that other NVM technologies will supersede current Flash memory. PCM for instance might be 1000 times faster and 1000 times more resilient. Some NVM technologies have been considered as a feasible replacement for SRAM \cite{noguchi2014highly,noguchi20153,ahn2014dasca}. Studies suggest that replacing SRAM with STT-RAM could save 60\% of LLC energy with less than 2\% performance degradation \cite{noguchi2014highly}. 
It is unclear when most of the new technologies may be mature enough and which of them will prevail. 

\subsubsection{Impact on hardware}
Memristors will deliver non-volatile memory which can be used potentially in addition to DRAM, or as a complete replacement. The latter will lead to a new Storage-Class Memory (SCM), i.e., a technology that blurs the distinction between memory and storage by enabling new data access modes and protocols that serve both ``memory'' and ``storage''. These new SCM types of non-volatile memory could be integrated on-chip with the microprocessor cores as they use CMOS-compatible sets of materials and require different device fabrication techniques from Flash. In a VLSI post-processing step they can be integrated on top of the last metal layer (see the note on Back-end of line service in section~\ref{sec-resistive}). One of the challenges for the next decade is the provision of appropriate interfacing circuits between the SCMs and the microprocessor cores. The benefits of memristor devices in integration density, energy consumption and access times may not get lost by costly interface circuitry. This holds in particular for exploiting the multi-level cell storage capability of NVMs for future HPDA systems, e.g., for big data applications. Moreover, memristors offer orders of magnitude faster read/write accesses and also much higher endurance. They are resistive switching memory technologies, and thus rely on different physics than that of storing charge on a capacitor as is the case for SRAM, DRAM and Flash \cite{eleftheriou2015future}. 

STT-RAM devices are also an important class of non-volatile memory that primarily targets the replacement of DRAM, e.g., in Last-Level Caches (LLC). However, the asymmetric read/write energy and latency of NVM technologies introduces new challenges in designing memory hierarchies. Spintronic allows integration of logic and storage at lower power consumption.
Also new hybrid PCM / Flash SSD chips could emerge with a processor-internal last-level cache (STT-RAM), main processor memory (PCRAM), and storage class memory (ReRAM) \cite{eleftheriou2015future}.

%% file: 4_3_2_photonics.tex
\subsection{Photonics}

The general idea of using photonics into computing systems is to replace electrons with photons in intra-chip, inter-chip, processor-to-memory connections and maybe even logic.  

\subsubsection{Introduction to photonics and integrated photonics}

An optical transmission link is composed by some key modules: laser light source, a modulator that converts electronic signals into optical ones, waveguides and other passive modules (e.g. couplers, photonic switching elements, splitters) along the link, a possible drop filter to steer light towards the destination and a photodetector to revert the signal into the electronic domain.  The term \emph{integrated photonics} refers to a photonic interconnection where at least some of the involved modules are integrated into silicon~\cite{SiliconIntegrated}. \emph{Active} components (lasers, modulators and photodetectors) cannot be trivially implemented in CMOS process as they require the presence of materials different from silicon and, typically, not exactly compatible with it. 

Optical communication nowadays features about 10-50 GHz modulation frequency and can support wavelength-division-multiplexing (WDM) up to 100+ \emph{colors} in fiber and 10+ (and more are expected in near future) in silicon. Propagation loss is relatively small in silicon and polymer materials so that optical communication can be regarded as quite insensitive to chip- and board-level distances. Where fiber can be employed (e.g. rack- and data centre levels) attenuation is no problem. Optical communication can rely on extremely fast signal propagation speed (head-flit latency): around 15 ps/mm in silicon and about 5.2 ps/mm in polymer waveguides that is traversing a 2\,cm x 2\,cm chip corner-to-corner in 0.6 and 0.2 ns, respectively. However, conversions to/from the optical domain can erode some of this intrinsic low-latency, as it is the case for network-level protocols and shared resource management. 

Manufacturing of \emph{passive} optical modules (e.g. waveguides, splitters, crossings, microrings) is relatively compatible with CMOS process and the typical cross-section of a waveguide (about 500 nm) is not critical, unless for the smoothness of the waveguide walls as to keep light scattering small. Turns with curvature of a few \textmu m and exposing limited insertion loss are possible, as well as grating couplers to introduce/emit light from/into a fiber outside of the chip. Even various 5x5 optical switches~\cite{gu2009cygnus} can be manufactured out of basic photonic switching elements relying on tunable micro-ring resonators. Combining these optical modules, various optical interconnection topologies and schemes can be devised: from all-to-all contention-less networks up to arbitrated ones which share optical resources among different possible paths.

In practice, WDM requires precision in microring manufacturing, runtime tuning (e.g. thermal), alignment (multiple microrings with the same resonant frequency) and make more complex both the management of multi-wavelength light from generation, distribution, modulation, steering up to photo-detection. The more complex a topology, the more modules can be found along the possible paths between source and destination, on- and off-chip, and more laser power is needed to compensate their attenuation and meet the sensitivity of the detector. For these reasons, relatively simple topologies can be preferable as to limit power consumption and, spatial division multiplexing (using multiple parallel waveguides) can allow to trade WDM for space occupation. 

Optical inter-chip signals are then expected to be conveyed also on different mediums to facilitate integrability with CMOS process, e.g., polycarbonate as in some IBM research prototypes and commercial solutions.

\subsubsection{Current status and current roadmaps}

Currently, optical communication is mainly used in HPC systems in the form of \emph{optical cables} which have progressively substituted shorter and shorter electronic links. From 10+ meters inter-rack communication down to 1+ meter intra-rack and sub meter intra-blade links. 

A number of industrial and research roadmaps are projecting and expecting this trend to arrive within boards and then to have optical technology that crosses the chip boundary, connects chips within silicon- and then in optical-interposers and eventually arriving to a complete integration of optics on a different layer of traditional chips by around 2025. For this reason, also the evolution of 2.5 - 3D stacking technologies is expected to enable and sustain this roadmap up to seamless integration of optical layers along with logic ones. The expected rated performance/consumption/density metrics are shown in the 2016 Integrated Photonic Systems Roadmap~\cite{photonicsroadmap} (see Table \ref{tbl-photonics}).

    \begin{table*}
        \caption{Expected performance evolution of optical interconnection~\cite{photonicsroadmap}.}
        \label{tbl-photonics}
        \tabularz[header]{p{2cm},X,X,X,X,X,X}{
        Time Frame          &   \textasciitilde 2000    &   \textasciitilde 2005    &   \textasciitilde 2010    &   \textasciitilde  2015   &   \textasciitilde  2020    &   \textasciitilde 2025 \\
        Interconnect        &   Rack                    &   Chassis                 &   Backplane               &   Board                   &   Module                   &   Chip    \\
        Reach               &   20 -- 100\,m            &   2 -- 4\,m               &   1 -- 2\,m               &   0.1 -- 1\,m             &   1 -- 10\,cm             &   0.1 -- 3\,cm \\
        Bandw. (Gb/s, Tb/s)  &   40 -- 200\,G        &   20 -- 100\,G            &   100 -- 400\,G           &   0.3 -- 1\,T             &   1 -- 4\,T                &   2 -- 20\,T  \\
        Bandw. Density (GB/s/cm\textsuperscript{2})  &   \textasciitilde 100 &   \textasciitilde 100 -- 400  &   \textasciitilde 400 &   \textasciitilde 1250 & > 10000   & > 40000   \\
        Energy (pJ/bit)     &   1000 $\rightarrow$ 200  &   400 $\rightarrow$ 50    &   100 $\rightarrow$ 25    &   25 $\rightarrow$ 5      &   1 $\rightarrow$ 0.1    &   0.1 $\rightarrow$ 0.01 \\
    }
\end{table*}

IBM, HPM, Intel, STM, CEA–LETI, Imec and Petra, to cite a few, essentially share a similar view on this roadmap and on the steps to increase bandwidth density, power consumption and cost effectiveness of the interconnections needed in the Exascale, and post-Exascale HPC systems. For instance, Petra labs demonstrated the first optical silicon interposer prototype~\cite{urino2014} in 2013 featuring 30 TB/s/cm\textsuperscript{2} bandwidth density and in 2016 they improved consumption and high-temperature operation of the optical modules~\cite{Urino2016}. HP has announced the Machine system which relies on the optical X1 photonic module capable of 1.5 Tbps over 50m and 0.25 Tbps over 50km. Intel has announced the Omni-Path Interconnect Architecture that will provide a migration path between Cu and Fiber for future HPC/Data Centre interconnections. Optical thunderbolt and optical PCI Express by Intel are other examples of optical cable solutions. IBM is shipping polymer + micro-pod optical interconnection within HPC blades since 2012 and it is moving towards module-to-module integration.

The main indications from current roadmaps and trends can be summarized as follows. Optical-cables (optical links) are evolving in capability (bandwidth, integration and consumption) and are getting closer to the chips, leveraging more and more photonics in an integrated form. Packaging problem of photonics remains a major issue, especially where optical signals need to traverse the chip package. Also for these reasons, interposers (silicon and optical) appear to be the reasonable first steps towards optically integrated chips. Then, full 3D processing and hybrid material integration are expected from the process point of view.

Conversion from photons to electrons is costly and for this reason there are currently strong efforts in improving the crucial physical modules of an integrated optical channel (e.g. modulators, photodetectors and thermally stable and efficiently integrated laser sources). 

\subsubsection{Alternate and emerging technologies around photonics}

Photonics is in considerable evolution, driven by innovations in existing components (e.g. lasers, modulators and photodetectors) in order to push their features and applicability (e.g. high-temperature lasers). Consequently, its expected potential is a moving target based on the progress in the rated features of the various modules. At the same time, some additional variations, techniques and approaches at the physical level of the photonic domain are being investigated and could potentially create further discontinuities and opportunities in the adoption of photonics in computing. For instance, we cite here a few:
\begin{itemize}
    \item Mode division multiplexing~\cite{huang2015}: where light propagates within a group of waveguides in parallel. This poses some criticalities but could allow to scale parallelism more easily than WDM and/or be an orthogonal source of optical bandwidth;
    \item Free-air propagation: there are proposals to exploit light propagation within the chip package without waveguides to efficiently support some interesting communication pattern (e.g. fast signaling)~\cite{malik2015free};
    \item Plasmonics: interconnect utilize surface plasmon polaritons (SPPs) for faster communication than photonics and far lower consumption over relatively short distances at the moment (below 1mm)~\cite{gao2015chip};
    \item Optical domain buffering: recent results~\cite{Tsakmakidis2017} indicate the possibility to temporarily store light and delay its transmission. This could enable the evolution of additional network topologies and schemes, otherwise impossible, for instance avoiding the reconversion to the electronic domain;
    \item Photonic non-volatile memory~\cite{IntegratedAllPhotonic}. This could reduce latencies of memory accesses by eliminating costly optoelectronic conversions while dramatically reducing the differences in speed between CPU and main memory in fully optical chips.
    \item Optics computing: Optalysys project\footnote{\url{www.optalysys.com}} for computing in the optical domain mapping information onto light properties and elaborating the latter directly in optics in an extremely energy efficient way compared to traditional computers~\cite{OpticalComputing}. This approach cannot suit every application but a number of algorithms, like linear and convolution-like computations (e.g. FFT, derivatives and correlation pattern matching), are naturally compatible~\cite{Optalysys}. Furthermore, also bioinformatics sequence alignment algorithms have been recently demonstrated feasible. 
\end{itemize}

\subsubsection{Optical communication close to the cores and perspectives}

As we highlighted, the current trend is to have optics closer and closer to the cores, from board-to-board, to chip-to-chip and up to within chips. The more optical links get close to the cores, the more the managed traffic becomes processor-specific. Patterns due to the micro-architectural behaviour of the processing cores become visible and crucial to manage, along with cache-coherence and memory consistency effects. This kind of traffic poses specific requirements to the interconnection sub-system which can be quite different from the ones induced by traffic at a larger scale. In fact, at rack or inter-rack level, the aggregate, more application-driven, traffic tends to smooth out individual core needs so that "average" behaviours emerge.

For instance, inter-socket or intra-processor coherence and synchronizations have been designed and tuned in decades for the electronic technology and, maybe, need to be optimized, or re-though, to take the maximum advantage from the emerging photonic technology.

Research groups and companies are progressing towards inter-chip interposer solutions and completely optical chips.  In this direction researchers have already identified the \emph{crucial importance of a vertical cross-layer design} of a computer system endowed with integrated photonics. A number of studies have already proposed various kinds of on-chip and inter-chip optical networks designed around the specific traffic patterns of the cores and processing chips~\cite{pan2010,Vantrease2008,pan2009,petracca2008,Grani2014,oconnor2012}.

These studies suggest also that further challenges will arise from inter-layer design interference, i.e. lower-layer design choices (e.g. WDM, physical topology, access strategies, sharing of resources) can have a significant impact in higher layers of the design (e.g. NoC-wise and up to memory coherence and programming model implications) and vice versa. This is mainly due to the scarce experience in using photonics technology for serving computing needs (close to processing cores requirements) and, most of all, due to the intrinsic end-to-end nature of an efficient optical channel, which is conceptually opposed to the well-established and mature know-how of ``store-and-forward'' electronic communication paradigm. Furthermore, the quick evolution of optical modules and the arrival of discontinuities in their development hamper the consolidation of layered design practices.

Lastly, intrinsic low-latency properties of optical interconnection (on-chip and inter-chip) could imply a re-definition of what is local in a future computing system, at various scales, and specifically in a perspective HPC system, as it has already partially happened within the HP \emph{Machine}. These revised locality features will then require modifications in the programming paradigms as to enable them to take advantage of the different organization of future HPC machines, including resource disaggregation. On this specific point, if other emerging technologies (e.g. NVM, in-memory computation, approximate, quantum computing, etc.) will appear in future HPC designs as it is expected to meet performance/watt objectives, it is highly likely that for the reasons above, photonic interconnections will require to be co-designed in integration with the whole heterogeneous HPC architecture. 

\subsubsection{Funding opportunities}

Photonic technology at the physical and module level is quite well funded in H2020 program~\cite{horizon2020photonics} as it has been regarded as strategic by the EU since years. For instance Photonics 21~\cite{photonics21} initiative gather groups and researchers from a number of enabling disciplines for the wider adoption of photonics in general and specifically also integrated photonics. Typically, funding instruments and calls focus on basic technologies and specific modules and in some cases towards point-to-point links as a final objective (e.g. optical cables).

Conversely, as photonics is coming close to the processing cores, which expose quite different traffic behaviour and communication requirements compared to larger scale interconnections (e.g. inter-rack or wide-area), it would be highly advisable to promote also a separate funding program for investigating the specific issues and approaches for the effective adoption of integrated photonics at the inter-chip and intra-chip scale. In fact the market is getting close to the cores \emph{from the outside} with an \emph{optical cable} model that will be less and less suitable to serve the traffic as the communication distance will decrease. Therefore, now could be just the right time to invest into chip-to-chip and intra-chip optical network research in order to be prepared to apply it effectively in the few years when current roadmaps expect optics to arrive there.

%% file: 4_4_alternative_ways.tex
\section{Disruptive Technology (Alternative Ways of Computing)}
\label{sec-disruptive-alt}

%% file: 4_4_1_resistive.tex
\subsection{Resistive Computing}
\label{sec-resistive}

Apart from using memristors as non-volatile memory, there are several
other ways to use memristors in computing
\cite{DiVentra2013,Pershin2012}. Using memristors as memristive
synapses in neuromorphic computing
\cite{Pershin2012,Pickett2013a,Jo2010} and using memristors in quantum
computing \cite{Pershin2012} are discussed in separate sections. In
this section, resistive (or memristive) computing is discussed in
which logic circuits are built by memristors \cite{Borghetti2010}.

Memristive gates have a lower leakage power, but switching is slower
than in CMOS gates \cite{Pershin2012}. However, the integration of
memory into logic allows to reprogram the logic, providing low power
reconfigurable components \cite{Borghetti2009} and can reduce energy
and area constraints in principle due to the possibility of computing
and storing in the same device (computing in memory).  Memristors can
also be arranged in parallel networks to enable massively parallel
computing~\cite{Pershin2011}.

Resistive computing is one of the emerging and promising computing
paradigms \cite{Borghetti2010,DiVentra2013a,Hamdioui2015}. It takes
the data-centric computing concept much further by interweaving the
processing units and the memory in the same physical location using
non-volatile technology, therefore significantly reducing not only the
power consumption but also the memory bottleneck. Resistive devices
such as memristors have been shown to be able to perform both storage
and logic functions \cite{Borghetti2010,Snider2005,Gao2013,Xie2015}.

Resistive computing provides a huge potential as compared with the
current state-of the art:

\begin{itemize}
\item It significantly reduces the memory bottleneck as it interweaves
  the storage, computing units and the communication
  \cite{Borghetti2010,DiVentra2013a,Hamdioui2015}.
\item It features low power leakage \cite{Pershin2012}. 
\item It enables maximum parallelism \cite{Hamdioui2015,Pershin2011}. 
\item It allows full configurability and flexibility
  \cite{Borghetti2009}.
\item It provides order of magnitude improvements for the energy-delay
  product per operations, the computation efficiency, and performance
  per area \cite{Hamdioui2015}.
\end{itemize}

Serial and parallel connections of memristors were proposed for the
realization of Boolean logic gates with memristors by the so-called
memristor ratio logic. In such circuits the ratio of the stored
resistances in memristor devices is exploited for the set-up of
Boolean logic. Memristive circuits realizing AND, OR gates and the
implication function were presented in
\cite{Yang2013,Kvatinsky2011,Tran2012}. Hybrid memristive computing
circuits consist of memristors and CMOS gates. The research of Singh
\cite{Singh2015}, Xia et.al. \cite{Xia2009}, and Rothenbuhler
et.al.~\cite{Tran2012} are representative for numerous proposals of
hybrid memristive circuits, in which most of the Boolean logic
operators are handled in the memristors and the CMOS transistors are
mainly used for level restoration to retain defined digital signals.

\begin{figure*}[t]
    \centering
    \begin{tikzpicture}
        {\sffamily\small
        \ifx\deliverable\undefined
        \fill[background] (-6.0,0.75) rectangle ++(16.25,-8.25);
        \else\fi

        \node[draw=black,fill=white,minimum width=4cm] (mem) at (0,0) {Memristive Computing};
        \node[draw=black,fill=white,minimum width=4cm] (ana) at (-2.5,-1.5) {Analog Computing};
        \node[draw=black,fill=white,minimum width=4cm] (dig) at (2.5,-1.5) {Digital Computing};

        \node[draw=black,fill=table,ellipse,minimum width=4cm,align=center,inner sep=-0.01cm] (neural) at (-2.5,-3.25) {Neural networks,\\ neuromorphic processing,\\ STDP};
        \node[draw=black,fill=white,minimum width=4cm,align=center] (hyb) at (7.5,-3.0) {Hybrid Approaches:\\ CMOS + Memristors};

        \node[anchor=south east,yshift=-0.0cm,xshift=-0.25cm] at (neural.north) {\footnotesize{large field}};

        \node[draw=black,fill=white,minimum width=2.5cm,align=center] (ratioed) at (-0.5,-5.5) {Ratioed Logic};
        \node[draw=black,fill=white,minimum width=2.5cm,align=center] (imply)   at (2.5,-5.5) {IMPLY Logic};
        \node[draw=black,fill=white,minimum width=2.5cm,align=center] (cmos)    at (6.5,-5.5) {CMOS-circuit like equivalent\\ memristor networks};

        \node[ellipse,draw=black,fill=table,fit=(ratioed)(imply)(cmos),inner sep=-1.0cm, minimum height=3cm] (ell) {};
        \node[anchor=south,yshift=0.25cm] at (ell.south) {\textbf{In-Memory Computing}};

        \node[draw=black,fill=white,minimum width=2.5cm,align=center] (ratioed) at (-0.5,-5.5) {Ratioed Logic};
        \node[draw=black,fill=white,minimum width=2.5cm,align=center] (imply)   at (2.5,-5.5) {IMPLY Logic};
        \node[draw=black,fill=white,minimum width=2.5cm,align=center] (cmos)    at (6.5,-5.5) {CMOS-circuit-like equivalent\\ Memristor Networks};

        \draw[-stealth,thick] (mem.south) -- (ana.north);
        \draw[-stealth,thick] (mem.south) -- (dig.north);
        \draw[-stealth,thick] (ana.south) -- (neural.north);
        \draw[-stealth,thick] (dig.south) -- (hyb.west);
        \draw[-stealth,thick] (dig.south) -- (ratioed.north);
        \draw[-stealth,thick] (dig.south) -- (imply.north);
        \draw[-stealth,thick] (dig.south) -- (cmos.north);

        }
    \end{tikzpicture}
    \caption{Summary of activities on resistive and memristive computing.}
    \label{fig-memristive}
\end{figure*}
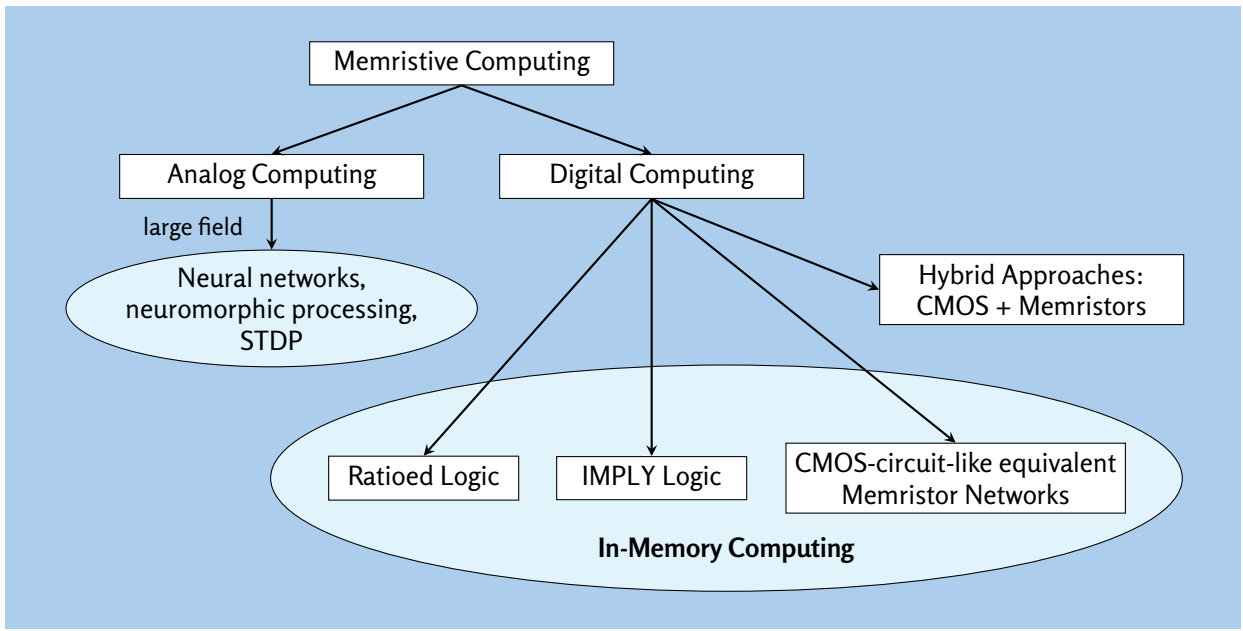

Figure \ref{fig-memristive} summarizes the activities on resistive or memristive computing. We have the large block of neuromorphic processing with memristors (see section~\ref{sec-neuromorphic}) and concerning the published papers probably smaller branch of digital memristive computing with several sub branches, like ratioed logic, imply logic or CMOS-like equivalent memristor circuits in which Boolean logic is directly mapped onto crossbar topologies with memristors.  These solutions refer to pure in-memory computing concepts. Besides that, exist proposals for hybrid solutions in which the the memristors are used as memory for CMOS circuits.

\subsubsection{Current state} A couple of start-up companies appeared in
2015 on the market who offer memristor technology as BEOL (Back-end of
line) service in which memristive elements are post-processed in CMOS
chips directly on top of the last metal layers. Also some European
institutes reported just recently at a workshop meeting ``Memristors:
at the crossroad of Devices and Applications'' of the EU cost action
1401 MemoCiS\footnote{\url{www.cost.eu/COST\_Actions/ict/IC1401}} the
possibility BEOL integration of their memristive technology to allow
experiments with such technologies. This offers new perspectives in
form of hybrid CMOS/memristor logic which use memristor networks for
high-dense resistive logic circuits and CMOS inverters for signal
restoration to compensate the loss of full voltage levels in
memristive networks.  Multi-level cell capability of memristive
elements can be used to face the challenge to handle the expected huge
amount of Zettabytes produced annually in a couple of years. Besides,
proposals exist to exploit the multi-level cell storing property for
ternary carry-free arithmetic \cite{El-Slehdar2015,Fey2014} or both
compact storing of keys and matching operations in future associative
memories realized with memristors \cite{Junsangsri2014}.

\subsubsection{Impact on hardware} Using NVM technologies for resistive
computing is a further step towards energy-aware measures for future
HPC architectures. It supports the realization of both near-memory and
in-memory computing concepts which are both an important brick for the
realization of more energy-saving HPC systems (see Section
\ref{Summary-of-Potential-Long-Term-Impacts-of-Disruptive-Technologies-for-HPC-Hardware}). Near-memory
is currently based on 3D stacking of a logic layer with DRAMs
extending HBM and may in future stack logic with NVMs. In-memory
computing could be based on resistive computing techniques combined
with resistive memory.

A further way to save energy, e.g. in near-memory computing schemes,
is to use hybrid non-volatile register cells, in which each SRAM
flip-flop cell is attached to a NVM cell and using NVM technology in
the memory hierarchy as SCM.

The NVMs, either as part of a flip-flop memristor register pair or as
pair of a complete SRAM cell array and a subsequent attached memristor
cell array are used to keep data in time periods in which this data is
not needed for computation. Other data, which have to be processed,
are stored in conventional faster SRAM/DRAM devices. Using pipeline
schemes, e.g. under control of the OS, parts of data are shifted from
NVM to SRAM/DRAM before they are accessed in the fast memory. The
latency for the data transfer from NVM to DRAM can be hidden by a
timely overlapping of data transfer with simultaneous processing of
other parts of the DRAM. The same latency hiding principle can happen
in the opposite direction. Data that are newly computed and that are not
needed in the next computing steps can be saved in NVMs. It is
to be expected that we will see in future HPC systems SCMs as
near- ad mid-term solution and in a possibly next step also hybrid
flip-flops as realization of registers.

\subsubsection{Perspective} Resistive computing, if successful, will be able to
significantly reduce the power consumption and enable massive
parallelism; hence, increase computing energy and area efficiency by
orders of magnitudes. This will transform computer systems into new
highly parallel architectures and associated technologies, and enable
the computation of currently infeasible big data and data-intensive
applications, fuelling important societal changes.

Research on resistive computing is still in its infancy stage, and the
challenges are substantial at all levels, including material and
technology, circuit and architecture, tools and compilers, and
algorithms. As of today most of the work is based on simulations and
small circuit designs. It is still unclear when the technology will be
mature and available. Nevertheless, some start-ups on memristor
technologies are emerging such as KNOWM\footnote{\url{www.knowm.org}}.

%% file: 4_4_2_neuromorphic.tex
\subsection{Neuromorphic Computing}
\label{sec-neuromorphic}

Neuromorphic Computing (NMC), as developed by Carver Mead in the late 1980s, describes the use of large-scale adaptive analog systems to mimic organizational principles used by the nervous system. Originally, the main approach was to use elementary physical phenomena of integrated electronic devices (transistors, capacitors, \dots) as computational primitives~\cite{Mead1990}. In recent times, the term neuromorphic has also been used to describe analog, digital, and mixed-mode analog/digital hardware and software systems that transfer aspects of structure and function from biological substrates to electronic circuits (for perception, motor control, or multisensory integration). Today, the majority of NMC implementations is based on CMOS technology. Interesting alternatives are, for example, oxide-based memristors, spintronics, or nanotubes~\cite{Pershin2012,Pickett2013,Jo2010}. Such kind of research is still in its infancy.

The basic idea of NMC is to exploit the massive parallelism of such circuits and to create low-power and fault-tolerant information-processing systems. Aiming at overcoming the big challenges of deep-submicron CMOS technology (power wall, reliability, and design complexity), bio-inspiration offers alternative ways to (embedded) artificial intelligence. The challenge is to understand, design, build, and use new architectures for nanoelectronic systems, which unify the best of brain-inspired information processing concepts and of nanotechnology hardware, including both algorithms and architectures~\cite{Rueckert2016}. A key focus area in further scaling and improving of cognitive systems is decreasing the power density and power consumption, and overcoming the CPU/memory bottleneck of conventional computational architectures~\cite{Eleftheriou2015}.

\subsubsection{Current State} Large scale neuromorphic chips exist based on CMOS technology, replacing or complementing conventional computer architectures by brain-inspired architectures. Mapping brain-like structures and processes into electronic substrates has recently seen a revival with the availability of deep-submicron CMOS technology. Large programs on brain-like electronic systems have been launched worldwide. At present, the largest programs are the SyNAPSE program (Systems of Neuromorphic Adaptive Plastic Scalable Electronics) in the US (launched in 2009,~\cite{SyNAPSE2013}) and the EC flagship Human Brain Project (launched in 2013,~\cite{HBP2017}). 
SyNAPSE is a DARPA-funded program to develop electronic neuromorphic machine technology that scales to biological levels. More simply stated it is an attempt to build a new kind of computer with similar form and function to the mammalian brain. Such artificial brains would be used to build robots whose intelligence matches that of mice and cats. The ultimate aim is to build an electronic microprocessor system that matches a mammalian brain in function, size, and power consumption. It should recreate 10 billion neurons, 100 trillion synapses, consume one kilowatt (same as a small electric heater), and occupy less than two litres of space~\cite{SyNAPSE2013}.

The ``Cognitive Computing via Synaptronics and Supercomputing'' (C2S2) project is a funded project from DARPA's SyNAPSE initiative. Headed by IBM the group will turn to digital special-purpose hardware for brain emulation. The TrueNorth chip is an impressive outcome of this project integrating a two-dimensional on-chip network of 4096 digital application-specific cores ($64 \times64$) and over 400 Mio. bits of local on-chip memory (\textasciitilde{}100 Kb SRAM per core) to store synapses and neuron parameters as well as 256 Mio. individually programmable synapses on-chip. One million individually programmable neurons can be simulated time-multiplexed per chip, sixteen-times more than the current largest neuromorphic chip. The chip with about 5.4 billion transistors is fabricated in a \SI{28}{\nm} CMOS process (\SI{4.3}{\cm\per\squared} die size, \SI{240}{\um} $\times$ \SI{390}{\um} per core). By device count, TrueNorth is the largest IBM chip ever fabricated and the second largest (CMOS) chip in the world. The total power, while running a typical recurrent network at biological real-time, is about \SI{70}{\milli\watt} resulting in a power density of about \SI{20}{\milli\watt\per\cm\squared} (about \SI{26}{\pico\joule} which is in turn comparable to the cortex but three to four orders-of magnitude lower compared to 50--100 \si{\watt\per\cm\squared} for a conventional CPU~\cite{Merolla2014}.

The Human Brain Project (HBP) is a European Commission Future and Emerging Technologies Flagship. The HBP aims to put in place a cutting-edge, ICT-based scientific research infrastructure that will allow scientific and industrial researchers to advance our knowledge in the fields of neuroscience, computing, and brain-related medicine. The project promotes collaboration across the globe, and is committed to driving forward European industry. Within the HBP the subproject SP9 designs, implements, and operates a Neuromorphic Computing Platform with configurable Neuromorphic Computing Systems (NCS). The platform provides NCS based on physical (analogue or mixed-signal) emulations of brain models, running in accelerated mode (NM-PM1, wafer-scale implementation of 384 chips with about 200.000 analog neurons on a wafer in \SI{180}{\nm} CMOS, 20 wafer in the full system), numerical models running in real time on a digital multicore architecture (NM-MC1 with 18 ARM cores per chip in \SI{130}{\nm} CMOS, 48 chips per board, and 1200 boards for the full system), and the software tools necessary to design, configure, and measure the performance of these systems. The platform will be tightly integrated with the High Performance Analytics and Computing Platform, which will provide essential services for mapping and routing circuits to neuromorphic substrates, benchmarking, and simulation-based verification of hardware specifications~\cite{HBP2017}. For both neuromorphic hardware systems new chip versions are under development within HBP (NM-PM2: wafer-scale integration based on a new mixed-signal chip in \SI{65}{\nm} CMOS; NM-MC2: 68 ARM M4 cores per chip in \SI{28}{\nm} CMOS with floating point support).

Closely related to the HBP is the Blue Brain Project~\cite{BBP2017}. The goal of the Blue Brain Project (EPFL and IBM, launched 2005): ``[...] is to build biologically detailed digital reconstructions and simulations of the rodent, and ultimately the human brain. The supercomputer-based reconstructions and simulations built by the project offer a radically new approach for understanding the multilevel structure and function of the brain.'' The project uses an IBM Blue Gene supercomputer (100 TFLOPS, 10TB) with currently 8,000 CPUs to simulate ANNs (at ion-channel level) in software~\cite{BBP2017}.

In the long run also the above mentioned memristor technology (see section~\ref{sec-nvm} and section~\ref{sec-resistive}) is heavily discussed in literature for future neuromorphic computing. The idea, e.g. in so-called spike-time-dependent plasticity (STDP) networks \cite{Snider08,memristor_STDP_13}, is to mimic directly the functional behaviour of a neuron. In STDP networks the strength of a link to a cell is determined by the time correlation of incoming signals to a neuron along that link and the output spikes. The shorter the input pulses are compared to the output spike, the stronger the input links to the neuron are weighted. In contrast, the longer the input signals lay behind the output spike, the weaker the link is adjusted. 
This process of strengthening or weakening the weight shall be directly mapped onto memristors by increasing or decreasing their resistance depending which voltage polarity is applied to the poles of a two-terminal memristive device. This direct mapping of an STDN network to an analogue equivalent of the biological cells to artificial memristor-based neuron cells shall emerge new extreme low-energy neuromorphic circuits. Besides this memristor-based STDP networks there are lots of proposals for neural networks to be realised with memristor-based crossbar and mesh architectures for cognitive detection and vision applications, e.g. \cite{Lim_2014}.

All above mentioned projects have in common that they model spiking neurons, the basic information processing element in biological nervous systems. A more abstract implementation of biological neural systems are Artificial Neural Networks (ANNs). Popular representatives are Deep Neural Networks (DNNs) as they have propelled an evolution in the machine learning field. DNNs share some architectural features of the nervous systems, some of which are loosely inspired by biological vision systems~\cite{LeCun1998}. DNNs are dominating computer vision today and observe a strong growing interest for solving all kinds of classification, function approximation, interpolation, or forecasting problems. Training DNNs is computationally intense. For example, Baidu Research\footnote{\url{www.baidu.com}} estimated that training one DNN for speech recognition can require up to 20 Exaflops ($10^{18}$ floating point operations per second); whereas the world's largest supercomputer delivers about 100 Petaflops ($10^{15}$ floating point operations per second). Companies such as Facebook and Google have a nearly unlimited appetite for performance, because increasing the available computational resources enables more accurate models as well as newer models for high-value problems such as autonomous driving and to experiment with more-advanced uses of artificial intelligence (AI) for digital transformation. Corporate investment in artificial intelligence is predicted to triple in 2017, becoming a \$100 billion market by 2025~\cite{Wellers2017}.
Hence, a variety of hardware and software solutions have emerged to slake the industry's thirst for performance. The currently most well-known commercial machines targeting deep learning are the TPUs of Google and the Nvidia Volta V100. 
A tensor processing unit (TPU) is an ASIC developed by Google specifically for machine learning. The chip has been specifically designed for Google's TensorFlow framework. The first generation of TPUs applied 8-bit integer MAC (multiply accumulate) operations. It is deployed in data centres since 2015 to accelerate the inference phase of DNNs. An in-depth analysis was recently published by Jouppi et al.~\cite{Jouppi2017}. The second generation TPU of Google was announced in May 2017. The individual TPU ASICs are rated at 45 TFLOPS and arranged into 4-chip 180 TFLOPS modules. These modules are then assembled into 256 chip pods with 11.5 PFLOPS of performance~\cite{Wikipedia2017}. The new TPUs are optimized for both training and making inferences.
Nvidia's Tesla V100 GPU contains 640 Tensor Cores delivering up to 120 Tensor TFLOPS for training and inference applications. Tensor Cores and their associated data paths are custom-designed to dramatically increase floating-point compute throughput with high energy efficiency. For deep learning inference, V100 Tensor Cores provide up to 6x higher peak TFLOPS compared to standard FP16 operations on Nvidia Pascal P100, which already features 16-bit FP operations~\cite{NvidiaCorporation2017}.

Matrix-Matrix multiplication operations are at the core of DNN training and inferencing, and are used to multiply large matrices of input data and weights in the connected layers of the network. Each Tensor Core operates on a $4 \times 4$ matrix and performs the following operation: $D = A \times B + C$, where $A$, $B$, $C$, and $D$ are $4\times4$ matrices. Tensor Cores operate on FP16 input data with FP32 accumulation. The FP16 multiply results in a full precision product that is then accumulated using FP32 addition with the other intermediate products for a $4 \times 4 \times4$ matrix multiply~\cite{NvidiaCorporation2017}. The new Nvidia DGX-1 system based on the Volta V100 GPUs will be delivered in the third quarter of 2017~\cite{Morgan2017}. It is the world's first purpose built system optimized for deep learning, with fully integrated hardware and software.

Many more options for DNN hardware acceleration are showing up~\cite{Chen2014}. AMD's forthcoming Vega GPU should offer 13 TFLOPS of single precision, 25 TFLOPS of half-precision performance, whereas the machine-learning accelerators in the Volta GPU-based Tesla V100 can offer 15 TFLOPS single precision and 120 TFLOPS for deep learning workloads. Microsoft has been using Altera FPGAs for similar workloads, though a performance comparison is tricky; the company has performed demonstrations of more than 1 Exa-operations per second~\cite{Bright2017}. Intel offers the Xeon Phi 7200 family and IBMs TrueNorth tackles deep learning as well~\cite{Gwennap2016}. 
Other chip and IP (Intellectual Property) vendors---including Cadence, Ceva, Synopsys, and Qualcomms zeroth---are touting DSPs for learning algorithms. Although these hardware designs are better than CPUs, none was originally developed for DNNs. Ceva's new XM6 DSP core\footnote{\url{www.ceva-dsp.com}} enables deep learning in embedded computer vision (CV) processors. The synthesizable intellectual property (IP) targets self-driving cars, augmented and virtual reality, surveillance cameras, drones, and robotics. The normalization, pooling, and other layers that constitute a convolutional-neural-network model run on the XM6's 512-bit vector processing units (VPUs). The new design increases the number of VPUs from two to three, all of which share 128 single-cycle $(16 \times 16)$-bit MACs, bringing the XM6's total MAC count to 640. The core also includes four 32-bit scalar processing units.

Examples for start-ups are Nervana Systems\footnote{\url{www.nervanasys.com}}, Knupath\footnote{\url{www.knupath.com}}, Wave Computing\footnote{\url{www.wavecomp.com}}. The Nervana Engine will combine a custom \SI{28}{\nm} chip with 32 GB of high bandwidth memory and replacing caches with software-managed memory. Kupath second generation DSP Hermosa is positioned for deep learning as well as signal processing. The \SI{32}{\nm} chip contains 256 tiny DSP cores operation at \SI{1}{\GHz} along with 64 DMA engines and burns \SI{34}{\watt}. The dataflow processing unit from Wave Computing implements \mbox{``tens of thousands''} of processing nodes and ``massive amounts'' of memory bandwidth to support TensorFlow and similar machine-learning frameworks. The design uses self-timed logic that reaches speeds of up to \SI{10}{\GHz}. The \SI{16}{\nm} chip contains 16 thousand independent processing elements that generate a total of 180 Tera 8-bit integer operations per second.

\subsubsection{Perspective} Brain-inspired hardware computing architectures have the potential to perform AI tasks better than conventional architecture by means of better performance, lower energy consumption, and higher resilience to defects. Neuromorphic Computing and Deep Neural Networks represent two approaches for taking inspiration from biological brains. Software implementations on HPC-clusters, multi-cores (OpenCV), and GPGPUs (NVidia cuDNN) are already commercially used. FPGA acceleration of neural networks is available as well. From a short term perspective these software implemented ANNs may be accelerated by commercial transistor-based neuromorphic chips or accelerators. Future emerging hardware technologies, like memcomputing and 3D stacking~\cite{Belhadj2014} may bring neuromorphic computing to a new level and overcome some of the restriction of Von-Neumann-based systems in terms of scalability, power consumption, or performance.

Particularly attractive is the application of ANNs in those domains where, at present, humans outperform any currently available high-performance computer, e.g., in areas like vision, auditory perception, or sensory motor control. Neural information processing is expected to have a wide applicability in areas that require a high degree of flexibility and the ability to operate in uncertain environments where information usually is partial, fuzzy, or even contradictory. This technology is not only offering potential for large scale neuroscience applications, but also for embedded ones: robotics, automotive, smartphones, IoT, surveillance, and other areas. Even more computational power may be obtained by emerging technologies like quantum computing, molecular electronics, or novel nano-scale devices (memristors, spintronics, nanotubes (CMOL, i.e. combining CMOS with nanowire crossbars)~\cite{Rueckert2016}.
Neuromorphic computing appears as key technology on several emerging technology lists. Hence, Neuromorphic technology developments are considered as a powerful solution for future advanced computing systems~\cite{FET2017}. Neuromorphic technology is in early stages, despite quite a number of applications appearing. To gain leadership in this domain there are still many important open questions that need urgent investigation (e.g. scalable resource-efficient implementations, online learning, and interpretability). 
There is a need to continue to mature the NMC system and at the same time to demonstrate the usefulness of the systems in applications, for industry and also for the society: more usability and demonstrated applications. More focus on technology access might be needed in Europe. Regarding difficulties for NMC in EC framework programmes, integrated projects were well fitting the needs of NMC in FP7, but are missing in H2020. For further research on neuromorphic technology the FET-OPEN scheme could be a good path as it requires several disciplines (computer scientists, material science, engineers in addition to neuroscience, modelling). One also needs support for many small-scale interdisciplinary exploratory projects to take advantage of newly coming out developments, and allow funding new generation developers having new ideas.

\subsubsection{Impact on Hardware} Creating the architectural design for NMC requires an integrative, interdisciplinary approach between computer scientists, engineers, physicists, and materials scientists. NMC would be efficient in energy and space and applicable as embedded hardware accelerator in mobile systems. The building blocks for ICs and for the Brain are the same at nanoscale level: electrons, atoms, and molecules, but their evolutions have been radically different. The fact that reliability, low-power, reconfigurability, as well as asynchronicity have been brought up so many times in recent conferences and articles, makes it compelling that the Brain should be an inspiration at many different levels, suggesting that future nano-architectures could be neural-inspired. The fascination associated with an electronic replication of the human brain has grown with the persistent exponential progress of chip technology. The present decade 2010–2020 has also made the electronic implementation more feasible, because electronic circuits now perform synaptic operations such as multiplication and signal communication at energy levels of 10 fJ, comparable to biological synapses. Nevertheless, an all-out assembly of $10^{14}$ synapses will remain a matter of a few exploratory systems for the next two decades because of several challenges~\cite{Rueckert2016}.

Up to now, there is little agreement on what a learning chip should actually look like. The companies withheld details on the internal architecture of their learning accelerators. Most of the designs appear to focus on high throughput for low-precision data, backed by high bandwidth memory subsystems. The effect of low-precision on the learning result has not been analysed in detail yet. Recent work on low-precision implementations of backprop-based neural nets~\cite{Gupta2015} suggests that between 8 and 16 bits of precision can suffice for using or training DNNs with backpropagation. What is clear is that more precision is required during training than at inference time, and that some forms of dynamic fixed point representation of numbers can be used to reduce how many bits are required per number. Using fixed point rather than floating point representations and using less bits per number reduces the hardware surface area, power requirements, and computing time needed for performing multiplications, and multiplications are the most demanding of the operations needed to use or train a modern deep network with backprop.

%% file: 4_4_3_quantum.tex
\subsection{Quantum Computing}

Today's computers, both in theory (Turing machines) and in practice (personal computers) are based on classical bits which can be either 0 or 1 to perform operations. Modern Quantum Computing systems operate differently as they make use of quantum bits (qubits) which can be in a superposition state and entangled with other qubits \cite{QuantumComputing101}. Superposition and entanglement are thus the two main phenomena that one tries to exploit in quantum computing. Superposition implies that a qubit is both in the ground and the excited state. Entanglement means that two (or more) qubits can be combined with each other such that their states have become inseparable. This gives rise to very interesting properties that can be exploited algorithmically.

The computational power of a quantum computer is directly related to these phenomena and the number of qubits. Two qubits can hold four values at any given time, namely (00, 01, 10, and 11). With each qubit that is added, the compute capacity of the quantum computer is doubled and thus increases exponentially. All these qubits states (in superposition and entangled with each other) can then be manipulated in parallel as, e.g., gates are applied on them which gives the exponential computing power. The problem is that building a qubit is an extremely difficult task as the quantum state that is needed is very fragile and decoheres (losing the state information due to dynamic coupling with the external environment) rapidly. In addition, it is impossible to read out the state of a qubit, which ultimately is necessary to get the answer of a computation, without destroying the superposition state, thus destroying information contained in the qubit state. Basically, it turns into a classical bit that houses only a single value \cite{Metz2015}.

\subsubsection{Current State} A well-known but highly debated example of a quantum computer is the \mbox{D-Wave} machine built by the Canadian company \mbox{D-Wave} \cite{Metz2015}. \mbox{D-Wave} is based on quantum annealing and thus only usable for specific optimization problems. \mbox{D-Wave's} qubits are much easier to build than the equivalent in more traditional quantum computers, but their quantum states are also more fragile, and their manipulation less precise \cite{Gibney2017}. \mbox{D-Wave's} latest processor already has 4,000 qubits.

An alternative direction is to build a universal quantum computer based on quantum gates, such as Hadamard, rotation gates and CNOT. Google, IBM and Intel have all initiated research projects in this domain and currently superconducting qubits seem to be the most promising direction \cite{Simonite2014, Simonite2015-IBM, Odeh2013, Clark2015}.

IBM has announced two new quantum computers as a continuation of the IBM~Q program. The first is a 16 qubit machine that will be used as a follow-on to the 5 qubit machine that is currently accessible through the IBM Quantum Experience program \cite{Quantumcomputingreport2017}. IBM~Q states to have successfully built and tested two of its most powerful universal quantum computing processors to date: 16 qubits for public use and a 17 qubit prototype commercial processor \cite{Ibmq}.

On the application side, Google, NASA, Lockheed Martin, Los Alamos National Lab, and Volkswagen have all focused on developing their own applications and software tools \cite{Hemsoth2017}. Volkswagen Group IT is cooperating with quantum computing company D-Wave on a research project for traffic flow optimization. The first research project is traffic flow optimization in the Chinese mega-metropolis of Beijing. Data scientists and AI specialists from Volkswagen have successfully programmed an algorithm to optimize the travel time of all public taxis in the city \cite{Volkswagen2017}.

Virginia Tech researchers are working on next-generation tools to fit the 4,000 qubit quantum systems D-Wave to help expand the application set and developer tools \cite{Hemsoth2017}.

A major threat on cybersecurity is that quantum computers could attack RSA und EEC encryption. Research towards Post-Quantum-Cryptography (PQC) is a concern even for companies like Infineon Technologies \cite{Arnold2017}. 

Currently, the European Commission is preparing the ground for the launch in 2018 of a €1 billion flagship initiative on quantum technologies \cite{Europa2016}.

\subsubsection{Perspective} Making use of Quantum Computing has the benefit to improve the speed-up of certain computations enormously, and even allows solving problems that are impossible for classical computing. Even though the challenges are substantial, they can be separated in physics-oriented and engineering-oriented ones. The physics challenges primarily have to address the lifetime of qubits and the fidelity of qubit gate operations. The engineering challenges go from identifying relevant algorithms and provide compiler and runtime support. It is also clear that a quantum computer will require a supercomputer to provide the necessary quantum error correction mechanisms as error rates of around $ 10^{-3} $ are not uncommon. As the quantum phenomena require mK (milli Kelvin) conditions, the control logic should be brought as close as possible to reduce the transfer of data up to room temperature computers. Understanding how conventional CMOS behaves under cryo-conditions is another challenge.

Quantum Computing might have the advantage to solve some problems that couldn't be solved with classical computers - one example is \textit{Shor's Algorithm} for decryption which, at least assuming that a large scale quantum computer can be built consisting of millions of qubits, could decrypt a 2,000 bit word in around one day which is completely impossible for conventional supercomputers.

In the short term, the \textit{Quantum Key Distribution Algorithm} (QKD) \cite{Odeh2013} can be used as a new encryption technology that relies on the fact that, when a third party tries to eavesdrop, the entangled state is immediately destroyed.

Further quantum algorithms are \cite{Kudrow2013}:

\begin{itemize}

\item \textit{Grover's Algorithm} is the second most famous result in quantum computing. Often referred to as ``quantum search'', Grover's Algorithm actually inverts an arbitrary function by searching n input combinations for an output value in $\sqrt n$ time.

\item \textit{Binary Welded Tree} is the graph formed by joining two perfect binary trees at the leaves. Given an entry node and an exit node, The Binary Welded Tree Algorithm uses a quantum random walk to find a path between the two. The quantum random walk finds the exit node exponentially faster than a classical random walk.

\item \textit{Boolean Formula Algorithm} can determine a winner in a two player game by performing a quantum random walk on a NAND tree.

\item \textit{Ground State Estimation Algorithm} determines the ground state energy of a molecule given a ground state wave function. This is accomplished using quantum phase estimation.

\item \textit{Linear Systems Algorithm} makes use of the quantum Fourier Transform to solve systems of linear equations.
Shortest Vector problem is an NP-Hard problem that lies at the heart of some lattice-based cryptosystems. The Shortest Vector Algorithm makes use of the quantum Fourier Transform to solve this problem. 

\item \textit{Class Number} computes the class number of a real quadratic number field in polynomial time. This problem is related to elliptic-curve cryptography, which is an important alternative to the product-of-two-primes approach currently used in public-key cryptography.

\end{itemize} 

It is expected that machine learning will be transformed into quantum learning - the prodigious power of qubits will narrow the gap between machine learning and biological learning \cite{Simonite2015-Google}.

In general, the focus is now on developing algorithms requiring a low number of qubits (a few hundred) as that seems to be the most likely reachable goal in the 10-15 year time frame.

\subsubsection{Impact on Hardware} An interesting point to investigate is a better hardware architecture supporting the power efficiency of quantum better. If this is too complex, it should be at least possible to provide a hybrid architecture of both systems enabling to run the simplest sequences of an application as usually on classical computers and the complex ones on quantum co-processors. By doing this, the system performance can be improved during runtime \cite{Clark2015}.

As pointed out earlier, a quantum computer will always be a heterogeneous computing platform where conventional supercomputing facilities will be combined with quantum processing units. How they interact and communicate is clearly a challenging line of research \cite{Kudrow2013}. Quantum Computing looks more and more as a viable technology for the future and Europe best starts developing some serious activities, as indicated by the flagship project on quantum technologies that starts in 2018.

%% file: 4_5_beyond_cmos.tex
\section{Beyond CMOS}
\label{sec-disruptive-beyond}

%% file: 4_5_1_nanotubes.tex
\subsection{Nanotubes and Nanowires}

Nano structures like Carbon Nanotubes (CNT) or Silicon Nanowires (SiNW) expose a number of special properties which make them attractive to build logic circuits or memory cells.

CNTs are tubular structures of carbon atoms.
These tubes can be single-walled (SWNT) or multi-walled nanotubes (MWNT).
Their diameter is in the range of a few nanometers.
Their electrical characteristics vary, depending on their molecular structure, between metallic and semiconducting \cite{wiki2017nanotube}.

A CNTFET consists of two metal contacts which are connected via a CNT.
These contacts are the drain and source of the transistor.
The gate is located next to or around the CNT and separated via a layer of silicon oxide \cite{rispal2009large}.
Also, crossed lines of appropriately selected CNTs can form a tunnel diode.
This requires the right spacing between the crossed lines.
The spacing can be changed by applying appropriate voltage to the crossing CNTs.

SiNWs can be formed in a bottom up self-assembly process.
This might lead to substantially smaller structures as those that can be formed by lithographic processes.
Additionally, SiNWs can be doped and thus, crossed lines of appropriately doped SiNW lines can form diodes.

Both, CNTs and SiNWs can be used to build nano-crossbars, which logically are similar to a PLA (programmable logic array).
They offer wired-AND conjunctions of the input signal.
Together with inversion/buffering facilities, they can create freely programmable logic structures.
The density of active elements is much higher as with individually formed CNTFETs.

\subsubsection{Current state} In September 2013, Max Shulaker from Stanford University published a computer with digital circuits based on carbon nanotubes.
It contains a 1 bit processor, consisting of 178 transistors and runs with a frequency of 1 kHz.\cite{shulaker2013carbon}

Nanotube-based RAM is a proprietary memory technology for non-volatile random access memory developed by Nantero (this company also refers to this memory as NRAM) and relies on crossing CNTs as described above.
An NRAM ``cell'' consists of a non-woven fabric matrix of CNTs located between two electrodes.
The resistance state of the fabric is high (representing \emph{off} or $0$ state) when (most of) the CNTs are not in contact and is low (representing \emph{on} or $1$ state) vice versa.
Switching the NRAM is done by adjusting the space between the layers of CNTs.
In theory NRAM can reach the density of DRAM while providing performance similar to SRAM \cite{wiki2017nanoram}.

Nano crossbars have been created from CNTs and SiNWs \cite{devisree2016nanoscale}.
In both cases, the fundamental problem is the high defect density of the resulting circuits.
Under normal semiconductor classifications, these devices would be considered broken.
In fact, usage of these devices is only possible, if the individual defects of the devices can be respected during the logic mapping stage of the HW synthesis \cite{zamani2011self}.

Currently, less research on nanowires and nanotubes is active than in the early 2000s.
Nevertheless, some groups are pushing the usage of nanowires and nanotubes for the creation of logic circuits.
At the same time, more research is going on to deal with the high defect density.

\subsubsection{Perspective} It will take an unknown number of years before NRAM drives might be in production stage \cite{compworld2017nanomem}.
It is unclear whether the defect density can be substantially lowered by better fabrication processes. 

\subsubsection{Impact on hardware} CNTs and SiNWs can be utilized for a lot of different applications in several areas of research.
The construction of carbon nanotube field-effect transistors (CNTFETs) and nanotube-based RAM (or Nano-RAM) are important for HPC.
CNTs are very good thermal conductors.
Thus, they could significantly improve conducting heat away from CPU chips \cite{extremtech2017conductive}.

Nano crossbar circuits are inherently programmable.
This leads to more freedom, if the programmability is taken into account during the HW design stage.
Potentially, customizable HW is available in each component once nano crossbars are employed as logic circuits.


%% file: 4_5_2_graphene.tex
\subsection{Graphene}

In 2010 two physicists at Manchester University in the U.K. shared a Nobel Prize in Physics for their work on a new wonder material: graphene, a flat sheet of carbon with the thickness of a single atom.
Konstantin Novoselov and Andre Geim discovered the material by applying plain old sticky tape to simple graphite \cite{moskvitch2015graphene}.

Graphene grows on semiconductor i.e. on the surface of a germanium crystal, which is seen as big step towards manufacturability, see \cite{benchoff2015graphene,jacobberger2015direct}.

\subsubsection{Current state} In 2010, IBM researchers demonstrated a radio-frequency graphene transistor with a cut-off frequency of 100 Gigahertz.
This is the highest achieved frequency so far for any graphene device.
In 2014, engineers at IBM Research have built the world's most advanced graphene-based chip, with performance that's 10,000 times better than previous graphene ICs.
The key to the breakthrough is a new manufacturing technique that allows the graphene to be deposited on the chip without it being damaged \cite{ibm2010made}.

Graphene Project is an EC Flagship project with considerable research efforts in making graphene useful, however, still focused more on the material science perspective than on its potential usage for future computer technology.
Graphene is among the strongest materials known and has attractive potential also outside of computer technology, e.g., as electrodes for solar cells, for use in sensors, as the anode electrode material in lithium batteries and as efficient zero-band-gap semiconductors \cite{rodewald2008researchers}.

The use of graphene in CMOS circuits has been demonstrated in different settings \cite{chen2010fully,lee2010low}.
Also, graphene has been used in digital circuits as an interconnect material for an FPGA \cite{lee2013demonstration}.
In its most advanced form, graphene is subject to electrostatic doping which results in a behaviour that resembles classical p-type and n-type semiconductors.
Thus, graphene layers doped in this way can form p-n junctions which in turn can be used to build so called Pass-XNOR gates \cite{tanachutiwat2010reconfigurable}.
Unfortunately, these gates require a clocked evaluation signal which results in a two-phased operation and limits the operating frequency of the logic.
By clever combination of several such Pass-XNOR gates, one can create a real PLA \cite{tenace2016graphene}.
Currently, the focus of this research is on low-power operation, and the resulting circuit is not very fast due to the two-phased logic operation.

\subsubsection{Perspective} Graphene is a promising technology in laboratory.
Due to the fact that the new graphene manufacturing method is actually compatible with standard silicon CMOS processes, it will probably be possible to realize commercial graphene computer chip in future \cite{anthony2014ibmbuilds}.

Graphene as an interconnect material offers many advantages which might play an important role in future chip architectures, since data transport over longer distances will be much faster and less power hungry than current metal based transmission structures.

Usage of graphene as active element in logic circuits is still in its infancy.
Electrostatically doped graphene layers can be used to build p-n junctions, Nevertheless, these junctions cannot yet be used to build high-speed, high-density logic circuits.
It is unclear whether other basic circuit design approaches will help to circumvent this drawback.

\subsubsection{Impact on hardware} Graphene has an excellent capacity for conducting heat and electricity.
New on-chip communication architectures might come up due to these good conductance values.
Using graphene as active element results in PLA structures.
Thus, similar opportunities and problems apply to these PLAs (programmability + defect density).

%% file: 4_5_3_diamond.tex
\subsection{Diamond Transistors}

Diamonds can be processed in a way that they act like a semiconductor.
Diamond based transistors can be fabricated. 

\subsubsection{Current state} Researchers at the Tokyo Institute of Technology fabricated a diamond junction field-effect transistors (JFET) with lateral p-n junctions.
The device shows excellent physical properties such as a wide band gap of 5.47 eV, a high breakdown field of 10 MV/cm (3–4 times higher than 4H-SiC and GaN), and a high thermal conductivity of 20 W/cm*K (4–10 times higher than 4H-SiC and GaN).
It has been found that this diamond transistor works with excellent electrical characteristics, up to 723 K \cite{twasaki2013high}.

\subsubsection{Perspective} Currently the gate length of the fabricated diamond transistors is in the single-digit micrometer range.
Compared with the current 22nm technology with gate lengths of about 25nm \cite{wiki201722nanometer}, a reduction in size is absolutely necessary in order to allow fast working circuits (limitation of the propagation delays). 

Producing reasonable diamond wafers for mass production could be possible with the method of \cite{aida2016fabrication}.
The time for producing diamond wafers is another factor that has to be reduced drastically to compete with other technologies.

\subsubsection{Impact on hardware} The high thermal conductivity of diamond, which is several magnitudes higher than that of conventional semiconductor material, allows faster heat dissipation.
This could solve the temperature problem of stacked dies.
Switching energy of a diamond based semiconductor is expected to be much smaller than silicon and the maximum operating temperature can be much higher.
It may "revive" the traditional Moore law.


%% file: 5_impact.tex
\chapter{Impact of Disruptive Technologies}
\label{sec-impact}

\section{Summary of Potential Long-Term Impacts of Disruptive Technologies for HPC Hardware}
\label{Summary-of-Potential-Long-Term-Impacts-of-Disruptive-Technologies-for-HPC-Hardware}

Potential long-term impacts of disruptive technologies could concern the processor logic, the processor-memory interface, the memory hierarchy, and future hardware accelerators.

\subsection{New Ways of Computing}
Processor logic could be totally different if materials like graphene, nanotube or diamond would replace classical integrated circuits based on silicon transistors, or could integrate effectively with traditional CMOS technology to overcome its current major limitations like limited clock rates and heat dissipation. 

A physical property that these materials share is the high thermal conductivity: Diamonds for instance can be used as a replacement for silicon, allowing diamond based transistors with excellent electrical characteristics. Graphene and nanotubes are highly electrically conductive and could allow a reduced amount of heat generated because of the lower dissipation power, which makes them more energy efficient. With the help of those good properties, less heat in the critical spots would be expected which allows much higher clock rates and highly integrated packages. Whether such new technologies will be suitable for computing in the next decade is very speculative.

Furthermore, Photonics, a technology that uses photons for communication, can be used to replace communication busses to enable a new form of inter- and intra-chip communication.

Current CMOS technology may presumably scale continuously in the next decade, down to 4 or 3 nm. However, scaling CMOS technology leads to steadily increasing costs per transistor, power consumption, and to less reliability. Die stacking could result in 3D many-core microprocessors with reduced intra core wire length, enabling high transfer bandwidths, lower latencies and reduced communication power consumption.

\subsection{New Processor-Memory Interfaces}
Near-memory computing and in-memory computing will change the interface between the processor and the memory: memory will in future not only be accessed by loads and stores respectively cache-line misses, but additionally provide a semantically stronger access pattern based on simple operations on a large number of memory cells.

\paragraph{Near-memory computing} to be considered as a near- and mid-term realizable concept, is characterized by logic, e.g. small cores, which are located directly to the memory in order to carry out pre-processing steps, like e.g. stencil operations, or vector operations on data either stored in memory, caches or so-called storage class memory (SCM). It is an acceptable fact that due to energy reasons it is preferable to process data in-situ directly where they are located before they are sent to the processor in particular if this pre-processing goes along with a reduction of data amount. 

\paragraph{In-memory computing} goes a step further in such a way that the NVM cell itself is not only a storage cell but it becomes an integral part of the processing step. This can help to reduce further the energy consumption and the area requirement in comparison to near-memory computing. However, this technology has to be improved and therefore it is considered at least as a mid-term or probably as a more long-term solution. 

In-memory computing uses the NVM cells not only for storing but as inherent part of the processing step itself, e.g. to pre-process data in pre-processing steps embedded as part of a much more complex processing task. This will help to face the challenging of processing and holding large data amount as we see in HPDA.

In-memory computing will also influence strongly Edge computing approaches in which new architectures have to be found that are characterized by processing data directly at sensors where the data is captured to reduce as described above the amount of data that has to be transferred to more-coarse grained cores for post-processing.

In-memory and near-memory computing concepts will rely on 3D stacking techniques for an efficient realization. In near-memory, computing refers to the coupling of logic cores and e.g. hybrid memory cells; in-memory concepts refer on the coupling of NVM cell arrays and conventional CMOS that will work as memory controller to control the NVMs. This coupling has to be realized in a so-called BEOL (back end on line) step, i.e. the NVM cells are deposited in a post-process step on the top metal layer of a CMOS chip.  

\subsection{New Memory Hierarchies}
3D stacking will also be used to scale Flash memories, because 2D NAND Flash technology does not further scale. In the long run even 3D Flash memories will probably be replaced by memristor or other non-volatile memory (NVM) technologies. These, depending on the actual type, allow higher structural density, less leakage power, faster read- and write access, more endurance and can nevertheless be more cost efficient.

However, the whole memory hierarchy may change in the upcoming decade. DRAM scaling will only continue with new technologies, in fact NVMs, which will deliver non-volatile memory potentially replacing or being used in addition to DRAM. Some new non-volatile memory technologies could even be integrated on-chip with the microprocessor cores and offer orders of magnitude faster read/write accesses and also much higher endurances than Flash. Intel demonstrated the possible fast memory accesses of the 3D-XPoint NVM Technology used in their Optane Technology. HP's computer architecture proposal called ``The Machine'' targets a machine based on new NVM memory and photonic busses. The Machine sees the memory instead of processors in the centre. This so called Memory-Centric Computing unifies the memory and storage into one vast pool of memory. HP proposes advanced photonic fabric to connect the memory and processors. Using light instead of electricity is the key to rapidly accessing any part of the massive memory pool while using much less energy.

\begin{figure*}[ht]
    \centering
    \begin{tikzpicture}
        {\sffamily\small
        \fill[background] (-4.5,0.75) rectangle ++(15.0,-8.5);
              \node (top) at (0,0) {};
              \node (left) at (-4.0,-7.25) {};
              \node (right) at (4.0,-7.25) {};
              \draw[fill=table] (left.center) -- (top.center) -- (right.center) -- (left.center);
              \begin{scope}
                  \clip (left.center) -- (top.center) -- (right.center) -- (left.center);
                  \draw (-5,-1.875) -- ++(10,0);
                  \draw (-5,-2.625) -- ++(10,0);
                  \draw (-5,-3.375) -- ++(10,0);
                  \draw (-5,-4.875) -- ++(10,0);
                  \draw (-5,-5.625) -- ++(10,0);
                  \draw (-5,-6.375) -- ++(10,0);
              \end{scope}
              \draw[ultra thick] (-4.3,-4.125) -- ++(7.8,0);
              \node[align=center,anchor=south] (regs) at (0,-1.85) {CPU\\ Registers};
              \node (l1) at (0,-2.25) {L1 \$};
              \node (l2) at (0,-3) {L2 \$};
              \node (l3) at (0,-3.75) {L3 \$};
              \node (mem) at (0,-4.5) {Memory};
              \node[align=center] (scmem1) at (0,-5.25) {Storage Class Memory SCM1};
              \node[align=center] (scmem2) at (0,-6.0) {Storage Class Memory SCM2};
              \node (bulk) at (0,-6.75) {Bulk Storage};

              \node[rotate=90,anchor=west] at (-4.05,-4.0) {Processor Chip};
              \node[rotate=90,anchor=east] at (-4.05,-4.25) {Off Chip};

              \node[anchor=west,font=\footnotesize] at (4.2,-2.625) {SRAM, expensive cost/mm\textsuperscript{2}};
              \node[anchor=west,font=\footnotesize] at (4.2,-3.75) {SRAM, STT-RAM};
              \node[anchor=west,font=\footnotesize] at (4.2,-4.5) {3D-RAM, high bandwidth memory / PCM};
              \node[anchor=west,font=\footnotesize] at (4.2,-5.25) {PCM or other NVM, fast read, less fast write};
              \node[anchor=west,font=\footnotesize] at (4.2,-6.0) {NAND Flash, low cost/mm\textsuperscript{2}};
              \node[anchor=west,font=\footnotesize] at (4.2,-6.75) {Disk, Cloud, low cost/mm\textsuperscript{2}};

        }
    \end{tikzpicture}
    \caption{Usage of NVM in a future complex supercomputer memory hierarchy.}
\end{figure*}
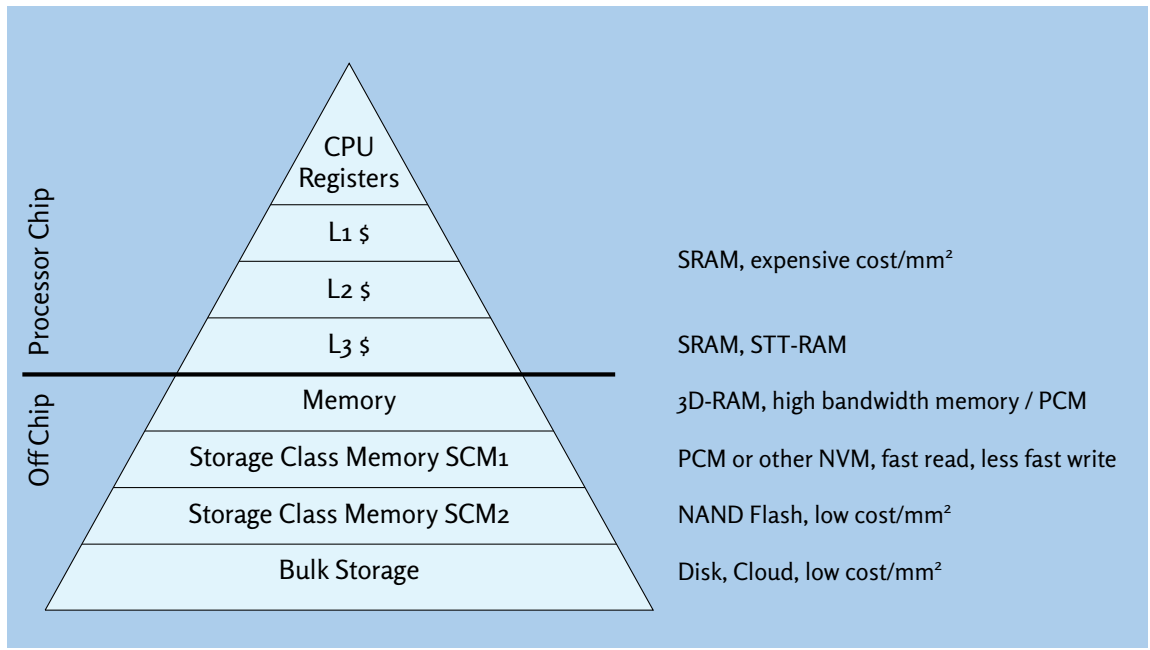

The Machine is a first example of the new Storage-class Memory (SCM), i.e., a non-volatile memory technology in between memory and storage, which may enable new data access modes and protocols that are neither ``memory'' nor ``storage''. It would particularly increase efficiency of fault tolerance check pointing, which is potentially needed for shrinking CMOS processor logic that leads to less reliable chips. There is a major impact from this technology on software and computing. SCM provides orders of magnitude increase in capacity with near-DRAM latency which would push software towards in-memory computing.

\subsection{New Hardware Accelerators}
Resistive Computing, Neuromorphic Computing and Quantum Computing are promising technologies that may be suitable for new hardware accelerators but less for new processor logic. Resistive computing promises a reduction in power consumption and massive parallelism. It could enforce memory-centric and reconfigurable computing, leading away from the Von-Neumann architecture. Humans can easily outperform currently available high-performance computers in tasks like vision, auditory perception and sensory motor-control. As Neuromorphic Computing would be efficient in energy and space for artificial neural network applications, it would be a good match for these tasks. More lack of abilities of current computers can be found in the area of unsolved problems in computer science. Quantum Computing might solve some of these problems, with important implications for public-key cryptography, searching, and a number of specialized computing applications. 

\section{Applying Disruptive Technologies More Aggressively}
A valuable way to evaluate potential disruptive technologies is to examine their impact on the fundamental assumptions that are made when building a system using current technology, in order to determine whether future technologies have the potential to change these assumptions, and if yes what the impact of that change is.

\subsection{Power is a First-Class Citizen when Committing to New Technology}
For the last decade, power and thermal management has been of high importance. The entire market focus has moved from achieving better performance through single-thread optimizations, e.g., speculative execution, towards simpler architectures that achieve better performance per watt, provided that vast parallelism exists. The problem with this approach is that it is not always easy to develop parallel programs and moreover, those parallel programs are not always performance portable, meaning that each time the architecture changes, the code may have to be rewritten.

Research on new materials, such as graphene, nanotubes and diamonds as (partial) replacements for silicon can turn the tables and help to produce chips that could run at much higher frequencies and with that may even use massive speculative techniques to significantly increase the performance of single threaded programs. A change in power density vs. cost per area will have an effect on the likelihood of dark silicon.

The reasons why such technologies are not state of the art yet are their premature state of research, which is still far from fabrication, and the unknown production costs of such high performing chips. But we may assume that in 10 to 20 years the technologies may be mature enough or other such technologies will be discovered.

Going back to improved single thread performance may be very useful for many segments of the market. Reinvestment in this field is essential since it may change the way we are developing and optimizing algorithms and code.

Dark Silicon (i.e. large parts of the chip have to stay idle due to thermal reasons) may not happen when specific new technologies ripen. New software and hardware interfaces will be the key for successfully applying future disruptive technologies.

\subsection{Locality of References}
Locality of references is a central assumption of the way we design systems. The consequence of this assumption is the need of hierarchically arranged memories, 3D stacking and more.

But new technologies, including optical networks on die and Terahertz based connections, may reduce the need for preserving locality, since the differences in access time and energy costs to local memory vs. remote storage or memory may not be as significant in future as it is today.

When such new technologies find their practical use, we can expect a massive change in the way we are building hardware and software systems and are organizing software structures.

The restriction here is purely the technology, but with all the companies and universities that work on this problem, we may consider it as lifted in the future.

\subsection{Digital and Analog Computation}
The way how today's computers are built is based on the digital world. This allows the user to get accurate results, but with the drawbacks of cost of time, energy consumption and loss of performance. But accurate results are not always needed. Due to this limitation the production of more efficient execution units, based on analog or even a mix between analog and digital technologies could be possible. Such an approach can revolutionize the way of the programming and usage of future systems. 

Currently the main problem is, that we have no effective way to reason at run time on the amount of inaccuracy we introduces to a system.

\subsection{End of Von Neumann Architecture}
The Von Neumann architecture assumes the use of central execution units that interface with different layers of memory hierarchies. This model, serves as the execution model for more than three decades. But this model is not effective in terms of performance for a given power. 

New technologies like memristors may allow an on-chip integration of memory which in turn grants a very tightly coupled communication between memory and processing unit.

Assuming that these technologies will be mature, we could change algorithms and data structures to fit the new design and thus allow memory-heavy ``in-memory'' computing algorithms to achieve significantly better performance.

We may need to replace the notion of general purpose computing with clusters of specialized compute solution. Accelerators will be ``application class'' based, e.g. for deep learning (such as Google's TPU and Fujitsu's DLU), molecular dynamics, or other important domains.

It is important to understand the usage model in order to understand future architectures/systems.

\subsection{Open Questions and Research Challenges}
The discussion above leads to the following principal questions und research challenges for future HPC hardware architectures and implicitly for software and applications as well:

\begin{itemize}
    \item Impact, if power and thermal will not be limiter anymore (frequency increase vs. many-cores)?
    \item Impact, if Dark Silicon can be avoided?
    \item Impact, if communication becomes so fast so locality will not matter?
    \item Impact, if data movement could be eliminated (and so data locality)?
    \item Impact, if memory and I/O could be unified and efficiently be managed?
\end{itemize}

Evolution of system complexity: will systems become more complex or less complex in future?

\section{Summary of Potential Long-Term Impacts of Disruptive Technologies for HPC Software and Applications}
New technologies will lead to new hardware structures with demands on system software and programming environment and also opportunities for new applications. 

CMOS scaling will require system software to deal with higher fault rate and less reliability. Also programming environment and algorithms may be affected, e.g., leading to specifically adapted approximate computing algorithms.

The most obvious change will result from changes in memory technology. NVM will prevail independent of the specific memristor technology that will win. The envisioned Storage-Class Memory (SCM) will influence system software and programming environments in several ways: 
\begin{itemize}
    \item Memory and storage will be accessed in a uniform way.
    \item Computing will be memory-centric.
    \item Faster memory accesses by the combination of NVM and photonics could lead either to an even more complex or to a shallower memory hierarchy envisioning a flat memory where latency does not matter anymore.
    \item Read accesses will be faster than write accesses, though, software needs to deal with the read/write disparity, e.g., by database algorithms that favour more reads over writes.
    \item NVM will allow in-memory checkpointing, i.e. checkpoint replication with memory to memory operations.
    \item Software and hardware needs to deal with limited endurance of NVM memory.
\end{itemize}

A lot of open research questions arise from these changes for software. 

Full 3D stacking may pose further requirements to system software and programming environments:
\begin{itemize}
    \item The higher throughput and lower memory latency when stacking memory on top of processing may require changes in programming environments and application algorithms.
    \item Stacking specialized (e.g. analog) hardware on top of processing and memory elements lead to new (embedded) high-performance applications.
    \item Stacking hardware accelerators together with processing and memory elements require programming environment and algorithmic changes.
    \item 3D multicores require software optimizations able to efficiently utilize the characteristics of 3rd dimension, .i.e. e.g., different latencies and throughput for vertical versus horizontal interconnects.
    \item 3D stacking may to new form factors that allow for new (embedded) high-performance applications.
\end{itemize}

Photonics will be used to speed up all kind of interconnects – layer to layer, chip to chip, board to board, and compartment to compartment with impacts on system software, programming environments and applications such that:
\begin{itemize}
    \item A flatter memory hierarchy could be reached (combined with 3D stacking and NVM) requiring software changes for efficiency redefining what is local in future.
    \item It is mentioned that energy-efficient Fourier-based computation is possible as proposed in the Optalysys project.
    \item The intrinsic end-to-end nature of an efficient optical channel will favour broadcast/multicast based communication and algorithms.
    \item A full photonic chip will totally change software in a currently rarely investigated manner.
\end{itemize}

A number of new technologies will lead to new accelerators. We envision programming environments that allow defining accelerator parts of an algorithm independent of the accelerator itself. OpenCL and OpenACC are such languages distinguishing ``general purpose'' computing parts and accelerator parts of an algorithm, where the accelerator part can be compiled to GPUs, FPGAs, or many-cores like the Xeon Phi. Such programming environment techniques and compilers have to be enhanced to improve performance portability and to deal with potentially new accelerators as, e.g., neuromorphic chips, quantum computers, in-memory resistive computing devices etc. System software has to deal with these new possibilities and map computing parts to the right accelerator.

Neuromorphic Computing is particularly attractive for applying artificial neural network and deep learning algorithms in those domains where, at present, humans outperform any currently available high-performance computer, e.g., in areas like vision, auditory perception, or sensory motor-control. Neural information processing is expected to have a wide applicability in areas that require a high degree of flexibility and the ability to operate in uncertain environments where information usually is partial, fuzzy, or even contradictory. The success of the IBM Watson computer is an example for such new application possibilities. It is envisioned that neuromorphic computing could help understanding the multi-level structure and function of the brain and even reach an electronic replication of the human brain at least in some areas such as perception and vision.

Quantum Computing potentially solves problems impossible by classical computing, but posts challenges to compiler and runtime support. Moreover, quantum error correction is needed due to high error rates (10-3). Applications of quantum computers could be new encryptions, quantum search, quantum random walk, etc.

Resistive Computing may lead to massive parallel computing based on data-centric and reconfigurable computing paradigms. In memory computing algorithms may be executed on specialised resistive computing accelerators.

Quantum Computing, Resistive Computing as well as Graphene and Nanotube-based computing are still highly speculative hardware technologies.

%% file: 6_vertical_callenges.tex
\chapter{Vertical Challenges: Green ICT, Energy and Resiliency}
\label{sec-vertical}

\section{GreenICT}
The term ``Green ICT'' refers to the study and practice of environmentally sustainable computing. The 2010 estimates put the ICT at 3\% of the overall carbon footprint, ahead of the airline industry~\cite{Smarr2010}. Modern large-scale data centres are already multiple of tens of MWs, on par with estimates for Exascale HPC sites. Therefore, computing is among heavy consumers of electricity and subject of sustainability considerations with high societal impact.

For the HPC sector the key contributors to electricity consumption are the computing, communication, and storage systems and the infrastructure including the cooling and the electrical subsystems. Power usage effectiveness (PUE) is a common metric characterizing the infrastructure overhead (i.e., electricity consumed in IT equipment as a function of overall electricity). Data centre designs taking into consideration sustainability~\cite{Shuja2016} have reached unprecedented low levels of PUE. Many EU projects have examined CO2 emissions in cloud-based services~\cite{ECO2017} and approaches to optimize air cooling~\cite{CoolEmAll2017}.

It is expected that the \mbox{(pre-)Exascale} IT equipment will use direct liquid cooling without use of air for the heat transfer~\cite{DEEP2017}. Cooling with temperatures of the liquid above 45°\,C open the possibility for ``free cooling'' in all European countries and avoid energy cost of water refrigeration. Liquid cooling has already been employed in HPC since the earlier Cray machines and continues to play a key role. The CMOSAIC project~\cite{CMOSAIC2017} has demonstrated two-phase liquid cooling previously shown for rack-, chassis- and board-level cooling to 3D-stacked IC as a way to increase thermal envelopes. The latter is of great interest especially for end of Moore's era where stacking is emerging as the only path forward in increasing density. Many vendors are exploring liquid immersion technologies with mineral-based oil and other material to enable higher power envelopes.

We assert that to reach Exascale performance an improvement must be achieved in driving the Total Power usage effectiveness (TUE) metric~\cite{TUE2017}. This metric highlights the energy conversion costs within the IT equipment to drive the computing elements (processor, memory, and accelerators). As a rule of thumb, in the pre-Exascale servers the power conversion circuitry consumes 25\% of all power delivered to a server. Facility targeting TUE close to one will focus the power dissipation on the computing (processor, memory, and accelerators) elements. 
The CMOS computing elements (processor, memory, accelerators) power dissipation (and therefore also the heat generation) is characterized by the leakage current. It doubles for every 10°\,C increase of the temperature~\cite{Wolpert2012}. Therefore the coolant temperature has influence on the leakage current and may be used to balance the overall energy effectiveness of the data centre for the applications. We expect that the \mbox{(pre-)Exascale} pilot projects, in particular funded by the EU, will address creation and usage of the management software for global energy optimization in the facility~\cite{Li2012}.

Beyond Exascale we expect to have results from the research related to the CMOS devices cooled to low temperatures~\cite{Ellsworth2001} (down to Liquid Nitrogen scale, 77\,K). The expected effect is the decrease of the leakage current and increased conductivity of the metallic connections at lower temperatures. We suggest that an operating point on this temperature scale can be found with significantly better characteristics of the CMOS devices. Should such operating point exist, a practical way to cool such computational device must be found. This may be one possible way to overcome the CMOS technology challenges beyond the feature size limit of \SI{10}{\nm}~\cite{Hu2017}. We suggest that such research funded in Europe may yield significant advantage to the European HPC position beyond Horizon 2020 projects.

The electrical subsystem also plays a pivotal role in Green ICT. Google has heavily invested in renewables and announced in 2017 that their data centres will be energy neutral. However, as big consumers of electricity, HPC sites will also require a tighter integration of the electrical subsystem with both the local/global grids and the IT equipment. Modern UPS systems are primarily designed to mitigate electrical emergencies. Many researchers are exploring the use of UPS systems as energy storage to regulate load on the electrical grid both for economic reasons, to balance the load on the grid or to tolerate the burst of electricity generated from renewables. The Net-Zero data centre at HP and GreenDataNet~\cite{GreenDataNet2017} are examples of such technologies.

\section{Resiliency}
Preserving data consistency in case of faults is an important topic in HPC. Individual hardware components can fail causing software running on them to fail as well. System software would take down the system if it experiences an unrecoverable error to preserve data consistency. At this point the machine (or component) must be restarted to resume the service from a well-defined state.
The traditional failure recovery technique is to restart the whole user application from a user-assisted coordinated checkpoint taken at synchronization point. The optimal checkpoint period is a function of time/energy spent writing the checkpoint and the expected failure rate~\cite{Plank2001}. The challenge is to guess the failure rate, since this parameter is not known in general. If a failure could be predicted, preventive action such as the checkpoint can be taken to mitigate the risk of the pending failure.

No deterministic failure prediction algorithm is known. However, collecting sensor data and Machine Learning (ML) on this sensor data yields good results~\cite{Turnbull2003}. We expect that the \mbox{(pre-)Exascale} machine design especially funded by the EU will incorporate sufficient sensors for the failure prediction and monitoring. This may be a significant challenge, as the number of components and the complexity of the architecture will increase. Therefore, also the monitoring data stream will increase, leading to a fundamental Big Data problem just to monitor a large machine.
We see this monitoring problem as an opportunity for the EU funding of fundamental research in ML techniques for real-time monitoring of hardware facilities in general. The problem will not yet be solved in the next round of the \mbox{(pre-)Exascale} machine development. Therefore, we advocate a targeted funding for this research to extend beyond Horizon 2020 projects.
The traditional failure recovery scheme with the coordinated checkpoint may be relaxed if fault-tolerant communication libraries are used~\cite{Fagg2000}. In that case the checkpoints do not need to be coordinated and can be done per node when the computation reaches a well-defined state. When million threads are running in a single scalable application, the capability to restart only a few communicating threads after a failure is important.

The non-volatile memories may be available for the checkpoints; it is a natural place to dump the HBM contents. We expect these developments to be explored on the time scale of \mbox{(pre-)Exascale} machines. It is clear that the system software will incorporate failure mitigation techniques and may provide feedback on the hardware-based resiliency techniques such as the ECC and Chipkill. The software-based resiliency has to be designed together with the hardware-based resiliency. Such design is driven by the growing complexity of the machines with a variety of hardware resources, where each resource has its own failure pattern and recovery characteristics.

On that note the compiler assisted fault tolerance may bridge the separation between the hardware-only and software-only recovery techniques~\cite{Herault2015}. This includes automation for checkpoint generation with the optimization of checkpoint size~\cite{Plank1995}. More research is needed to implement these techniques for the Exascale and post-Exascale architectures with the new levels of memory hierarchy and increased complexity of the computational resources. We see here an opportunity for the EU funding beyond the Horizon 2020 projects.

Stringent requirements on the hardware consistency and failure avoidance may be relaxed, if an application algorithm incorporates its own fault detection and recovery. Fault detection is an important aspect, too. Currently, applications rely on system software to detect a fault and bring down (parts of) the system to avoid the data corruption. There are many application environments that adapt to varying resource availability at service level---Cloud computing works in this way. Doing same from within an application is much harder. Recent work on the ``fault-tolerant'' message-passing communication moves the fault detection burden to the library, as discussed in the previous section. Still, algorithms must be adopted to react constructively after such fault detection either by ``rolling back'' to the previous state (i.e. restart from a checkpoint) or ``going forward'' restoring the state based on the algorithm knowledge. The forward action is subject of a substantial research for the \mbox{(pre-)Exascale} machines and typically requires algorithm redesign. For example, a possible recovery mechanism is based on iterative techniques exploited in Linear Algebra operations~\cite{Langou2007}.

The Algorithm Based Fault Tolerance (ABFT) may also use fault detection and recovery from within the application. This requires appropriate data encoding, algorithm to operate on the encoded data and the distribution of the computation steps in the algorithm among (redundant) computational units~\cite{Huang1984}. We expect these aspects to play a role with NMP. The ABFT techniques will be required when running applications on machines where the strong reliability constraint is relaxed due to the subthreshold voltage settings. Computation with very low power is possible~\cite{Gupta2015} and opens a range of new ``killer app'' opportunities. We expect that much of this research will be needed for post-Exascale machines and therefore is an opportunity for EU funding beyond the Horizon 2020 projects.


%% file: 7_system_software.tex
\chapter{System Software and Programming Environment}
\label{sec-system}

\section{Scope}
The system software is the part of the HPC software stack that is optimized by the HPC vendor and managed by the system's operator, and it includes the Operating System (OS), cluster management tools, distributed file systems, and resource management software (job scheduler). It is essential for an operational HPC system to have an efficient system software stack below the end user's application.  The programming environment comprises the development tools used to build the end user's application (compilers, IDEs, debuggers, and performance analysis tools) along with the associated abstractions (e.g. programming models), as well as the runtime components: libraries and runtime systems. Workflow management tools and commonly pre-installed application libraries such as BLAS and LAPACK are also in the scope of this section.

\section{Current Research Trends}

\subsection{Sustained Increases in System Complexity, Specialization, and Heterogeneity}
An important role of the system software and programming environment is to provide the application developers with common standardized abstractions. Such abstractions greatly improve programmer productivity and portability across systems. Today's dominant abstractions include Fortran, C, MPI, POSIX-style file systems, threads and locking, which are all relatively low-level. By 2030, disruptive technologies may have forced the introduction of new and currently unknown low-level abstractions that are very different from these, and this topic is addressed below. Nevertheless, today's abstractions will continue to evolve incrementally and probably increase in their level of abstraction, and will continue to be used well beyond 2030, since scientific codebases have very long lifetimes, on the order of decades. Developers are unwilling to adopt a new programming language or API until they are convinced that it will be supported for a long time.

Continuous CMOS scaling and 3D stacking are pointing towards increasingly complex hardware. High-bandwidth (3D integrated) and non-volatile memories (memristors, etc.) will lead to different memory hierarchies. Increasing performance per watt demands accelerators (many-core, GPU, vector, dataflow, and their successors), heterogeneous processors (big and small cores) and potentially reconfigurable logic (FPGA).  The choice of processor cores will likely become increasingly heterogeneous (within a system) and varied (across systems). Certain techniques for energy efficiency (near threshold, DVFS, energy-efficient interconnects) increase timing variability among the processes in an HPC application.  Virtualization, if adopted, will also increase timing variability. In addition to hardware complexity, execution environments will also increase in complexity, through interactive use (which will require workloads to adjust to dynamically variable numbers of nodes, cores, memory capacities, and so on).  

Hiding or mitigating this increasingly complex and varied hardware requires more and more intelligence across the programming environment.  Manual optimization of the data layout, placement, and caching will become uneconomic and time consuming, and will, in any case, soon exceed the abilities of the best human programmers.  There needs to be a change in mentality from programming ``heroism'' towards trusting the compiler and runtime system (as in the move from assembler to C/Fortran). Automatic optimization requires advanced techniques in the compiler and runtime system.  In the compiler, there is opportunity for both fully automated transformations and the replacement of manual refactoring by automated program transformations under the direction of human programmers (e.g. Halide [14]).  Advanced runtime and system software techniques, e.g., task scheduling, load balancing, malleability, caching, energy proportionality are needed. 

Increasing complexity also requires an evolution of the incumbent standards such as OpenMP, in order to provide the right programming abstractions.  There is as yet no standard language for GPU-style accelerators (CUDA is controlled and only well supported by a single vendor and OpenCL provides portability). Domain-specific languages (e.g. for partial differential equations, linear algebra or stencil computations) allow programmers to describe the problem in terms much closer to the original scientific problem, and they provide greater opportunities for automatic optimization.  In general there is a need to raise the level of abstraction. In some domains (e.g. embedded) prototyping is already done in a high-level environment similar to a DSL (Matlab), but the implementation still needs to be ported to a more efficient language. 

A different opinion expressed the need to continue to provide a (simple) cost model, in similar terms to the correspondence of the programming language C to a von Neumann CPU, so that programmers could have an intuition about the effect on performance.  There is scope for ways to express non-functional properties of software, as commonly done in embedded systems, in order to trade various metrics, e.g., performance vs. energy or accuracy vs. cost, both of which may become more relevant with near threshold, approximate computing or accelerators (quantum/neuromorphic). 

There is a need for global optimization across all levels of the software stack, including OS, runtime system, application libraries, and application.  Examples of global problems that span multiple levels of the software stack include a) support for resiliency (system/application-level checkpointing), b) data management transformations, such as data placement in the memory hierarchy, c) minimising energy (sleeping and controlling DVFS), d) constraining peak power consumption or thermal dissipation, and e) load balancing. Different software levels have different levels of information, and must cooperate to achieve a common objective subject to common constraints, rather than competing or becoming unstable.  

\subsection{Complex Application Performance Analysis and Debugging}
Performance analysis and debugging are particularly difficult problems beyond Exascale.  The problems are two-fold. The first problem is the enormous number of concurrent threads of execution (millions), which provides a scalability challenge (particularly in performance tools, which must not unduly affect the original performance) and in any case there will be too many threads to analyse by hand.  Secondly, there is an increasing gap between (anomalous) runtime behaviour and the user's changes in the source code needed to fix it, due to libraries, runtime systems and system software, and potentially disaggregated resources, that the application programmer would know little or nothing about.

Spotting anomalous behaviour, such as the root cause of a performance problem or bug, will be a \emph{big data} problem, requiring techniques from data mining, clustering and structure detection, as well as high scalability through summarized data, sampling and filtering and special techniques like spectral analysis.  As implied above, the tools need to be interoperable with programming abstractions, so that problems in a loop in a library or dynamic scheduling of tasks can be translated into terms that the programmer can understand.

\section{Potential Implications of Disruptive Technologies}

\subsection{Disruptive Hardware Models of Computation}
Many of the fundamental abstractions used in computing in general, and high-performance computing in particular, have evolved steadily since their introduction decades ago:
\begin{itemize}
    \item Fortran programming language (introduced in the 1950s)
    \item C programming language (1973)
    \item Sockets communications (1983)
    \item File system in terms of files, directories, POSIX API (1988)
    \item POSIX threads, locks, condition variables, etc. (1988)
    \item MPI message passing API (1994)
    \item OpenMP (1997)
\end{itemize}

An important question is whether and to what degree these fundamental abstractions may be broken by new technologies, especially disruptive technologies.  The above abstractions have stood the test of time and will endure in HPC, given the long lifetimes of scientific codebases.  Nevertheless, certain disruptive technologies on the horizon have the potential to challenge certain basic assumptions. 

\subsection{Convergence Between Storage and Memory}
All existing computing systems make a strong distinction between memory and storage. Random-access memory is fast (in both bandwidth and latency), it is byte addressable and randomly accessible by the processor, it has high cost-per-bit, and its contents are volatile. Storage is slow, in both bandwidth and latency, data is accessed through at I/O device in 512-byte (or larger) blocks, it has lower cost-per-bit, and the data is persistent. 

This (hardware) correspondence between persistence on the one hand and speed, addressability and granularity on the other is the basis for the different roles of memory and storage. Temporary data structures are held in memory, and manipulated using random accesses. Data that must be persistent and/or passed among programs is serialized to a file as a byte stream. 

Storage-class memory, including HPE's Persistent Memory, has similar speed, addressability and cost as DRAM with the non-volatility of storage.  In the context of HPC, such memory can reduce the cost of checkpointing or eliminate it entirely.  There is also work on persistent objects, e.g., NV-Heaps, and further work is needed.

\subsection{Neuromorphic, Resistive and Quantum Computing}
The adoption of neuromorphic, resistive computing and/or quantum computing may have a dramatic effect on the system software and programming model.  It is currently unclear whether it will be sufficient to offload tasks, as on GPUs, or whether more dramatic changes will be needed.